\documentclass{JHEP3}
\usepackage{epsfig}\input{epsf}
\usepackage{amssymb,latexsym}
\usepackage{graphicx}

\newcommand{\fig}[1]{Fig.~\ref{#1}}
\newcommand{\beq}{\begin{equation}}

\newcommand{\eeq}{\end{equation}}

\newcommand{\ba}{\begin{array}}

\newcommand{\ea}{\end{array}}

\newcommand{\bea}{\begin{eqnarray}}

\newcommand{\eea}{\end{eqnarray}}

\def \ang {\frac{2\pi}{k}}
\def \ci{\cite}

\def\p{\partial}

\def\bL{\bar L}
\def \bm{\bar m}
\def \br{\bar r}
\def\tr{{\rm Tr\;}}

\def\mod{{\rm mod}}
\def \R {\mathbb{R}}
\def \Z {\mathbb{Z}}
\def \N {\mathbb{N}}

\def \ha {{1 \over 2}}
\def\ch{{\rm{ch}}}

\def\CC{{\cal C}}
\def\DD{{\cal D}}

\def\JJ{{\cal J}}

\def\NN{{\cal N}}
\def\OO{{\cal O}}

\def\QQ{{\cal Q}}

\def\SS{{\cal S}}

\def\ZZ{{\cal Z}}
\def\Zi{{\Z}}
\def\llangle{{\langle \langle}}
\def\rrangle{{\rangle \rangle}}

\preprint{\\ \\}

\title{D-branes and extended characters in $SL(2,\R)/U(1)$}

\author{Angelos Fotopoulos$^a$
\footnote{Also at the Centre de Physique Theorique, Ecole Polytechnique,
Palaiseau 91128, France.} , Vasilis Niarchos$^b$ and Nikolaos Prezas$^c$
\\
\\$^a$ Theoretical Physics Group, Imperial College,
\\Prince Consort Road, London SW7 2BZ, UK.
\\{\tt angelos.fotopoulos@cpht.polytechnique.fr}
\\
\\$^b$ Department of Physics, University of Chicago,
\\ 5640 S. Ellis Av. Chicago, IL 60637, USA.
\\{\tt vniarcho@theory.uchicago.edu}
\\
\\$^c$ Institut f\"ur Physik, Humboldt-Universit\"at zu Berlin,
\\ Newtonstra\ss e 14, D-12489 Berlin, Germany.
\\{\tt prezas@physik.hu-berlin.de}
}

\preprint{
Imperial/TP/3-04/11
\\EFI-04-18
\\HU-EP-04/26
\\hep-th/0406017}

\abstract{We present a detailed study of D-branes in
the axially gauged $SL(2,\R)_k/U(1)$ coset conformal field theory for integer level $k$.
Our analysis is based on the modular bootstrap approach and utilizes
the extended $SL(2,\R)/U(1)$ characters and
the embedding of the parafermionic coset algebra in the $\NN=2$ superconformal algebra.
We propose three basic classes of boundary states corresponding to D0-, D1- and D2-branes.
We verify that these boundary states satisfy the Cardy consistency conditions
and discuss their physical properties. The D0- and D1-branes agree with those
found in earlier work by Ribault and Schomerus using different methods
(descent from the Euclidean $AdS_3$ model). The D2-branes are new. They
are not, in general, space-filling
but extend from the asymptotic circle at infinity
up to a minimum distance $\rho_{min}\geq 0$ from the tip of the cigar.}

\begin{document}

\section{Introduction}\label{INTRO}

Understanding the D-brane spectrum of string theory is a problem
of paramount importance for well-known reasons.
For string backgrounds that admit an exact conformal
field theory description, this spectrum can be uncovered by utilizing
techniques from boundary conformal field theory.
In this formalism, the physical properties of a D-brane are encoded in a ``boundary
state''. The boundary states are generalized coherent states on the Hilbert
space of first quantized closed strings and
they are specified by a set of gluing conditions between
the left- and right-moving chiral algebra generators.

In general, the classification of the full set of boundary states
is a difficult open problem. Boundary states need to obey various
consistency conditions, which are mostly understood in the
case of rational conformal field theories. An important condition of this sort
comes from worldsheet duality and is known
as the Cardy consistency condition \cite{Cardy:ir}. This condition is the following statement.
The general amplitude between a pair of boundary states is
a tree-level cylinder amplitude in closed string theory describing the
emission, propagation and absorption of a closed string, and it
can be written in terms of the characters that appear in the torus
partition sum. After a worldsheet duality transformation the cylinder amplitude
can be restated
as a one-loop open string amplitude and one should be able to express
it in terms of open string characters with positive integer multiplicities.
There is a general way to solve this set of constraints in rational
conformal field theories \cite{Cardy:ir}.
The resulting boundary states are in one-to-one correspondence
with the conformal blocks of the theory and can be written
as appropriate linear combinations of a set of coherent states (Ishibashi states)
satisfying a prescribed set of gluing conditions \cite{Ishibashi:1988kg}.
The coefficients of this expansion are dictated by the $\SS$-modular
transformation matrix elements, and the fact that the Cardy consistency condition
is satisfied is a consequence of the Verlinde formula \cite{Verlinde:sn}.

The above construction, usually referred to as the Cardy ansatz,
has been employed with great success in a variety of solvable
rational conformal field theories,
like Wess-Zumino-Witten (WZW) models based on compact Lie groups
and their cosets (c.f.\ \cite{Schomerus:2002dc} for a review and further references).
For example, D-branes in the parafermion
theory $SU(2)/U(1)$ were studied in detail in \cite{Maldacena:2001ky} with special emphasis
on the interplay between their geometrical interpretation
and their algebraic conformal field theory treatment.
It was found that there are two types of parafermionic branes,
which were called A-type and B-type in analogy to the branes of $\NN=2$
theories \cite{Ooguri:1996ck}.
The A-type includes D0-branes sitting at special points
on the boundary of the disc-shaped target space and D1-branes stretching between them.
The B-type contains (mostly unstable) D0- and D2-branes
located at the center of the disc.

The Cardy conditions are supplemented by
a further set of ``sewing'' constraints \cite{Lewellen:1991tb, Cardy:tv},
which again are better understood only in rational theories.
In more general (possibly non-rational) conformal field theories it is mostly
unclear how to implement the above constraints in a useful way.
In certain cases, however, considerable progress has been made in the last
few years. The Cardy consistency conditions and disk 1-point function factorization constraints
have been analyzed successfully in
the bosonic Liouville theory \cite{Fateev:2000ik, Teschner:2000md, Zamolodchikov:2001ah}
and the $\NN=1$ Liouville theory \cite{Fukuda:2002bv,Ahn:2002ev}. Similar work
on the $AdS_3$ WZW model has appeared in \cite{Giveon:2001uq, Parnachev:2001gw,
Lee:2001gh, Ponsot:2001gt, Ribault:2002ti}.

In this paper we would like to study D-branes in a non-compact, non-rational
version of the $SU(2)/U(1)$ parafermion theory.
This is the $SL(2,\R)/U(1)$ coset conformal field theory
described by a gauged WZW model
with axial $U(1)$ gauging \cite{Bardakci:1987ee, Altschuler:1987zb,
Rocek:1991vk, Mandal:1991tz, Elitzur:cb, Giveon:1991sy, Dijkgraaf:1991ba}.
The Lorentzian version of this theory
describes strings propagating on a two-dimensional black hole \cite{Witten:1991yr}.
Its Euclidean counterpart, with a cigar-like target space geometry,
has a great wealth of applications in string theory;  it is
a building block for exact string backgrounds with $\NN=4$ superconformal
symmetry \cite{Kounnas:1993ix, Antoniadis:1994sr}, it
appears in the holographic description of
Little String Theories \cite{Giveon:1999zm, Giveon:1999px},
and it plays an important role in
the worldsheet description of Calabi-Yau compactifications near
singularities \cite{Ooguri:1995wj}.
At the same time, it provides a useful testing ground
for ideas that might generalize to other non-rational conformal field theories.

The range of applications of $SL(2,\R)/U(1)$ renders a
good understanding of its D-brane spectrum obligatory.
Further motivation
comes from the recent re-interpretation of matrix models
for non-critical strings as effective field theories living on Liouville branes
\cite{McGreevy:2003kb, Klebanov:2003km, McGreevy:2003ep, Martinec:2003ka}.
It would be very interesting to see if a similar story
can be said for the $SL(2,\R)/U(1)$ theory and
the corresponding matrix model \cite{Kazakov:2000pm}.

The exact algebraic analysis of D-branes in $SL(2,\R)/U(1)$
is hampered, however, by the usual difficulties that
one encounters in non-rational conformal field theories.
Namely, the existence of an infinite (and usually continuous) set
of primaries in conjunction with the absence of an analogue of the
Verlinde formula, implies that one should go beyond the
Cardy construction. Nevertheless, the recent progress in bosonic and $\NN=1$ Liouville theory
\cite{Fateev:2000ik, Teschner:2000md, Zamolodchikov:2001ah, Fukuda:2002bv,Ahn:2002ev}
suggested that the naive extension of the Cardy ansatz
to non-rational theories may still yield consistent boundary states.
This idea was further explored in \cite{Eguchi:2003ik, Ahn:2003tt},
where it was used to determine branes in the $\NN=2$ Liouville theory.
Our goal in this paper is to apply a similar modular bootstrap method to the
$SL(2,\R)/U(1)$ theory. We formulate boundary states using the Cardy ansatz
and subsequently verify that all possible overlaps between these boundary states
satisfy the Cardy consistency conditions.

Part of this work has been also motivated
by a recent study on extended characters
for cosets $SL(2,\R)_k/U(1)$ with
integer level $k$ \cite{Israel:2004xj}.
These characters are direct analogues of the
extended characters of the  $\NN=2$ superconformal algebra,
which were the basic ingredient in the modular bootstrap analysis
of \cite{Eguchi:2003ik}. The extended characters are useful because they
exhibit simple modular transformation properties and reorganize the
representations of the theory in a compact way.
In some sense, this feature ``rationalizes'' the boundary state analysis
and exhibits various similarities with the corresponding analysis of the
$SU(2)/U(1)$ case.

D-branes on the cigar have been studied previously by several authors.
A semiclassical analysis based on the Dirac-Born-Infeld (DBI) action
appeared in \cite{Fotopoulos:2003vc}. More recently, \cite{Ribault:2003ss}
proposed a set of boundary states for $SL(2,\R)/U(1)$,
departing from a boundary conformal field theory analysis of
the $H_3^+$ (Euclidean $AdS_3$) model  \cite{Ponsot:2001gt}.
The latter is an Euclidean continuation of the
$SL(2,\R)$ WZW model and it reduces to $SL(2,\R)/U(1)$ upon gauging a $U(1)$ subgroup.
Extending standard techniques for the construction of boundary
states in rational coset models \cite{Fredenhagen:2001kw},
the authors of \cite{Ribault:2003ss} obtained $SL(2,\R)/U(1)$ boundary states from the
boundary states of \cite{Ponsot:2001gt} by descent.
Three families of branes were obtained in this way:
D0-, D1-, and D2-branes. The analysis of the respective
boundary states vindicated the semiclassical expectations and
several of the corresponding cylinder amplitudes were shown to satisfy the Cardy consistency
condition.

Our approach can be considered as complementary
to that of \cite{Ribault:2003ss} and is relevant for cosets $SL(2,\R)_k/U(1)$
at fractional level $k$. We find four classes of branes.
Two of them are the same as the D0- and D1-branes of \cite{Ribault:2003ss}.
For the D0-branes, in particular, we find
only one boundary state instead of the infinite multitude that appears in
\cite{Ribault:2003ss}. The authors of that paper noticed several puzzling
features for most of these branes and conjectured that only one of them
is consistent. This single D0-brane is precisely the one
we obtain with our methods.
Moreover, we find D1-branes and D2-branes that differ from those
of  \cite{Ribault:2003ss}. This extra class of D1-branes, which arises
from a direct application of the Cardy ansatz, exhibits
some puzzling semiclassical features and it is not entirely
clear if it satisfies further factorization constraints.
Our D2-branes cover the cigar only partially and are marginally stable.
They correspond to a set of extra solutions of the DBI action
\cite{Fotopoulos:2003vc}, which were not considered in
\cite{Ribault:2003ss}.
The D2-branes of \cite{Ribault:2003ss} cover the whole cigar
and can be contrasted with a
different set of  D2-branes in our construction, which turn out to be
inconsistent. Comments on this comparison appear in
section 5.

The layout of this paper is as follows. In Section 2,
we discuss several aspects of the  $SL(2,\R)/U(1)$
theory that will be relevant for our subsequent analysis.
We present the standard and extended coset characters and
briefly review the embedding of the $SL(2,\R)/U(1)$ parafermion algebra
into the $\NN=2$ superconformal algebra. This will be useful in the definition of the
coset Ishibashi states later in the paper. In addition, we discuss in detail
the torus partition function of the coset in terms of
the standard and extended characters.
In Section 3 we review some useful generalities on the Cardy
ansatz and the modular bootstrap method, lay out our notation, and then apply everything to
the $SL(2,\R)/U(1)$ case. First, we
define the appropriate coset Ishibashi states and then
consider various boundary states that follow from
the application of the Cardy ansatz.
In Section 4 we consider the Cardy consistency conditions.
We analyze the overlaps between our boundary states
and derive the corresponding open string densities and multiplicities.
We find that the majority of the boundary states of Section 3
satisfy the Cardy consistency conditions. Only one class is found to
be inconsistent and we discuss possible alternatives.
In Section 5 we summarize the results of Sections 3 and 4,
analyze the properties of each of our boundary states and comment on their geometrical interpretation.
More specifically, we discuss the corresponding
wavefunctions, the open string spectrum, and
the relation of our results with semiclassical
expectations. We also compare our findings to those of
\cite{Ribault:2003ss}. We conclude in
Section 6 with a brief discussion of our results and various ideas for
further applications and open problems. There are also three Appendices.
In Appendix A we demonstrate in detail how the coset partition sum
can be written in terms of the standard and extended  $SL(2,\R)/U(1)$
characters.
In Appendix B we derive the $\SS$-modular transformation properties
for the extended identity characters of the coset. This is a small add-on to
the analysis of \cite{Israel:2004xj}.
Finally, in Appendix C we provide several explicit expressions for
the extended identity characters, which are useful in the
analysis of the open string spectrum of the D0-brane.

\

\noindent {\bf Note added:} While this paper was being prepared for submission,
two interesting preprints appeared in the hep-th archive discussing
related issues. In \cite{Israel:2004jt} the
D-brane spectrum of the supersymmetric coset and its $\NN=2$ Liouville mirror
is analyzed
by extending the methods
of \cite{Ribault:2003ss}. In \cite{Ahn:new} conformal boundary conditions
for the $\NN=2$ Liouville theory are studied using  conformal
and modular bootstrap methods.

\section{$SL(2,\R)/U(1)$ characters and the torus partition sum}\label{ECC}

\subsection{Representations and standard characters}

The $SL(2,\R)_k/U(1)$ coset is a two-dimensional
conformal field theory with central charge
\beq
\label{central}
c_{cs}=\frac{3k}{k-2}-1
~.
\eeq
It can be obtained from the $SL(2,\R)$ WZW model at level
$k$ by gauging a $U(1)$ subgroup of $SL(2,\R)$. Different
choices of the $U(1)$ subgroup yield different
target space geometries. A non-compact $U(1)$ subgroup results in a manifold
with Lorentzian signature, the 2D black hole \cite{Witten:1991yr}.
A compact $U(1)$ leads instead to a manifold with Euclidean
signature. Moreover, for a given $U(1)$ subgroup there are
two possible anomaly free $U(1)$ gauge
symmetries: the axial $g \rightarrow hgh$, and the vector $g\rightarrow hgh^{-1}$,
where $g(z,\bar z)$ are group elements of $SL(2,\R)$ and $h$ an element
of the $U(1)$ subgroup that is being gauged.

For a compact $U(1)$ the axial gauging produces a two-dimensional manifold with the
cigar geometry:
\beq
\label{cigar}
ds^2=d\rho^2+\tanh^2 \rho \;d\theta^2, ~ ~ e^{-\Phi}=e^{-\Phi_0} \cosh \rho
~.
\eeq
$\rho$ is a positive real number and $\theta$ is periodic, i.e.\ $\theta \sim
\theta +2\pi$. The tip of the cigar lies at $\rho=0$, where
the dilaton reaches its maximum value $\Phi=\Phi_0$.
In the asymptotic region $\rho \rightarrow \infty$ we obtain
the geometry of a cylinder in the presence of a linear dilaton with
background charge $Q^2=\frac{2}{k-2}$.

The vector gauging yields a two-dimensional manifold with trumpet geometry
\beq
\label{trumpet}
ds^2=d\rho^2+\coth^2 \rho \;d\psi^2, ~ ~ e^{-\Phi}=e^{-\Phi_0} \sinh \rho
~.
\eeq
In this case, $\psi$ is a periodic coordinate with period $\frac{2\pi}{k}$.
Unlike the cigar, the trumpet geometry is singular and the dilaton
diverges at $\rho=0$. It is believed, however, that the two backgrounds
are T-dual to each other when one
considers world-sheet instanton corrections to the trumpet geometry
\cite{ Giveon:1999px, Kiritsis:1991zt}.
In this paper, we focus on the cigar geometry.

As a non-compact analog of the $SU(2)/U(1)$ parafermion model,
the coset $SL(2,\R)/U(1)$ defines a non-compact parafermion algebra,
which is known to possess the following unitary
representations\footnote{Corresponding
unitary representations can also be found in $SL(2,\R)$.} and characters
\cite{Griffin:1990fg, Sfetsos:1991wn, Bakas:1991fs, Itoh:mt, Pakman:2003kh,Ribault:2003ss}:
\begin{enumerate}
\item[(i)] $Identity$ $representation$ $\DD^I_{j,r}$
: $(j=0,\frac{k}{2}$, $r\in \Z$)
\beq
\label{charid}
\lambda^I_r(\tau)=\eta(\tau)^{-2}
q^{-\frac{1}{4(k-2)}}q^{|r|+\frac{r^2}{k}}
\bigg [ 1+\sum_{s=1}^{\infty}(-1)^s q^{\frac{1}{2}(s^2+(2|r|+1)s-2|r|)}
(1+q^{|r|})\bigg] ~,
\eeq
$with$ $charge$ $J_0^3=r$,
\item[(ii)] $Discrete$ $representations$ $\DD^{\pm}_{j,r}$
: $(0<j<\frac{k}{2}$, $r\in \Z)$
\beq
\label{chardiscr}
\lambda^d_{j,r}(\tau)=\eta(\tau)^{-2}
q^{-\frac{(j-\frac{1}{2})^2}{k-2}}q^{\frac{(j+r)^2}{k}}
\sum_{s=0}^{\infty}(-1)^s q^{\frac{1}{2}s(s+2r+1)},
\eeq
$with$ $charge$ $J_0^3=j+r$,
\item[(iii)] $Continuous$ $representations$ $\CC_{s,\alpha+r}$
: $(s \in \R$,$0\leq \alpha<1$, $r\in \Z)$
\beq
\label{charconti}
\lambda^c_{\frac{1}{2}+is,\alpha+r}=\eta(\tau)^{-2}
q^{\frac{s^2}{k-2}}q^{\frac{(\alpha+r)^2}{k}}.
\eeq
$with$ $charge$ $J_0^3=\alpha+r$.
\end{enumerate}
The above characters are defined, as usual, by
the traces $\tr q^{L_0-\frac{c_{cs}}{24}}$ over the corresponding coset modules
and $q=e^{2\pi i \tau}$. The Dedekind eta function reads
$\displaystyle \eta(\tau)=q^{\frac{1}{24}}\prod_{n=1}^{\infty} (1-q^n)$
and $J_0^3$ is the $U(1)$ generator of the parent $SL(2,\R)$ theory,
which is being gauged.

The torus partition sum of the axially-gauged (cigar) coset can be written
explicitly in terms of the above continuous and discrete characters. We will
return to this issue later in this Section to discuss it further and various important
details will be provided in Appendix A.
For now, we make a small detour to remind the reader of the well-known connection
between the $SL(2,\R)/U(1)$ parafermion algebra and the $\NN=2$
superconformal algebra. This will help establish our notation and later
define and analyze the extended coset characters, which will be the cornerstone of our
boundary conformal field theory analysis.

\subsection{Embedding the $SL(2,\R)/U(1)$ parafermion theory in $\NN=2$}

One way to obtain the Virasoro characters appearing in eqs.\
(\ref{charid}), (\ref{chardiscr}), (\ref{charconti})
is by embedding the $SL(2,\R)/U(1)$ parafermion algebra into
an $\NN=2$ superconformal algebra.
This embedding will be central in the following
considerations, so we take a moment to remind the reader of a few
relevant details. For further discussion see for example  \cite{Lykken:1988ut, Dixon:1989cg,
Bakas:1991fs}.

On the level of currents
there is a well-known connection between $SL(2,\R)$
and the $\NN=2$ superconformal algebra. Indeed, we may introduce
a free time-like chiral boson $Y$ and write the $SL(2,\R)$ currents
in the form
\beq
\label{JYY}
J^3=-i \sqrt\frac{k}{2} \p Y, ~ ~ ~ J^{\pm}=\sqrt {k} \psi^{\pm}
e^{\pm i \sqrt{\frac{2}{k}} Y} ~.
\eeq
In this definition $Y(z)$ is a canonically normalized scalar with the
OPE
\beq
\label{YOPE}
Y(z)Y(w) \sim \log(z-w)
~.
\eeq
Then, we can decompose an affine $SL(2,\R)$
primary field $\Phi_{j,m}$
into an $SL(2,\R)/U(1)$ parafermion
$\hat{\Phi}_{j,m}$ times a $U(1)$ primary by writing
\beq
\label{PYY}
\Phi_{j,m}=e^{i \sqrt{\frac{2}{k}} m Y} \hat{\Phi}_{j,m}
~.
\eeq
By replacing the $time$-like boson $Y$ with a new $space$-like
boson $\phi$, again canonically normalized as
$\phi(z) \phi(w) \sim -\log(z-w)$, we are able to construct
the currents of an $\NN=2$ superconformal algebra
\beq
\label{N2currents}
J=i\sqrt{\hat c} \p \phi, ~ ~ G^{\pm}=\sqrt{2\hat c} \psi^{\mp} e^{\pm \frac{i}{\sqrt{\hat c}}\phi}
~, ~ ~ T=T_{SL(2,\R)/U(1)}-\frac{1}{2}(\p \phi)^2 ~.
\eeq

For $SL(2,\R)$ primaries in the $\hat {\cal D}^+_j$ discrete representation
the $J_0^3$ charge is $m=j+r$ with $r\geq 0$.
Then, corresponding $\NN=2$ primary fields can be constructed as
\beq
\label{N2primaries}
Z_{j,j+r}=e^{i\frac{2(j+r)}{k-2}\frac{1}{\sqrt {\hat c}}\phi} \hat{\Phi}_{j,j+r}
~.
\eeq
In each of the above relations
\beq
\label{hatcdef}
\hat{c}= \frac{k}{k-2}
~
\eeq
denotes the $SL(2,\R)$ central charge at level $k$ divided by 3.
The dimension and $U(1)_R$ charge of the $Z_{j,j+r}$ operators are
\beq
\label{dimcharge}
h_{j,j+r}=\frac{-j(j-1)+(j+r)^2}{k-2}, ~ ~ Q_{j+r}=\frac{2(j+r)}{k-2}
~.
\eeq

In order to account for the full discrete spectrum of the $\NN=2$ algebra,
we also have to consider affine primaries belonging in the  $\hat {\cal D}^-_j$
representation of $SL(2,\R)$ or, equivalently, affine descendants
of $\hat {\cal D}^+_j$.
In that case we have $m=j+r+1$ with $r<0$ and
the conformal weight and $U(1)_R$ charge of
the corresponding $\NN=2$ representation reads \cite{Israel:2004xj}
\beq
\label{dimcharge1}
h_{j,j+r}=\frac{-j(j-1)+(j+r)^2}{k-2}-r-\frac{1}{2}, ~ ~ Q_{j+r}=\frac{2(j+r)}{k-2}-1
~.
\eeq

For the principal continuous series of $SL(2,\R)$ with $j=\frac{1}{2}+is,\; s\geq 0$
and $2m \in \Z$, the corresponding $\NN=2$ representations have
\beq
\label{dimchargecont}
h_{j,m}=\frac{-j(j-1)+m^2}{k-2}, ~ ~ Q_{m}=\frac{2m}{k-2}
~.
\eeq

The above discussion makes clear that there is an intimate
connection between $\NN=2$ and $SL(2,\R)/U(1)$ representations. Indeed,
the unitary highest weight representations of the
$\NN=2$ algebra and their characters fall also into three classes
\cite{Dobrev:1986hq, Kiritsis:1986rv}
(it is enough to consider only the NS sector of the $\NN=2$ theory)\footnote{In
\cite{Eguchi:2003ik}
these representations are called graviton, massless matter and massive
respectively.}:
\begin{enumerate}
\item[(i')] $Identity$ $(vacuum)$ $representation$: $(h=Q=0)$
\beq
\label{graviton}
\ch_I(\tau,z)=q^{-(\hat c-1)/8}\frac{1-q}{(1+yq^{1/2})(1+y^{-1}q^{1/2})}
\frac{\theta_3(\tau,z)}{\eta^3(\tau)}
~.
\eeq
This representation and its spectral flowed analogs to be discussed in
the next subsection have two null vectors. The corresponding $SL(2,\R)/U(1)$ representations
in (\ref{charid}) have $j=0$ and $j=\frac{k}{2}$.
\item[(ii')] $Discrete$ $representations$:
$(h_{j,\pm j}=\frac{|Q_{\pm j}|}{2}$, $0<j<\frac{k}{2})$
\beq
\label{massless}
\ch_d(h_{j,\pm j},Q_{\pm j};\tau,z)=q^{h_{j,\pm j}-(\hat c-1)/8} y^{Q_{\pm j}}
\frac{1}{1+y^{\pm}q^{1/2}}\frac{\theta_3(\tau,z)}{\eta^3(\tau)}
~.
\eeq
These representations are built upon the primary vertex operators $Z_{j,\pm j}$ and they
have one null vector. The states $Z_{j,j}$
and $Z_{j,-j}$ are respectively chiral and anti-chiral. The corresponding $SL(2,\R)/U(1)$ representations
are given by the discrete series $\DD^{\pm}_{j,r}$. The extra index $r$ coincides
with the $\NN=2$ spectral flow parameter (see below).
\item[(iii')] $Continuous$ $representations$: $(h_{j,m}>\frac{|Q_m|}{2})$
\beq
\label{massive}
\ch_c(h_{j,m},Q_m;\tau,z)=q^{h_{j,m}-(\hat c-1)/8} y^{Q_m} \frac{\theta_3(\tau,z)}
{\eta^3(\tau)}
~.
\eeq
For $j=\frac{1}{2}+is$, $s \geq 0$, $m=\alpha+r$ with $0\leq \alpha<1$ and $r\in \Z$
these representations correspond to the continuous $SL(2,\R)/U(1)$ series
$\CC_{s,\alpha+r}$.
\end{enumerate}
In the above expressions we use the standard theta function notation
\beq
\label{theta3}
\theta_3(\tau,z)=\prod_{n=1}^{\infty} (1-q^n)(1+yq^{n-\frac{1}{2}})
(1+y^{-1}q^{n-\frac{1}{2}})=
\sum_{n=-\infty}^{\infty} y^n q^{\frac{n^2}{2}}
~,
\eeq
where $y=e^{2\pi i z}$.

These $\NN=2$ characters decompose naturally into a sum of
products between appropriate $U(1)$ and $SL(2,\R)/U(1)$ characters. One can
show the following decomposition formulae. For the
identity representation we have:
\beq
\label{iddecom}
\ch_I(\tau,z)=\eta^{-1}(\tau) \sum_{n\in \Z}
y^n q^{\frac{k-2}{2k}n^2}
\lambda^I_{-n}(\tau)
~,
\eeq
for the discrete representations
\beq
\label{discdecom}
\ch_d(h_{j,\pm j},Q_{\pm j};\tau,z)=\eta^{-1}(\tau) \sum_{n\in \Z}
y^{\pm \frac{2j}{k-2}+n}
q^{\frac{k-2}{2k}(\pm\frac{2j}{k-2}+n)^2} \lambda^d_{j,-n}(\tau)
~,
\eeq
and for the continuous representations
\beq
\label{contdecom}
\ch_c(h_{\frac{1}{2}+is,\alpha+r},Q_{\alpha+r};\tau,z)=
\eta^{-1}(\tau) \sum_{n\in \Z}
y^{\frac{2(\alpha+r)}{k-2}+n}
q^{\frac{k-2}{2k}(\frac{2(\alpha+r)}{k-2}+n)^2} \lambda^c_{\frac{1}{2}+is,n-(\alpha+r)}(\tau)
~.
\eeq
These decompositions are helpful for defining extended coset characters
and for deriving their modular transformation properties using known
facts about the $\NN=2$ extended characters. The extended coset characters
will be the basic building block of our boundary conformal field theory analysis
in Section 3 and we now turn to their definition.

\subsection{Extended characters}

The standard $\NN=2$ characters presented above have
an obvious drawback. Under modular transformations
they generate a continuous spectrum of $U(1)_R$ charges.
This is an undesirable feature, especially if
one is interested in the formulation of superstring theory
on a background, part of which is described on the
worldsheet by the $\NN=2$ theory at hand. The GSO
projection requires integral $U(1)_R$ charges and the modular
transformations would spoil this requirement.
Therefore, it is desirable to find a different set of ``extended'' characters that
possess integral $U(1)_R$ charges and which form a closed set
under modular transformations
\cite{Eguchi:1987wf, Eguchi:1988af, Eguchi:1988vr, Odake:1988bh, Odake:1989dm}.
Such characters can be defined in the following
way \cite{Eguchi:2003ik, Israel:2004xj}.

Consider a general $\NN=2$ theory with rational central charge
\beq
\label{N2central}
\hat c=1+\frac{2K}{N}, ~ ~ (K,N \in \N)
~.
\eeq
In such cases, extended characters can be defined
by summing the standard characters over
integer spectral flows. It is well known that the spectral flow generators
$U_{\eta}$ define an automorphism of the $\NN=2$
superconformal algebra with the form
\bea
\label{flow}
U^{-1}_{\eta} L_m U_{\eta}&=&L_m+\eta J_m+\frac{\hat c}{2}
\eta^2 \delta_{m,0} ~,\\
U^{-1}_{\eta} J_m U_{\eta}&=&J_m+\hat c \eta \delta_{m,0} ~, \\
U^{-1}_{\eta} G^{\pm}_r U_{\eta}&=&G^{\pm}_{r\pm\eta}
~
\eea
and they act on a general $\NN=2$ character
to produce the new characters
\beq
\label{echara}
\ch_*(*,\eta;\tau,z) \equiv q^{\frac{\hat c}{2}\eta^2} y^{\hat c \eta}
\ch_*(*;\tau,z+\eta \tau)
~.
\eeq
In the notation of the previous subsection and
in the special case of integer flow parameter $\eta=n\in \Z$
we simply get
\beq
\label{intecha}
\ch_*(h_{j,m},Q_m,n;\tau,z)=
\ch_*(h_{j,m+n},Q_{m+n};\tau,z)
~.
\eeq

Extended $\NN=2$ characters could be defined from
the standard ones by summing over all integers $n$ in (\ref{intecha}),
but, as it turns out, this is not completely satisfying. Such extended characters would
include theta functions at fractional levels and they would still behave
badly under modular transformations \cite{Eguchi:2003ik}. This problem
can be avoided by taking ``mod $N$'' partial sums over spectral
flows.

In our case, the central charge is given by $\hat c=\frac{k}{k-2}$ and
it is fractional only for fractional level $k$.
This is not a  terrible compromise
because most (if not all) physically interesting cases correspond to fractional
levels. For example, critical bosonic string theory
on the coset $SL(2,\R)/U(1)$ demands $k=\frac{9}{4}$.
Also, in many applications that involve the supersymmetric version of
the $SL(2,\R)/U(1)$ coset in superstring theory (e.g. for applications
relevant to the physics of string propagation near Calabi-Yau singularities
or string propagation in the near horizon limit of various NS5-brane
configurations) the level $k$ is an integer. In what follows, we focus
our discussion only on integer levels $k$. More general fractional
levels can be handled with appropriate modifications.

Consequently,
let us set $N=k-2$ and $K=2$ in (\ref{N2central}).
Then, following \cite{Eguchi:2003ik}, for each of the previously presented
representations\footnote{We consider only
the positive discrete representations here. The negative discrete
representations can be treated in an analogous manner.} we introduce the
$\NN=2$ extended characters
\bea
\label{idextend}
\chi_I(r;\tau)&=&\sum_{n\in r+(k-2) \Z} \ch_I(n;\tau)
~, \\
\label{discextend}
\chi_d(j,r;\tau)&=&\sum_{n\in r+(k-2) \Z} \ch_d(h_{j,j+n},Q_{j+n};\tau)
~, \\
\label{contextend}
\chi_c(s,m;\tau)&=&\sum_{n\in m+(k-2)\Z} \ch_c(h_{\frac{1}{2}+is,n},Q_n;\tau)
~,
\eea
where we restrict to $z=0$ for simplicity.
Note that in these definitions $r \in \Z_{k-2}$ and $2m \in \Z_{2(k-2)}$.
Our notation for the continuous extended characters $\chi_c(s,m;\tau)$ is
related to that of \cite{Eguchi:2003ik} by the equation
\beq
\label{egucont}
\chi_c(s,m;\tau) = \chi(\sqrt{\frac{2}{k-2}}s,2m;\tau)
~.
\eeq

Through the decomposition formulae (\ref{iddecom}), (\ref{discdecom}) and
(\ref{contdecom}) the extended $\NN=2$ characters naturally give rise
to extended characters in the $SL(2,\R)/U(1)$ coset. These extended
characters have been considered recently in \cite{Israel:2004xj} and it
is straightforward to derive their form. For example, let us consider in detail
the identity characters. From (\ref{idextend}) we obtain
\bea
\label{ididex}
\chi_I(r;\tau)&=&\sum_{n\in \Z} \ch_I(r+(k-2)n;\tau) = \nonumber\\
&=& \eta^{-1}(\tau)\sum_{n,s\in \Z} q^{\frac{k-2}{2k}(\frac{2r+2(k-2)n}{k-2}+s)^2}
\lambda^I_{-s+r+(k-2)n}(\tau) = \nonumber\\
&=& \eta^{-1}(\tau)\sum_{n,s \in \Z} q^{\frac{k-2}{2k}(\frac{2r}{k-2}+s)^2}
\lambda^I_{-s+r+kn}(\tau)
~.
\eea
With the extended coset character definition
\beq
\label{idcoset}
\Lambda^I_r(\tau) \equiv \sum_{n\in \Z}\lambda^I_{r+kn}(\tau)
~,
\eeq
and the index $r$ now restricted in $\Z_k$, we obtain the decomposition
\bea
\label{idcosetdecom}
\chi_I(r;\tau)&=&\eta^{-1}(\tau)\sum_{n\in \Z}
q^{\frac{k-2}{2k}(\frac{2r}{k-2}+n)^2} \Lambda^I_{-n+r} = \nonumber\\
&=&\eta^{-1}(\tau) \sum_{n\in \Z_k}
\Theta_{2r+n(k-2),\frac{k(k-2)}{2}}(\tau) \Lambda^I_{-n+r}(\tau)
~.
\eea
We use the standard definition of the classical theta function
\beq
\label{Theta}
\Theta_{m,k}(\tau)=\sum_{n\in \Z} q^{k(n+\frac{m}{2k})^2}
~.
\eeq
We can also bring (\ref{idcosetdecom}) into the
form
\beq
\label{idddd}
\chi_I(r;\tau)=\eta^{-1}(\tau)\sum_{n\in \Z}
\Theta_{2r+n(k-2),\frac{k(k-2)}{2}}(\tau) \lambda^I_{-n+r}(\tau)
~.
\eeq

A similar analysis can be performed for the discrete and continuous
representations. For the relevant details we refer the reader to
\cite{Israel:2004xj}. The extended discrete coset characters are defined by the equation
\beq
\label{disccoset}
\Lambda^d_{j,r}(\tau)\equiv \sum_{n\in \Z} \lambda^d_{j,r+kn}(\tau)
~,
\eeq
with $r\in \Z_k$ and they satisfy the decomposition formulae
\bea
\label{disccosetdecom}
\chi_d(j,r)(\tau)&=&\eta^{-1}(\tau)\sum_{n\in \Z}
\Theta_{2(j+r)+n(k-2),\frac{k(k-2)}{2}}(\tau) \lambda^d_{j,-n+r}(\tau) =\nonumber\\
&=&\eta^{-1}(\tau)\sum_{n\in \Z}
q^{\frac{k-2}{2k}(\frac{2(j+r)}{k-2}+n)^2} \Lambda^d_{j,-n+r}(\tau) =\nonumber\\
&=&\eta^{-1}(\tau)\sum_{n\in \Z_k}
\Theta_{2(j+r)+n(k-2),\frac{k(k-2)}{2}}(\tau) \Lambda^d_{j,-n+r}(\tau)
~.
\eea
Notice that as a consequence of the identity
$\lambda^d_{j,r}=\lambda^d_{\frac{k}{2}-j,-r}$ for the ordinary coset characters,
one can obtain a similar relation for the extended ones:
$\Lambda^d_{j,r}=\Lambda^d_{\frac{k}{2}-j,-r}$. This will be useful
for our computations in Section 4. We should add that the respective
representations are conjugate but not equivalent.
This fact
is easy to see by checking, for example, that they have different
$\SS$-transformation matrix elements (c.f. subsection 2.5)
\cite{Gepner:1989jq}.

Finally, the continuous characters are defined by the
equation
\beq
\label{conticoset}
\Lambda^c_{\frac{1}{2}+is,m}(\tau)=\sum_{n\in \Z}\lambda^c_{\frac{1}{2}+is,m+kn}(\tau)
~,
\eeq
where $m=\alpha+r$ with $\alpha=0,\frac{1}{2}$ and $r\in \Z_k$.
They satisfy the decomposition formulae
\bea
\label{conticosetdecom}
\chi_c(s,m)(\tau)&=&\eta^{-1}(\tau)\sum_{n\in \Z}
\Theta_{2m+n(k-2),\frac{k(k-2)}{2}}(\tau) \lambda^c_{\frac{1}{2}+is,-n+m}(\tau) =\nonumber\\
&=&\eta^{-1}(\tau)\sum_{n\in \Z}
q^{\frac{k-2}{2k}(\frac{2m}{k-2}+n)^2} \Lambda^c_{\frac{1}{2}+is,-n+m}(\tau) =\nonumber\\
&=&\eta^{-1}(\tau)\sum_{n\in \Z_k}
\Theta_{2m+n(k-2),\frac{k(k-2)}{2}}(\tau) \Lambda^c_{\frac{1}{2}+is,-n+m}(\tau)
~.
\eea

The above decompositions involve theta functions at half-integer
level $\frac{k(k-2)}{2}$. These functions have good modular
transformation properties only for integer level, i.e.\ $k$ even. For odd $k$ we can
use appropriate theta function identities to rewrite everything in
terms of integer level theta functions. We will not discuss this case here
and in what follows we simply restrict ourselves to even integers $k$.
Again, we want to stress that this is done only for purposes of
simplicity and that the present technology can be used to analyze
any case of fractional level $k$.

In the next Section we intend to use
the above defined extended coset characters as the basic building blocks
for the construction of boundary states in the coset $SL(2,\R)/U(1)$ CFT. Our motivation
for using the extended characters is based on the following observations:
\begin{enumerate}
\item[(1)] The extended coset characters have nice modular
transformation properties. For the continuous and
discrete representations these properties have been derived recently
in \cite{Israel:2004xj}. For the identity representation we derive them
in Appendix B.
The extended coset characters are closed under $\SS$-modular transformations
and this situation bears many common characteristics with the
case of the rational conformal field theories. In particular,
the infinity of standard coset characters for a given Casimir,
is organized into a finite set parameterized by a half-integer periodic
charge taking values in $\Z_k$. In that respect,
the extended characters are expected to be better suited for
the application of the Cardy construction
in the case of irrational conformal field theories.
\item[(2)] We are focusing on the case of
integer (or fractional) levels $k$. In many respects, our analysis
is similar to that of a compact boson at integer (or fractional)
squared radius, say $R^2=2k$, where the standard $U(1)$
symmetry is enhanced to the affine $U(1)_k$, or the case
of the bosonic parafermion theory given by the coset $SU(2)_k/U(1)_k$.
In both cases the extended characters play a vital role
in organizing the (Virasoro) representations of the theory
into finite sets
according to the enhanced symmetries. The $SL(2,\R)_k/U(1)$ coset
with integer $k$ shares many similarities with
the above cases. Moreover, in the next subsection we show that when $k$ is
integer, the torus partition sum of the cigar CFT can also be written
in terms of the extended characters.
\end{enumerate}

\subsection{Character decomposition of the torus partition function}

It is instructive to express the torus partition function of the axially-gauged
(cigar) coset in terms of the above standard and extended characters. This will
clarify certain aspects of the closed string spectrum in this theory and
will provide a clear basis for the use of the above extended characters
in the case of integer level $k$. We present two results: the expression of the torus partition
sum in terms of the standard characters appearing in eqs.\ (\ref{chardiscr})
and (\ref{charconti}) and the torus partition sum written in terms
of the extended characters appearing in eqs.\ (\ref{disccoset}) and
(\ref{conticoset}). The first expression was already implicit in \cite{Hanany:2002ev}.
Here we clarify and elucidate certain aspects of that analysis. The
supersymmetric version of the second expression
has appeared recently in refs.\ \cite{Eguchi:2004yi, Israel:2004ir} and it
shares many  common characteristics with our
bosonic analysis.

In Appendix A, we determine that the torus partition sum of the cigar CFT
can be recast in terms of the standard continuous and discrete characters
as
\bea
\label{MTtorus}
\ZZ (\tau,\bar{\tau})&=&16\sum_{n, w \in \Z} \int_0^{\infty}
ds ~ \rho(s;m,\bar m) \lambda^c_{\frac{1}{2}+is,m}(\tau)
\lambda^c_{\frac{1}{2}+is,\bar m}(\bar{\tau})
\nonumber\\
&+&8\sum_{\frac{1}{2}<j<\frac{k-1}{2}}
\sum_{w,r,\bar r \in \Z} \delta(2j+r+\bar r+kw)
\lambda^d_{j,r}(\tau) \lambda^d_{j,\bar r}(\bar{\tau})
~,
\eea
where $m$ and $\bar m$ are quantum numbers related to the integers $n$ and $w$
by eq.\ (\ref{Mmmnw}) below.
The spectral density of the continuous representations is given by
\beq
\label{rhosmm}
\rho(s;m,\bar m)=2\Bigg(
\frac{1}{2\pi}\log\epsilon+\frac{1}{2\pi i}\frac{d}{4ds}\log
\frac{\Gamma(-is+\frac{1}{2}-m)
\Gamma(-is+\frac{1}{2}+\bar m )}
{\Gamma(+is+\frac{1}{2}-m)
\Gamma(+is+\frac{1}{2}+\bar m)}\Bigg)
~.
\eeq
The result (\ref{MTtorus}) can be derived from another expression of the torus
partition sum, which follows from a direct path integral computation
in the corresponding gauged WZW theory.

We should mention that
eq.\ (\ref{MTtorus}) represents only part of the full answer. It
includes the full contribution of the discrete representations, but
only part of the continuous. Roughly speaking, the continuous contribution
includes two pieces. One that has an infrared divergence and needs to be regularized
and another that is finite. Eq.\ (\ref{MTtorus}) includes only the contribution
of the first diverging term. The remaining finite part cannot be written
in terms of the standard characters and has not been included in
(\ref{MTtorus}). The interpretation of these extra states
is unclear \cite{Israel:2004ir}, as they do not seem to fit in any of the
known $SL(2,\R)/U(1)$ representations. For more details we refer the reader to Appendix A.

In (\ref{MTtorus}) only continuous and discrete representations appear.
We can think of the continuous quantum number $s$ as the momentum
in the radial direction $\rho$ of the cigar, and the quantum numbers $m$ and
$\bar m$ as the $J_0^3$ and $\bar J_0^3$ charges under the unbroken
global $U(1) \subset SL(2,\R)$ symmetry that remains after gauging the axial $U(1)$.
These can be re-expressed
in terms of the angular momentum $n$ and
winding number $w$ in the compact $U(1)$ direction $\theta$ of the cigar. The precise
relation is the following
\beq
\label{Mmmnw}
m=\frac{n-kw}{2},\;\;\;
\bar m =-\frac{n+kw}{2}.
\eeq
The reason for the different identification of charges between
the left and right movers can be traced back to the form of the BRST
constraint of the axially gauged
WZW theory \cite{Dijkgraaf:1991ba}. This will be important later, when we
discuss the different types of boundary conditions on $\theta$.
The continuous part of the spectrum corresponds to
wave-like modes propagating along
the cigar, whereas the discrete part consists of states
localized near the tip.

The conformal weights of the primary fields of the cigar CFT
can be read off directly from the torus partition sum (\ref{MTtorus}):

$Continuous:$

\beq
\label{contcw}
h^j_{m}=\frac{s^2+\frac{1}{4}}{k-2}+\frac{m^2}{k}, \;\;\;\;\;
\bar{h}^j_{\bar m}=\frac{s^2+\frac{1}{4}}{k-2}+\frac{{\bar m}^2}{k}
\eeq

$Discrete:$

\beq
\label{discw}
h^j_{m}=\left\{
\begin{array}{lll}
\displaystyle -\frac{j(j-1)}{k-2}+\frac{m^2}{k}&,& m\geq j \\
\displaystyle -\frac{j(j-1)}{k-2}+\frac{m^2}{k}+|m-j|&,& m<j
\end{array}
\right., \;\;
{\bar h}^j_{\bar m}=
\left\{
\begin{array}{lll}
\displaystyle -\frac{j(j-1)}{k-2}+\frac{{\bar m}^2}{k}&,& {\bar m}\geq j\\
\displaystyle -\frac{j(j-1)}{k-2}+\frac{{\bar m}^2}{k}+|{\bar m}-j|&,& \bar m<j
\end{array}
\right.
\eeq
The origin of the extra terms in the scaling weights of the primary fields
with $m$, $\bar m<j$ can be traced to the leading order behavior of
the function
\beq
\label{MSr}
S_r=\sum_{s=0}^{\infty} (-)^s q^{\frac{1}{2}s(s+2r+1)}
\eeq
which appears in the definition of the discrete characters $\lambda^d_{j,r}$
(see eq.\ (\ref{chardiscr})). This behavior
depends crucially on the sign of $r$: for $r>0$ we get
$S_r=1-q^{1+r}+\ldots$, but for $r<0$ we have instead $S_r=q^{-r}-q^{-2r+1}+\ldots$.

The presence of discrete representation primary fields with $m$, $\bar m<j$
in (\ref{discw}) might appear to be in contradiction with the
constraints on the allowed values of $n$ and $w$
appearing in \cite{Dijkgraaf:1991ba, Aharony:2004xn}\footnote{We would like to thank
E. Kiritsis, D. Kutasov, and J.Troost for helpful discussions on these issues.}.
The authors of \cite{Dijkgraaf:1991ba} considered coset primary fields
that can be written in the form
\beq
\label{cosetDVV}
V(z,\bar z)=T_{SL(2,\R)}(\rho(z,\bar z),\theta_L(z,\bar z),\theta_R(z,\bar z))
e^{iq_L Y(z)+iq_R Y(\bar z)}
~
\eeq
and by analyzing the normalizability properties of the
$SL(2,\R)$ wavefunctions $T_{SL(2,\R)}(z,\bar z)$ they concluded that
the physical discrete coset primaries should satisfy the constraint $|n|<|kw|$.
In particular, this implies that no discrete states with zero winding
should be allowed. The authors of \cite{Aharony:2004xn} derived this
bound in the exact theory by looking at the LSZ poles of non-normalizable
vertex operators of the form (\ref{cosetDVV}). They argued that such poles
correspond to normalizable states of the theory and showed that they have
to satisfy the constraint $|n|<|kw|$.

We want to emphasize that this bound does not, in fact, contradict the result appearing
in (\ref{discw}). The states appearing in (\ref{cosetDVV}) correspond to affine
primaries of the parent $SL(2,\R)$ WZW theory. For such states the bound
$|n|<|kw|$ is, of course, correct. In the torus partition sum the contributing
coset states descend from $SL(2,\R)$ states that belong either
to ${\hat {\cal D}}^+ \otimes {\hat {\cal D}}^+$ or
to ${\hat {\cal D}}^- \otimes {\hat {\cal D}}^-$ \cite{Hanany:2002ev}.
Here ${\hat {\cal D}}^+$ and  ${\hat {\cal D}}^-$ denote affine $SL(2,\R)$ modules
based on lowest-weight and highest-weight discrete representations of $SL(2,\R)$ respectively.
For the special coset primaries (\ref{cosetDVV}) that descend from affine $SL(2,\R)$ primaries
the charges $m$ and ${\bar m}$ are given by (\ref{Mmmnw})
and they should have the same sign, i.e.\ they can be either
$m=j+r, ~ {\bar m}=j+{\bar
r} $ with $r,\bar r >0$
or $m=-j-r, {\bar m}=-j-{\bar r} $ with $r,\bar r >0$.
In both cases, this results to the constraint $|n|<|wk|$.

The torus partition sum (\ref{MTtorus}) implies, however, that the
coset conformal field theory has more primary fields.
Indeed, there are coset primary fields descending
from affine $SL(2,\R)$ descendants.
This is precisely what we see in the discrete part of (\ref{MTtorus}).
States with $m<j$ appear in the physical spectrum, but they cannot
be descending from affine $SL(2,\R)$ primaries in $\hat {\cal D}^+$, because
the latter have necessarily
$m\geq j$. For example, the coset primary state with $m=j-1$ would
come from the affine descendant state $J_{-1}^-|j,j\rangle$, which can easily
be shown to be a Virasoro primary (for both $SL(2,\R)$ and the coset).
Here $|j,j\rangle$ denotes the lowest-weight state of  $\hat {\cal D}^+$. The vertex operator
for this state contains derivatives of the target space fields and
does not fall into the class of primaries considered by
\cite{Dijkgraaf:1991ba, Aharony:2004xn} (see eq.\ (\ref{cosetDVV})).
Finally, notice that we can also interpret the extra $m<j$ states as descending from
$\eta=-1$ spectral flowed affine primaries of
$\hat {\cal D}^-_{\frac{k}{2}-j}$.

For integer level $k$ it is possible to rewrite (part of) the above partition sum
(\ref{MTtorus}) in terms of the extended coset characters. This is straightforward
for the discrete part. Starting from
(\ref{MTtorus}) we can write
\bea
\label{Mdiscreteextended}
\ZZ_d(\tau,\bar{\tau})&=& 8 \sum_{\frac{1}{2}<j<\frac{k-1}{2}}
\sum_{w,n,\bar n\in \Z} \sum_{r,\bar r\in \Z_k}
\delta(2j+r+\bar r+k(n+\bar n+w)) \lambda^d_{j,r+kn}(\tau)
\lambda^d_{j,\bar r+k\bar n}(\bar{\tau}) \nonumber\\
&=&8 \sum_{\frac{1}{2}<j<\frac{k-1}{2}}
\sum_{w,n,\bar n\in \Z} \sum_{r,\bar r\in \Z_k}
\delta(2j+r+\bar r+kw) \lambda^d_{j,r+kn}(\tau)
\lambda^d_{j,\bar r+k\bar n}(\bar{\tau}) \nonumber\\
&=&8 \sum_{\frac{1}{2}<j<\frac{k-1}{2}}
\sum_{w\in \Z} \sum_{r,\bar r\in \Z_k}
\delta(2j+r+\bar r+kw) \Lambda^d_{j,r}(\tau)
\Lambda^d_{j,\bar r}(\bar{\tau})
~.
\eea
The continuous contribution involves a series of subtle steps,
which are being explained in detail in Appendix A.
Again, the continuous contribution that can be written in terms
of the extended characters comes from that part of the full partition
sum that is infrared divergent and needs to be regularized.
The final result takes the form
\bea
\label{TORUS}
\ZZ(\tau,\bar{\tau})&=&16 \int_0^{\infty} ds \sum_{g\in \Z_{2k}}
\rho(s;g) \Lambda^c_{\frac{1}{2}+is,\frac{g}{2}}(\tau)
\Lambda^c_{\frac{1}{2}+is,-\frac{g}{2}}(\bar{\tau}) \nonumber\\
&+& 8 \sum_{\frac{1}{2}<j<\frac{k-1}{2}}
\sum_{w\in \Z} \sum_{r,\bar r\in \Z_k}
\delta(2j+r+\bar r+kw) \Lambda^d_{j,r}(\tau)
\Lambda^d_{j,\bar r}(\bar{\tau})
~,
\eea
with spectral density
\beq
\rho(s;g)=2\Bigg(
\frac{1}{2\pi}\log\epsilon+\frac{1}{2\pi i}\frac{d}{4ds}\log
\frac{\Gamma(-is+\frac{1}{2}-\frac{g}{2})
\Gamma(-is+\frac{1}{2}-\frac{g}{2} )}
{\Gamma(+is+\frac{1}{2}-\frac{g}{2})
\Gamma(+is+\frac{1}{2}-\frac{g}{2})}\Bigg) .
\eeq

\subsection{Modular transformations of the extended coset characters}

The $\SS$-modular transformation properties of the continuous and
discrete extended coset characters have been derived
recently in \cite{Israel:2004xj}. The identity representation
does not appear in the torus partition sum (\ref{TORUS}) and
lies outside the physical spectrum of the coset.
It can appear, however, in the open string spectrum and for that reason
we need to know its $\SS$-modular transformation matrix as well.
The necessary modular identity
has been derived in detail in Appendix B. For quick reference, we
summarize here the $\SS$-modular transformation properties of all the
extended characters.

For the identity representation we have:
\bea
\label{eq:graviton}
\Lambda^I_{r}\big (-\frac{1}{\tau}\big) &=&
\frac{1}{\sqrt{k(k-2)}} \sum_{m \in \Z_{2k}}
\int_0^{\infty} ds'
e^{2 \pi i \frac{r m}{k}}
\frac{\sinh(\frac{2\pi s'}{k-2}) \sinh(2\pi s')}
{\big | \cosh\pi( s'+i \frac{m}{2}) \big|^2}
\Lambda^c_{\frac{1}{2}+is',\frac{m}{2}}(\tau)
\nonumber\\
&+& \frac{2}{\sqrt{k(k-2)}}
 \sum_{2j'=2}^{k-2} \sum_{r'\in \Z_k}
\sin \frac{\pi (2j'-1)}{k-2}  e^{-2 \pi i \frac{r (2r'+2j')}{k}}
\Lambda^d_{j',r'}(\tau)
~.
\nonumber\\
\eea
For the discrete representations we get
\bea
\label{eq:discrete}
\Lambda^d_{j,r}\big (-\frac{1}{\tau}\big) &=&
\frac{1}{\sqrt{k(k-2)}} \sum_{m \in \Z_{2k}}
e^{2 \pi i \frac{(j+r) m}{k}}
\int_0^{\infty} ds'
\frac{\cosh\pi(s'\frac{k-4j}{k-2}+i\frac{m}{2})}{\cosh\pi(s'+i \frac{m}{2})}
\Lambda^c_{\frac{1}{2}+is',\frac{m}{2}}(\tau)
\nonumber\\
&+& \frac{i}{\sqrt{k(k-2)}}
 \sum_{2j'=2}^{k-2} \sum_{r'\in \Z_k}
e^{-\frac{4 \pi i}{k} (j+r) (j'+r')}
e^{\frac{4 \pi i}{k-2}(j-\frac{1}{2})(j'-\frac{1}{2})}
\Lambda^d_{j',r'}(\tau) \nonumber\\
&+&
\frac{i}{2\sqrt{k(k-2)}}  \sum_{r'\in \Z_k}
\Big(
e^{-\frac{4 \pi i}{k} (j+r) (\frac{1}{2}+r')}
\Lambda^d_{\frac{1}{2},r'}(\tau)-
e^{-\frac{4 \pi i}{k} (j+r) (-\frac{1}{2}+r')}
\Lambda^d_{\frac{k-1}{2},r'}(\tau) \Big)
~,
\nonumber\\
\eea
and for the continuous
\beq
\label{eq:continuous}
\Lambda^c_{\frac{1}{2}+is,\alpha+r}\big (-\frac{1}{\tau}\big) =
\frac{2}{\sqrt{k(k-2)}} \sum_{m \in \Z_{2k}}
e^{- 2\frac{\pi i}{k} m(r+\alpha)}
\int_0^{\infty} ds'
\cos(\frac{4 \pi s s'}{k-2})
\Lambda^c_{\frac{1}{2}+is',\frac{m}{2}}(\tau)
~.
\nonumber\\
\eeq

Notice that the identity character does not appear on the r.h.s.\ of any of the above modular transforms.
Hence, for a generic extended character labeled by $\xi$ we have an
$\SS$-modular identity of the form
\beq
\label{eq:Sgeneral}
\Lambda_{\xi}\big
(-\frac{1}{\tau}\big)= \int_0^{\infty} ds'
 \sum_{m' \in \Z_{2k}} \SS^c(s',\frac{m'}{2}|\xi)
\Lambda^c_{\frac{1}{2}+is',\frac{m'}{2}}(\tau) +
 \sum_{2j'=1}^{k-1} \sum_{r'\in \Z_k}
\SS^{d}(j',r'|\xi)
\Lambda^d_{j',r'}(\tau)
~.
\eeq

\section{Modular bootstrap for $SL(2,\R)/U(1)$}\label{MB}

\subsection{Generalities on the modular bootstrap}

D-branes in string theory can be formulated with the use of
the boundary state formalism. The boundary states
are specified by a set of gluing conditions that relate the left- and right-moving
generators of the chiral algebras (Virasoro or extended) of the
bulk conformal field theory. These conditions
are intimately related to the boundary conditions imposed
at the ends of the open strings.

The construction of a boundary state begins with the
formulation of Ishibashi states \cite{Ishibashi:1988kg}.
In rational conformal field theories the spectrum of
bulk primary fields is finite and discrete, and for each
primary field we construct one Ishibashi state as a coherent state
satisfying appropriate gluing conditions.
The general gluing conditions respect conformal invariance and
they read
\beq
(L_{n}- \bL_{-n})|R \rrangle=0.
\eeq
$|R\rrangle$ is the Ishibashi state built upon the primary field $R$.
In general, one can impose more symmetric boundary conditions
that preserve in addition some affine symmetry.

In generic conformal field theories
the spectrum typically contains an infinite set of fields, some of which
are labeled by a continuous parameter. In that case, one would like to construct
analogous Ishibashi states that will be labeled either by a continuous index
$s$ or by a discrete index $i$. An orthonormal set of such Ishibashi states
would satisfy the following cylinder amplitudes:
\bea
\llangle s| e^{-\pi T H^{(c)}} | s' \rrangle &=& \delta(s-s') \chi_{s}(iT) ~,\nonumber\\
\llangle i| e^{-\pi T H^{(c)}} | i' \rrangle &=& \delta_{i,i'} \chi_{i}(iT) ~,\nonumber\\
\llangle s| e^{-\pi T H^{(c)}} | i'\rrangle&=&0 ~,
\eea
$H^{(c)}=L_0+ \bL_0 - {c \over 12}$ is the closed
string Hamiltonian, with $T$ being the length of the cylinder and $\chi_{s},
\chi_{i}$ denote respectively the continuous and discrete characters
of the chiral algebra.

A boundary state can be constructed as a linear combination of the
above Ishibashi states. Let us denote the general boundary state
as $|X;\xi \rangle$. $X$ characterizes the boundary conditions
and $\xi$ specifies what linear combinations of Ishibashi states appear in
the boundary state. The problem of determining $|X;\xi \rangle$ reduces to
the problem of determining the consistent choices of $\xi$.
In rational conformal field theories, this problem has a well-known solution.
The cylinder amplitude between any two boundary states $|X;\xi_1 \rangle$
and $|X;\xi_2 \rangle$ can be written in terms of a finite discrete set
of characters $\chi_i$ appearing in the torus partition function
consistently with the gluing conditions. Worldsheet duality transforms
the cylinder amplitude into an annulus amplitude, where it can be viewed
as a 1-loop open string partition sum. As such, it should take the form
\beq
\langle X;\xi_1|e^{-\pi T H^{(c)}}|X;\xi_2\rangle=
\sum_{i} N(i;\xi_1|\xi_2) \chi_i(it)
~,
\eeq
with $t=1/T$ the annulus modulus and $N(i;\xi_1|\xi_2)$ positive
integer multiplicities. This form is quite restrictive and it gives rise to
a very useful consistency condition. This condition can be satisfied
by the Cardy ansatz, which employs the $\SS$-transformation matrix elements
to determine the coefficients of the Ishibashi state expansion of the boundary states
(see below for more detailed expressions).

In non-rational conformal field theories the annulus amplitude receives
extra contributions from an infinite set of characters
\beq
\label{cardycond}
\langle X;\xi_1|e^{-\pi T H^{(c)}}
|X;\xi_2\rangle= \int ds \;\rho(s;\xi_1| \xi_2) \chi_s(it) +
\sum_{i} N(i;\xi_1|\xi_2) \chi_i(it)
~.
\eeq
The generalization of the above condition would demand
positive definite spectral densities $\rho(s;\xi_1|\xi_2)$ and positive integer
multiplicities $N(i;\xi_1|\xi_2)$. Satisfying these conditions
is a highly non-trivial problem and it is not known if this modular bootstrap programme
can be implemented successfully in general.
Recent progress
\cite{Fateev:2000ik, Teschner:2000md, Zamolodchikov:2001ah, Fukuda:2002bv, Ahn:2002ev}
in bosonic and $\NN=1$ Liouville theory,
however, has led to a set of boundary states that pass the Cardy consistency conditions
(\ref{cardycond}) as well as a
variety of other consistency conditions (various factorization constraints on the disk).
This provides further motivation for studying
the solutions of the Cardy conditions in other non-rational conformal
field theories. In this spirit, the authors of \cite{Eguchi:2003ik, Ahn:2003tt}
formulated recently various classes of boundary states in $\NN=2$
Liouville theory.

In this paper, we consider the Cardy consistency
conditions in the $SL(2,\R)_k/U(1)$ coset conformal field theory
for integer levels $k$. As we mentioned in Section 2, this situation
shares many common characteristics with the rational $SU(2)_k/U(1)$
case and the use of the extended coset representations leads to
a more controlled application of the modular bootstrap method.
For example, the extended Ishibashi states
appearing in this Section have a finite range of charges and
the whole construction stands in close resemblance with that of
\cite{Maldacena:2001ky}
for the $SU(2)_k/U(1)$ parafermionic model.
Further confidence for the validity of our results is obtained
by the recent analysis of \cite{Ribault:2003ss}, where coset
boundary states were determined from consistent $H_3^+$ boundary states
\cite{Ponsot:2001gt} by descent.

\subsection{Extended coset Ishibashi states from $\NN=2$ decomposition}

There are two special types of boundary conditions in a theory with
$\NN=2$ superconformal symmetry \cite{Ooguri:1996ck}:
\bea
\label{bc}
{\rm A-type}&:& \;\;(J_{n} - \tilde{J}_{-n}) |B \rangle=0 ~, \
\ (G_r^{\pm}- i \tilde{G}_{-r}^{\mp})|B \rangle=0 ~, \\
{\rm B-type}&:& \;\;(J_{n} + \tilde{J}_{-n}) |B \rangle=0 ~, \ \
(G_r^{\pm}- i \tilde{G}_{-r}^{\pm}) |B \rangle=0
~.
\eea
Both are compatible with the diagonal $\NN=1$ superconformal symmetry
\beq
(L_{n} - \tilde{L}_{-n}) |B \rangle=0, \ \ (G_r- i \tilde{G}_{-r}) |B
\rangle=0~,
\eeq
where $G= G^+ + G^-$. In these equations $|B\rangle$ denotes a generic
$\NN=2$ boundary state.
The A-type/B-type boundary conditions
impose Dirichlet/Neumann boundary conditions
on the free boson associated to the $U(1)$ current $J$.

Extended $\NN=2$ Ishibashi states for each of the above boundary conditions
have been presented recently in \cite{Eguchi:2003ik}.
For example, let us consider in detail the A-type boundary conditions
and let us set in (\ref{N2central}) $N=k-2$ and $K=1$.
As we mentioned in the previous Section,
these values will be relevant for the analysis of the $SL(2,\R)_k/U(1)$ coset.
We need to consider only Ishibashi states for
the continuous and discrete representations. These representations
form a maximal subset closed under modular transformations and they
will be the only ones relevant for the subsequent analysis of the coset theory.

The A-type Ishibashi states will be denoted as
\bea
\label{N2ishibashi}
& &|A;s, \frac{m}{2}\rrangle, \;\;s \geq 0, ~ m \in \Z_{2(k-2)}, \\
& &|A;j,r \rrangle,  \;\;2\leq 2j \leq k-2, \ r \in \Z_{k-2}
\eea
and they satisfy the orthogonality conditions:
\bea
\label{N2OC}
\llangle A;s, {m \over 2}| e^{-\pi T H^{c}}
|A;s', {m' \over 2} \rrangle &=& \delta(s-s') \delta_{m,m'}^{2(k-2)}
\chi_c(s, {m \over 2}; iT) ~, \nonumber \\
\llangle A;j, r| e^{-\pi T H^{c}} |A;j', r' \rrangle &=&
\delta_{j,j'} \delta_{r,r'}^{k-2} \chi_d(j,r;iT) ~, \nonumber\\
\llangle A;j, r| e^{-\pi T H^{c}} |A;s, {m \over 2} \rrangle &=&0
~.
\eea
$\delta^{M}_{n,n'}$ denotes the usual Kronecker delta modulo $M$.
A-type $\NN=2$ boundary states can be written as appropriate linear
combinations of the above Ishibashi states.
The specific form of these combinations has been determined
in \cite{Eguchi:2003ik} by imposing the following modular bootstrap equations:
\bea
\label{N2MB}
\langle
A;0|e^{-\pi T H^{c}}|A;\xi \rangle &=& \chi_{\xi}(it), \\
\langle A;0|e^{-\pi T H^{c}}|A;0 \rangle &=& \chi_{I}(r=0, it).
\eea
$\xi$ is a label that parameterizes the $\NN=2$ boundary states
which solve these modular bootstrap equations.
A special set of solutions employs what is known as
the Cardy ansatz. This ansatz leads to boundary states of the form
\bea
\label{N2Cardy}
|A;\xi \rangle = \int_0^{\infty} ds \sum_{m \in
\Z_{2(k-2)}} \Psi_{\xi}(s,m) | A;s,{m \over 2} \rrangle + \sum_{r\in
\Z_{k-2}} \sum_{2j=2}^{k-2} C_{\xi} (j,r) |A;j,r \rrangle
~,
\eea
with wavefunctions
\bea
\label{Cardyansatz}
\Psi_\xi (s,m)&=&\frac{\SS^c(s,\frac{m}{2}|\xi)}{\sqrt{\SS^c(s,\frac{m}{2}|0)}},
\\
\label{Cardyansatz2}
C_{\xi}(j,r)&=&\frac{\SS^d(j,r|\xi)}{\sqrt{\SS^d(j,r|0)}},
\eea
directly expressible in terms of the $\SS$-transformation matrix elements of the
representation labeled by $\xi$. In total,
this ansatz yields three classes of consistent
branes corresponding to the continuous, discrete, and identity
representations. For more details we refer the reader to \cite{Eguchi:2003ik}.

\subsubsection{Ishibashi state decompositions: $\NN=2$ A-type $\rightarrow$ coset B-type}

Now let us examine how the above $\NN=2$ Ishibashi states can be decomposed
into appropriate coset Ishibashi states. The A-type $\NN=2$ Ishibashi states
have equal left- and right-moving $U(1)_R$ charges. This implies that the left- and right-moving
coset charges are also equal, i.e.\ $m=\bar m$ (see e.g. eqs.\ (\ref{dimcharge}),
(\ref{dimcharge1}), (\ref{dimchargecont})).\footnote{We should point out that the
whole discussion is algebraic.
The {\cal N}=2 theory has two realizations depending on the
right moving charge $\bar{Q}= \pm {2\bar{m} \over k-2}$
identification to the $SL(2)/U(1)$ charges. The plus sign
corresponds to the ${\cal N}=2$ Liouville theory and the minus sign
to its supercoset cigar mirror. Nevertheless the discussion in
this paper is totally independent of the choice of the ${\cal N}=2$ theory. We
therefore choose the plus sign.} Given the
relation of those charges with the left- and right-moving momenta
along the compact direction $\theta$ of the cigar (\ref{Mmmnw}),
we conclude that such boundary conditions impose Neumann boundary conditions on $\theta$.
By convention we call the corresponding coset Ishibashi states B-type.
On the other hand, coset Ishibashi states with Dirichlet boundary conditions on $\theta$
will be called A-type.
{\em From now on we will concentrate mostly on boundary conditions
or Ishibashi states of the coset theory and the A/B-type terminology will refer
to the above conventions}.

A more precise relation between extended $\NN=2$ and extended coset Ishibashi
states follows from the character decomposition formulae
(\ref{disccosetdecom}) and (\ref{conticosetdecom}).
In the case of $\NN=2$ A-type boundary conditions, we may
write\footnote{In these relations, we present explicitly the left- and right-moving
charges of the Ishibashi states to make our discussion as transparent
as possible.}
\bea
\label{gendecomcont}
|A;s,{m \over 2}, {m \over 2} \rrangle_{\NN=2}
=\sum_{n\in \Z_k}&&|A;m +n(k-2), m +n(k-2)\rrangle_{U(1)} \otimes\nonumber\\
&&|B;s,-n+{m \over 2},-n+{{m} \over 2}\rrangle
~
\eea
for the continuous representations, and
\bea
\label{gendecomdisc}
|A;j,r,r \rrangle_{\NN=2}
= \sum_{n\in \Z_k} &&|A; 2(j+r)+n(k-2), 2(j+r)+n(k-2)\rrangle_{U(1)} \otimes \nonumber\\
&&|B;j,-n+r,-n+r \rrangle
\eea
for the discrete representations. The boundary states appearing on the  r.h.s.\ of
these decompositions are respectively
$\widehat{U(1)}_{\frac{k(k-2)}{2}}$ Ishibashi states and extended coset Ishibashi states. The
overlaps of the latter will be discussed shortly.
Notice that we label the Ishibashi coset charges in the same way that
we did for the corresponding characters in Section 2.
The actual (extended) coset charges appearing on the r.h.s.\ of the
above equations are $(-n+\frac{m}{2},-n+\frac{m}{2})$
for the continuous representations and
$(j+r-n,j+r-n)$ for the discrete representations.
The definition of the $\widehat{U(1)}_{\frac{k(k-2)}{2}}$ Ishibashi states can be found, among
other places, in a related context in
\cite{Maldacena:2001ky}.

The above decompositions result in a certain set of B-type
coset Ishibashi states of which only a subset is physical. The physical
boundary states are allowed to couple only to closed string modes that appear in the torus
partition sum (\ref{TORUS}). This requirement yields the following constraints
on the charges:
\beq
\label{physcont}
m+\bm=0 \ {\rm mod} \ 2k
\eeq
for the continuous representations, and
\beq
\label{physdisc}
2j+r+\br=0 \ {\rm mod} \ k
~
\eeq
for the discrete. Therefore, we are led to conclude that
an Ishibashi state $|B;s,\frac{m}{2},\frac{\bar{m}}{2}\rrangle$
is acceptable only when $m=\bm=0 \ {\rm mod} \ k$ with $m \in \Z_{2k}$.
Similarly, an Ishibashi state $|B;j,r,\bar r\rrangle$ is acceptable only when
$2(j+r)=0 \ {\rm mod} \ k$ and $r=\br$.
To summarize, we find that the only
allowed B-type Ishibashi states are:
\bea
\label{BCONT}
&Continuous:& ~ ~ |B;s,0,0\rrangle ~, ~ ~ |B;s,{k \over 2},{k
\over 2} \rrangle ~, \;\; s \geq 0,\\
\label{BDISC}
&Discrete:& ~ ~ |B;j,-j,-j \rrangle ~, ~ ~ |B;j,{k \over 2}-j,{k
\over 2}-j \rrangle, \;\;1\leq j \leq \frac{k}{2}-1
~.
\eea
Notice that in the discrete B-type Ishibashi states, $j$ takes only integer values.
This is a consequence of the
relation $2(j+r)=0 \ {\rm mod} \ k$ and the fact that $k$ has been assumed to be even.

The only  non-zero overlaps of these boundary states are the following:
\bea
\label{Boverlaps}
&&\llangle B;s_1,0,0|e^{-\pi TH^c}|B; s_2,0,0
\rrangle = \delta(s_1-s_2)
\Lambda_{\frac{1}{2}+is_1,0}(iT), \\
&&\llangle B;s_1,{k \over 2},{k \over 2}|e^{-\pi TH^c}|B;s_2,{k
\over 2},{k \over 2} \rrangle =
\delta(s_1-s_2) \Lambda^c_{\frac{1}{2}+is_1,{k \over 2}}(iT),\\
&&\llangle B;j_1,-j_1,-j_1|e^{-\pi TH^c}|B;j_2,-j_2,-j_2\rrangle =\delta_{j_1,j_2}
\Lambda^d_{j_1,-j_1}(iT), \\
&&\llangle B;j_1,\frac{k}{2}-j_1,{k \over 2}-j_1|e^{-\pi
TH^c}|B;j_2,\frac{k}{2}-j_2, {k \over 2}-j_2\rrangle
=\delta_{j_1,j_2} \Lambda^d_{j_1,\frac{k}{2}-j_1}(iT).
\eea

\subsubsection{Ishibashi state decompositions: $\NN=2$ B-type $\rightarrow$ coset A-type}

The construction of $discrete$ $\NN=2$ B-type Ishibashi states and their
decomposition into coset Ishibashi states is slightly more tricky.
The $\NN=2$ B-type gluing requires $Q_{j+r}=-{\bar Q}_{j + \bar r}$.\footnote{We use
the same quantum number $j$ for left- and right-moving modes. This
choice is dictated by the physical spectrum of the coset theory, where $j=\bar j$.}
Together with the condition of conformal invariance a gluing of this sort can be satisfied only
for $j=k/4$ and $\bar r = -r -1, r \geq 0$. For these values,  however,
the discrete $\NN=2$ Ishibashi states decompose into coset Ishibashi states
based on representations that do not appear in the physical
spectrum of the cigar. Indeed, the relevant coset charges are
$m=j+r=\frac{k}{4}+r$ and $\bar m = j+\bar r + 1 = \frac{k}{4}-r$,
and they yield $m+\bar m=\frac{k}{2}$. On the other hand,
the left-right coupling dictated by the torus partition sum (\ref{MTtorus})
requires $m+\bar m=-wk$ with $w$ integer. This implies the full absence
of the corresponding discrete Ishibashi states on the cigar.

An analogous problem does not exist for the continuous $\NN=2$ B-type Ishibashi states.
The $\NN=2$ B-type gluing
translates into A-type (Dirichlet) boundary conditions on the $\theta$
direction of the cigar, i.e.\ $m=-\bar m$,  and the corresponding boundary states couple
to closed string modes with zero winding $w$. The relevant Ishibashi state decomposition reads
\bea
\label{gendecomcont}
|B;s,{m \over 2}, -{m \over 2} \rrangle_{\NN=2}
=\sum_{n\in \Z_k}&& |B;m +n(k-2), -m -n(k-2)\rrangle_{U(1)} \otimes \nonumber\\
&&|A;s,-n+{m \over 2},n-{{m} \over 2}\rrangle
\eea
and the only acceptable A-type coset Ishibashi states are the following
\bea
\label{Aishibashi}
&Continuous:& ~ ~|A;s,{m \over 2},-{m
\over 2} \rrangle,\;\; s \geq 0, ~ m \in \Z_{2k}
~.
\eea
These boundary states satisfy the requirement (\ref{physcont}) coming from
the torus partition sum automatically and their overlaps are given by
\bea
\label{OC}
& &\llangle A;s,{m \over 2},-{m \over 2}| e^{-\pi TH^{c}}
|A;s', {m' \over 2},-{m' \over 2}\rrangle = \delta(s-s')\delta_{m,m'}^{2k}
\Lambda^c_{1/2+is, {m \over 2}}(iT) ~.
\eea

\subsection{Modular bootstrap and the Cardy ansatz for coset boundary states}
\label{ModBoot}

\subsubsection{A-type boundary states}

The A-type Ishibashi states presented above are in one-to-one correspondence with the
continuous physical primaries of the coset theory and it might be useful
to think of them as a direct analogue of the $\NN=2$ A-type Ishibashi states
that appeared in (\ref{N2ishibashi}). In order to determine Cardy consistent linear combinations
of these boundary states it would be tempting to impose the direct analogue of
the modular bootstrap equations (\ref{N2MB})
\bea
\label{ACMB}
\langle A;0| e^{-\pi T H^{c}} |A; \xi \rangle &=& \Lambda_\xi(it)  ~, \\
\langle A;0| e^{-\pi T H^{c}} |A; 0 \rangle &=& \Lambda^I_0(it), \ (T=1/t)
~
\eea
but the basic element of this approach is now missing.
Because of the absence of discrete A-type Ishibashi states in the
coset, we cannot construct a boundary state $|A;0 \rangle$ based on the identity representation
and we cannot implement the above equations directly.

Nevertheless, it is interesting to ask whether we can still formulate Cardy
consistent branes with the use of the Cardy ansatz appearing in eq.\ (\ref{Cardyansatz}).
Such boundary states would be the analogues of the A-type, class 2 boundary states
of the $\NN=2$ Liouville analysis of \cite{Eguchi:2003ik} and
they would take the form
\beq
\label{defCardy}
|A;s,\frac{m}{2} \rangle= \int_{0}^{\infty} ds' \sum_{m' \in \Z_{2k}}
\Psi_{s,m}(s',m')|A;s',{m' \over 2},-{m' \over 2} \rrangle
~,
\eeq
with wavefunctions
\beq
\label{WFansatz}
\label{ansatz}
\Psi_{s,m}(s',m')= e^{i\delta(s',m')}
\frac{S^{c}(s', {m' \over 2}|s,\frac{m}{2})}{\sqrt{S^c(s',\frac{m'}{2}|r=0)}}
~.
\eeq
The precise form of the $\SS$-transformation matrix elements can be found in
Section \ref{ECC}.
Note that we have explicitly allowed for an extra phase $e^{i\delta(s',m')}$, which
may also depend on the variables $s'$ and $m'$. This extra phase will be fixed partially in a moment
in a way that accomodates the Cardy consistency conditions of the next Section.
More generally, a phase ambiguity of this sort
always exists in the modular bootstrap approach \cite{Eguchi:2003ik} and extra information
is needed to fix it.

Besides the Cardy inspired wavefunctions (\ref{WFansatz})
we would like to consider
also a slight variant motivated by the analysis
of the D1 boundary states of \cite{Ribault:2003ss} (in Section 5
we provide a detailed comparison of our boundary states
with those of \cite{Ribault:2003ss}). These boundary states
are not of the Cardy form, i.e.\ they cannot be written in terms
of the $\SS$-matrix elements. They have been derived in \cite{Ribault:2003ss}
from corresponding boundary states in the $H_3^+$ conformal field
theory by descent and they possess the expected semiclassical properties.
It is useful to consider them also in the context of the present discussion.

In the next Section, we check the Cardy consistency of these
two classes of boundary states by studying all possible overlaps
among them. In the process, we find that consistency
between the two classes fixes the
phase factor in (\ref{ansatz}) to $e^{i \delta(m')}= e^{\frac{\pi i}{2} (m' {\rm mod } ~ 2)}$.
To summarize, we propose the following classes of A-type
boundary states on $SL(2,\R)/U(1)$:

{\bf Class 2} ~ $|A; s,\frac{m}{2}\rangle, ~ s \geq 0, ~ m \in \Z_{2k}$
\beq
\label{C2A}
\Psi_{s,m}(s',m')=
\frac{2}{(k (k-2))^{1/4}} e^{i \delta(m')}
e^{-\pi i {m m' \over k}}
\cos ({4 \pi s s' \over k-2})
\sqrt{\frac{|\cosh\pi (s'+i\frac{m'}{2})|^2}{\sinh\big(\frac{2\pi s'}{k-2}\big) \sinh(2\pi s')}}
~,
\eeq

{\bf Class 2'} ~ $|A; s,\frac{m}{2}\rangle', ~ s \geq 0, ~ m \in \Z_{2k}$
\beq
\label{C2Ap}
\Psi'_{s,m}(s',m')=
\frac{2}{(k(k-2))^{1/4}}
e^{-\pi i {m m' \over k}}
\frac{e^{4 \pi i s s' \over k-2}+ (-1)^{m'} e^{-4 \pi i s s' \over k-2}}{2}
\sqrt{\frac{|\cosh\pi (s'+i\frac{m'}{2})|^2}{\sinh\big(\frac{2\pi s'}{k-2}\big) \sinh(2\pi s')}}
~.
\eeq
Notice that
for even $m'$ the above wavefunctions are the same. For odd $m'$,
however, the factor $\cos ({4 \pi s s' \over k-2})$ in class 2
is replaced by  $\sin ({4 \pi s s' \over k-2})$ in class 2'.

\subsubsection{B-type boundary states}

In a previous subsection we derived various B-type Ishibashi states. Some
of them are based on the continuous representations (\ref{BCONT})
and others on the discrete representations (\ref{BDISC}). None of
them, however, is in one-to-one correspondence with the respective physical
primaries of the coset conformal field theory and for that reason
it is not straightforward to impose directly the modular bootstrap
eqs.\ (\ref{ACMB}). Instead, we can directly apply the Cardy
ansatz (\ref{Cardyansatz}), (\ref{Cardyansatz2}) and then check that all the Cardy consistency
conditions are satisfied.

There is a well-known analogue of this situation in the rational $SU(2)/U(1)$ case
of \cite{Maldacena:2001ky}.
In the analysis of that paper the B-type $SU(2)/U(1)$ boundary states
were also constructed with the use of the Cardy ansatz and the Cardy
consistency conditions were checked. For example,
the self-overlap of the ``identity'' ($j=0$) B-type boundary state
was found to be (see eq.\ (3.21) in \cite{Maldacena:2001ky}):
\beq
\label{bbmin}
\langle B;0 |e^{-\pi T H^c}|B;0 \rangle = \sum_{n'=0}^{2k-1}
\chi_{0,n'}(it) ~,
\eeq
where $\chi_{0,n'}(q)$ are the appropriate $SU(2)/U(1)$ characters.
In the present $SL(2,\R)/U(1)$ case we find a very similar result.
The ``identity'' boundary state will be denoted as $|B \rangle$ and its
self-overlap takes the form
\beq
\label{BCMB}
\langle B|e^{-\pi T H^c}|B \rangle = \sum_{r \in
\Z_k} \Lambda^I_r(it) ~.
\eeq

More generally, we consider boundary states of the form
\bea
\label{BCardy}
|B;\xi\rangle &=& \int_0^{\infty} ds \Phi_{\xi}(s,0)|B;s,0\rrangle +
\int_0^{\infty} ds \Phi_{\xi}(s,k) |B;s,\frac{k}{2} \rrangle + \nonumber\\
&+&\sum_{j=1}^{k/2-1}D_{\xi}(j,-j) |B;j,-j\rrangle +
\sum_{j=1}^{k/2-1}D_{\xi}(j,\frac{k}{2}-j)
|B;j,\frac{k}{2}-j\rrangle ~
\eea
and wavefunctions of the Cardy type
\beq
\Phi_{\xi}(s',m')= \frac{S^{c}(s', {m' \over 2}|\xi)}{\sqrt{S^c(s',\frac{m'}{2}|r=0)}}, \;\;\;
D_{\xi}(j',r')= \frac{S^{d}(j', r'|\xi)}{\sqrt{S^d(j',r'|r=0)}}
~,
\eeq
with $m'=0,k$ for the continuous representations and $r'=-j,\frac{k}{2}-j$
for the discrete representations.
$\xi$ is a label that parametrizes the possible B-type boundary states
of the above form.

In this way, we find the following three candidate classes of B-type boundary states
(an extra normalization factor $\sqrt k$ is inserted for later consistency):

{\bf Class 1} ~ $|B\rangle$,
\bea
\label{Bphi}
&&\Phi(s',0)=\sqrt{k}
\sqrt{S^c(s',0|r=0)}=\frac{\sqrt k}{(k(k-2))^{1/4}}
\sqrt{\frac{\sinh\big(\frac{2\pi s'}{k-2}\big)\sinh(2\pi s')}
{|\cosh(\pi s')|^2}}
~,\\
&&\Phi(s',k)=\sqrt{k}
\sqrt{S^c(s',\frac{k}{2}|r=0)}=
\frac{\sqrt k}{(k(k-2))^{1/4}}
\sqrt{\frac{\sinh\big(\frac{2\pi s'}{k-2}\big)\sinh(2\pi s')}
{|\cosh(\pi (s'+i\frac{k}{2}))|^2}}
~,\\
&&D(j,-j)=\sqrt{k}
\sqrt{S^d(j,-j|r=0)}=\frac{\sqrt{2k}}{(k(k-2))^{1/4}}
\sqrt{\sin\frac{\pi (2j-1)}{k-2}}~, \\
&&D(j,\frac{k}{2}-j)=\sqrt{k}
\sqrt{S^d(j,\frac{k}{2}-j|r=0)}=\frac{\sqrt{2k}}{(k(k-2))^{1/4}}
\sqrt{\sin\frac{\pi (2j-1)}{k-2}}~. \eea

{\bf Class 2} ~ $|B;s,\alpha \rangle,   ~ s \geq 0, ~ \alpha = 0,\frac{1}{2}$
\bea
\label{Bsa}
&&\Phi_{s,\alpha}(s',0)=\frac{4\sqrt
k}{(k(k-2))^{1/4}} \cos \big (\frac{4\pi s s'}{k-2}\big)
\sqrt{\frac{|\cosh(\pi s')|^2}{\sinh\big( \frac{2\pi s'}{k-2}\big)\sinh(2\pi s')}}~,\\
&&\Phi_{s,\alpha}(s',k)=\frac{4\sqrt k}{(k(k-2))^{1/4}}
e^{2\pi i \alpha} \cos \big (\frac{4\pi s s'}{k-2}\big)
\sqrt{\frac{|\cosh(\pi (s'+i \frac{k}{2}))|^2}{\sinh\big( \frac{2\pi s'}{k-2}\big)\sinh(2\pi s')}}~,\\
&&D_{s,\alpha}(j,-j)=0,\\
&&D_{s,\alpha}(j,\frac{k}{2}-j)=0~.
\eea

{\bf Class 3} ~ $|B;j_1\rangle,  ~ 1\leq 2j_1 \leq k-1$
\bea
\label{BJ}
&&\Phi_{j_1}(s',0)= \frac{\sqrt{k}}{(k(k-2))^{1/4}}
\sqrt{\frac{|\cosh \pi s'|^2}{\sinh\big (\frac{2\pi s'}{k-2}\big)
\sinh(2\pi s')}} \frac{\cosh \pi \big (
s'\frac{k-4j_1}{k-2}\big)}{\cosh \pi s'},\\
\label{BJ2}
&&\Phi_{j_1}(s',k)=\frac{\sqrt{k}}{(k(k-2))^{1/4}}
\sqrt{\frac{|\cosh (\pi (s'+i\frac{k}{2}))|^2}{\sinh\big (\frac{2\pi s'}{k-2}\big)
\sinh(2\pi s')}} \frac{e^{2\pi i (j_1-\frac{k}{4})}\cosh \pi \big(
s'\frac{k-4j_1}{k-2}+i\frac{k}{2}\big)}{\cosh (\pi (s'+i\frac{k}{2}))},\\
\label{BJ3}
&&D_{j_1}(j,-j)=\frac{i}{\sqrt{2(k-2)}}
\frac{1}{\sqrt{\sin \frac{\pi (2j-1)}{k-2}}} e^{\frac{4\pi
i}{k-2} (j_1-\frac{1}{2})(j-\frac{1}{2})} ,\\
\label{BJ4}
&&D_{j_1}(j,\frac{k}{2}-j)=\frac{i}{\sqrt{2(k-2)}}
\frac{1}{\sqrt{\sin \frac{\pi (2j-1)}{k-2}}} e^{-2\pi i
j_1}e^{\frac{4\pi i}{k-2} (j_1-\frac{1}{2})(j-\frac{1}{2})}.
\eea

We would like to emphasize that we obtain only one class 1 brane with the above ansatz.
A priori, one might expect to obtain a larger set of class 1 boundary
states $|B;r\rangle$ (with $r\in \Z_k$) by using the
more general matrix elements $S^c(s,0|r)$, $S^c(s,\frac{k}{2}|r)$,
$S^d(j,-j|r)$, $S^d(j,\frac{k}{2}-j|r)$. It is straightforward
to check, however, that such boundary states will be identical to
$|B\rangle\equiv |B;r=0\rangle$.
We find a similar result for the class 3 branes.

The class 2 branes fall into two categories depending on whether the
parameter $m$ in $S^c(s',0|s,m)$ and  $S^c(s',k/2|s,m)$ is even or odd.
It is straightforward to verify that the functions $S^c(s',0|s,m)$
are independent of $m$, whereas
$S^c(s',k/2|s,m)$ depend on $m$ only through a phase
$e^{2\pi i a}$ with $a=0, ~ \ha$ according to the parity of the parameter $m$.

Finally, notice that our B-type class 3 branes are parametrized
by a single discrete label. The reader familiar with
 \cite{Eguchi:2003ik} may recall that the authors of that paper
proposed class 3 boundary states parametrized by two
discrete parameters. The underlying reason
for the appearance of a second discrete label is
the fact that the $\SS$-transformation property of discrete representations
includes the extra characters with $j=\frac{1}{2}$ and $j=\frac{k-1}{2}$,
which lie outside the range of the allowed Ishibashi states.
An appropriate linear combination of two discrete boundary
states allows for the cancellation of the extra characters.
This cancellation turns out to be automatic for B-type boundary conditions
in the $SL(2,\R)/U(1)$ coset
and there is no apriori reason to include a second discrete label.
The possibility of a second discrete label will be discussed further
in Section 4, where we determine which of the above boundary states
pass the Cardy consistency checks.

\section{Cardy consistency conditions}\label{CCC}

In this Section we compute several amplitudes between the
above boundary states. We are going to focus only on
A-A and B-B overlaps. Mixed overlaps between A-type and B-type branes
involve a different set of characters, whose properties will not be discussed here.

The modular transformation of these amplitudes
from the closed string channel (parameter $T$) to the open
(parameter $t=1/T$) yields the explicit forms of the spectral densities and
degeneracies of the open strings streching between the various branes.
In general, we find annulus amplitudes of the form:
\bea
\label{cardy}
\langle\xi_1| e^{-\pi T H^{(c)}}|\xi_2 \rangle &=&
\int_0^{\infty} ds \sum_{m \in \Z_{2k}}
\rho(s,m;\xi_1|\xi_2) \Lambda^c_{\ha+is, \frac{m}{2}}(it) + \nonumber\\
&+&
\sum_{2j=1}^{k-1}  \sum_{r \in \Z_k} N(j,r;\xi_1|\xi_2)
\Lambda^d_{j,r}(it) +  \sum_{r \in \Z_k} M(r;\xi_1|\xi_2)
\Lambda^I_{r}(it)
\eea
with positive definite real spectral densities $\rho(s,m|\xi_1,\xi_2)$
and positive integer multiplicities $N(j,r|\xi_1,\xi_2)$ and $M(r|\xi_1,\xi_2)$.
The identity representation and the discrete
representations with $j=\frac{1}{2},\frac{k-1}{2}$ do not appear
in the closed string spectrum of the coset, but they can
appear in the open string channel of the above amplitudes and
they have been included in (\ref{cardy}).

\subsection{A-A overlaps}

\noindent $\bullet$ {\em class 2 - class 2}

For the overlap of two class 2 boundary states we find:
\bea
\label{CD14}
& &\langle A; s_1,\frac{m_1}{2} |e^{-\pi T H^c} | A; s_2,\frac{m_2}{2}
\rangle = \nonumber\\
& &
\int_0^{\infty} ds' \bigg [ \rho_1(s';s_1|s_2)
\Lambda^c_{\frac{1}{2}+is',
\frac{m_1-m_2}{2}}(it)+
\rho_2(s';s_1|s_2) \Lambda^c_{\frac{1}{2}+is',
\frac{m_1-m_2+k}{2}}(it)
\bigg ],
\eea
with spectral densities
\bea
\label{CD15}
\rho_1(s';s_1|s_2)&=&\frac{8}{k-2}
\int^{\infty}_0 ds \cos\big (\frac{4\pi ss_1}{k-2} \big )
\cos\big(\frac{4\pi ss_2}{k-2}\big)\cos\big(\frac{4\pi s s'}{k-2}\big)
\frac{\cosh(2\pi s)}{\sinh\big(\frac{2\pi s}{k-2}\big)\sinh\big(2\pi s)}
\nonumber\\
&=&\sqrt{\frac{2}{k-2}}
\int_0^{\infty} dp \frac{\cos(2\pi pp')}
{\sinh(\pi \QQ p)\sinh\big(\frac{2\pi p}{\QQ}\big)}
\sum_{\epsilon_1,\epsilon_2 =\pm 1}\cosh\big(2\pi p\big ( \frac{1}{\QQ}+i\epsilon_1p_1+
i\epsilon_2p_2\big)\big),
\nonumber\\
\\
\label{CD16}
\rho_2(s';s_1|s_2)&=&\frac{8}{k-2}
\int_0^{\infty} ds \cos\big(\frac{4\pi ss_1}{k-2}\big)
\cos\big(\frac{4\pi ss_2}{k-2}\big) \cos\big(\frac{4\pi ss'}{k-2}\big)
\frac{1}{\sinh\big(\frac{2\pi s}{k-2}\big) \sinh(2\pi s)}
\nonumber\\
&=& 2\sqrt{\frac{2}{k-2}}
\int_0^{\infty} dp \frac{\cos(2\pi p p')}
{\sinh(\pi \QQ p)\sinh(\frac{2\pi p}{\QQ})}\sum_{\epsilon=\pm 1}
\cos(2\pi p (p_1+\epsilon p_2))
~,\eea
where $\QQ^2=\frac{2}{k-2}$.
We made use of the change of variables $p=s\QQ$
and the trigonometric identity
\beq
\label{trigo1}
\cosh A \cosh B \cosh \Gamma = \frac{1}{4} \sum_{\epsilon_1,\epsilon_2=\pm 1}
\cosh(A+\epsilon_1 B+\epsilon_2 \Gamma)
~
\eeq
to bring the spectral densities into a form that is already familiar from the analysis of
the boundary states in the $\NN=2$ Liouville theory \cite{Eguchi:2003ik}.

Let us briefly sketch the derivation of the above expressions. From the
wavefunctions given in (\ref{C2A})
we can easily obtain the overlap
\bea
\label{CD17}
& &\langle A; s_1,\frac{m_1}{2} |e^{-\pi T H^c} |A; s_2, \frac{m_2}{2} \rangle =
\int_0^{\infty} ds \sum_{m\in Z_{2k}} \frac{4}{\sqrt{k(k-2)}}
\cos\big(\frac{4\pi ss_1}{k-2}\big) \cos\big( \frac{4\pi ss_2}{k-2}\big)
\nonumber\\
& &e^{\pi i m(m_1-m_2)/k} \frac{\big |\cosh \pi (s+i\frac{m}{2} ) \big |^2}
{\sinh\big(\frac{2\pi s}{k-2}\big ) \sinh\big(2\pi s\big)}
\Lambda^c_{\frac{1}{2}+is,\frac{m}{2}}(iT)
~.\eea
Then, with the use of the modular transformation property (\ref{eq:continuous})
and the trigonometric identity
\beq
\label{coshidentity}
\big | \cosh\pi (s+i \frac{ m}{2} )\big|^2=
\frac{1}{2}\bigg ( \cosh(2\pi s) + \frac{1}{2} \big (e^{\pi i m}+e^{-\pi i m}
\big)\bigg)
\eeq
we can derive (\ref{CD14}).

As we said, the spectral densities appearing here are similar to those of
\cite{Eguchi:2003ik}. Strictly speaking, these densities are divergent because
of the diverging integrands near the point $p=0$.  There are two ways to deal
with this infrared divergence. The first, which will be adopted here and was also used in
 \cite{Eguchi:2003ik}, is to regularize
the densities by explicitly subtracting the divergent piece. Another
approach considers relative spectral densities, i.e.\ one is instructed to
subtract the amplitude of a reference boundary state with fixed
parameters $s_1$ and $s_2$ (e.g.\ $s_1=s_2=0$).
This method is based on the universal nature of the divergence
and has been used previously in refs.\
\cite{Ponsot:2001gt, Ribault:2003ss}.
Further comments on
this approach will appear in Section 5, where we also show that the relative densities
following from $\rho_1(s';s_1|s_2)$ and $\rho_2(s';s_1|s_2)$
above are equivalent to those appearing in \cite{Ribault:2003ss}.

The finite piece of the above densities can be written
in terms of the so-called q-gamma function $S_b(x)$, which is
defined as \cite{Fateev:2000ik, Eguchi:2003ik}:
\beq
\label{qgamma}
\log S_b(x) =\int_0^{\infty} \frac{dt}{2t}
\Bigg[ \frac{\sinh\Big(t(b^2+1-2 b x)\Big)}{\sinh(b^2 t) \sinh(t)}
-\frac{b^2+1-2 b x }{b^2 t}\Bigg].
\eeq
In our case the parameter $b$ is given by
$b=\frac{\QQ}{\sqrt{2}}=\frac{1}{\sqrt{k-2}}$.
Hence, the finite part of
$\rho_1(s';s_1|s_2)$ reads
\bea
\rho_1(s';s_1|s_2)|_{{\rm fin}}&=&\QQ \frac{i}{2\pi} \sum_{\epsilon_0,\epsilon_1,\epsilon_2=\pm 1}
\epsilon_0 \partial_{p'} \log S_b
\Big(\frac{b}{2}+\frac{i}{\sqrt{2}}(\epsilon_0 p'+\epsilon_1 p_1+\epsilon_2
p_2)\Big)\\
&=& \frac{i}{2\pi} \sum_{\epsilon_0, \epsilon_1,\epsilon_2=\pm 1}\epsilon_0
\partial_{s'} \log S_b
\Big(\frac{b}{2}+i b (\epsilon_0 s'+\epsilon_1 s_1+\epsilon_2
s_2)\Big),
\eea
while that of $\rho_2(s';s_1|s_2)$ is
\bea
\rho_2(s';s_1|s_2)|_{{\rm fin}}&=&\QQ \frac{2i}{\pi} \sum_{\epsilon_1,\epsilon_2=\pm 1}
\partial_{p'} \log S_b
\Big(\frac{1}{2b}+\frac{b}{2}+\frac{i}{\sqrt{2}}(p'+\epsilon_1 p_1+\epsilon_2
p_2)\Big)\\
 &=& \frac{2i}{\pi} \sum_{\epsilon_1,\epsilon_2=\pm 1}\partial_{s'} \log S_b
\Big(\frac{1}{2b}+\frac{b}{2}+i b (s'+\epsilon_1 s_1+\epsilon_2
s_2)\Big).
\eea
We can also write an explicit integral representation
of these finite densities with the use of the defining relation (\ref{qgamma}):
\bea
\label{rho1}
& &\rho_1(s';s_1|s_2)|_{{\rm fin}}= \nonumber\\
& &\frac{1}{2\pi i} \sum_{\epsilon_1,\epsilon_2=\pm 1} 2 \partial_{s'} \Bigg( i
\int_0^{\infty} \frac{dt}{t}
\Bigg[ \frac{\cosh(t) \sin\Big(2tb^2(s'+\epsilon_1 s_1+\epsilon_2 s_2)\Big)}{2\sinh(b^2 t) \sinh(t)}
-\frac{s'+\epsilon_1 s_1+\epsilon_2 s_2}{t}\Bigg] \Bigg), \nonumber\\
& &\\
\label{rho2}
& &\rho_2(s';s_1|s_2)|_{{\rm fin}}= \nonumber\\
& &\frac{1}{2\pi i}  \sum_{\epsilon_1,\epsilon_2=\pm 1} 4 \partial_{s'} \Bigg( i
\int_0^{\infty} \frac{dt}{t}
\Bigg[ \frac{\sin\Big(2tb^2(s'+\epsilon_1 s_1+\epsilon_2 s_2)\Big)}{2\sinh(b^2 t) \sinh(t)}
-\frac{s'+\epsilon_1 s_1+\epsilon_2 s_2}{t}\Bigg] \Bigg). \nonumber\\
&&
\eea

\noindent $\bullet$ {\em class 2 - class 2'}

This overlap can be computed along the same lines as the previous one. We obtain
\bea
\label{CD17}
& &\langle A; s_1,\frac{m_1}{2} |e^{-\pi T H^c} | A; s_2,\frac{m_2}{2}
\rangle' = \nonumber\\
& &
\int_0^{\infty} ds' \bigg [ \rho_3(s';s_1|s_2)
\Lambda^c_{\frac{1}{2}+is',
\frac{m_1-m_2}{2}}(it)+
\rho_4(s';s_1|s_2) \Lambda^c_{\frac{1}{2}+is',
\frac{m_1-m_2+k}{2}}(it)
\bigg ],
\eea
with spectral densities
\bea
\label{CD18}
& &\rho_3(s';s_1|s_2) = \nonumber\\
& & \frac{8}{k-2}
\int^{\infty}_0 ds
\cos\big (\frac{4\pi ss_1}{k-2} \big )
\cos\big(\frac{4\pi s s'}{k-2}\big)
\frac{\cosh^2(\pi s) \cos\big(\frac{4\pi ss_2}{k-2}\big) +
\sinh^2(\pi s) \sin\big(\frac{4\pi ss_2}{k-2}\big)}
{\sinh\big(\frac{2\pi s}{k-2}\big)\sinh\big(2\pi s)} ~, \nonumber\\
\\
\label{CD19}
& &\rho_4(s';s_1|s_2) = \nonumber\\
& &\frac{8}{k-2}
\int^{\infty}_0 ds
\cos\big (\frac{4\pi ss_1}{k-2} \big )
\cos\big(\frac{4\pi s s'}{k-2}\big)
\frac{\cosh^2(\pi s) \cos\big(\frac{4\pi ss_2}{k-2}\big) -
\sinh^2(\pi s) \sin\big(\frac{4\pi ss_2}{k-2}\big)}
{\sinh\big(\frac{2\pi s}{k-2}\big)\sinh\big(2\pi s)} ~.\nonumber\\
\eea

\noindent $\bullet$ {\em class 2'- class 2'}

Similarly we find
\bea
\label{CD20}
& & '\langle A; s_1,\frac{m_1}{2} |e^{-\pi T H^c} | A; s_2,\frac{m_2}{2}
\rangle' = \nonumber\\
& &
\int_0^{\infty} ds' \bigg [ \rho_5(s';s_1|s_2)
\Lambda^c_{\frac{1}{2}+is',
\frac{m_1-m_2}{2}}(it)+
\rho_6(s';s_1|s_2) \Lambda^c_{\frac{1}{2}+is',
\frac{m_1-m_2+k}{2}}(it)
\bigg ],
\eea
with spectral densities
\bea
\label{CD21}
\rho_5(s';s_1|s_2)&=&
\frac{8}{k-2}
\int^{\infty}_0 ds
\cos\big(\frac{4\pi s s'}{k-2}\big)
\Big [ \frac{\cosh^2(\pi s) \cos\big (\frac{4\pi ss_1}{k-2} \big ) \cos\big(\frac{4\pi ss_2}{k-2}\big)}
{\sinh\big(\frac{2\pi s}{k-2}\big)\sinh\big(2\pi s)}
\nonumber\\
&+&
\frac{\sinh^2(\pi s)  \sin\big(\frac{4\pi ss_1}{k-2}\big) \sin\big(\frac{4\pi ss_2}{k-2}\big)}
{\sinh\big(\frac{2\pi s}{k-2}\big)\sinh\big(2\pi s)} \Big] ~,
\\
\label{CD22}
\rho_6(s';s_1|s_2)&=& \frac{8}{k-2}
\int^{\infty}_0 ds
\cos\big(\frac{4\pi s s'}{k-2}\big) \Big[
\frac{\cosh^2(\pi s) \cos\big (\frac{4\pi ss_1}{k-2} \big ) \cos\big(\frac{4\pi ss_2}{k-2}\big)}
{\sinh\big(\frac{2\pi s}{k-2}\big)\sinh\big(2\pi s)} \nonumber\\
&-&
\frac{\sinh^2(\pi s)  \sin\big(\frac{4\pi ss_1}{k-2}\big) \sin\big(\frac{4\pi ss_2}{k-2}\big)}
{\sinh\big(\frac{2\pi s}{k-2}\big)\sinh\big(2\pi s)} \Big]
~.
\eea
These spectral densities can be written more compactly in terms of
the previously defined densities $\rho_1$ and $\rho_2$
as
\bea
\rho_5(s';s_1|s_2) &=&
\frac{1}{2}\Big(\rho_1(s';s_1-s_2|0)+\rho_2(s';s_1+s_2|0)\Big) ~,\\
\rho_6(s';s_1|s_2) &=&
\frac{1}{2}\Big(\rho_1(s';s_1+s_2|0)+\rho_2(s';s_1-s_2|0)\Big)
~.
\eea

The Cardy consistency conditions demand that the spectral densities are
positive functions of their arguments. For the densities appearing
in this subsection this is indeed true since the
leading divergence in the defining
integrals goes as ${1 \over s^2}$ with a positive coefficient.

\subsection{B-B overlaps}

\noindent $\bullet$ {\em class 1 - class 1}

There is a single B-type, class 1 boundary state.
The self-overlap of this boundary state in the open string channel can be computed
easily and yields the following answer:
\beq
\label{BB}
\langle B|e^{-\pi  TH^c}|B\rangle = \sum_{r\in \Z_k} \Lambda^I_r(it)
~.
\eeq
As we mentioned before, the form of this amplitude is very similar to that of
the B-type parafermion branes in \cite{Maldacena:2001ky} (see e.g.\ (\ref{bbmin})).

\

\noindent $\bullet$ {\em class 1 - class 2}

This overlap reads:
\bea
\label{BBc}
\langle B|e^{-\pi TH^c}|B;s,\alpha\rangle &=& \sum_{m\in \Z_{2k}} (1+e^{\pi i (m+2\alpha)})
\Lambda^c_{\frac{1}{2}+is,\frac{m}{2}} (it) ~.
\eea

\

\noindent $\bullet$ {\em class 1 - class 3}

The overlaps between class 1 and class 3 boundary states are
\beq
\label{BBjj}
\langle B|e^{-\pi TH^c}|B;j\rangle =
\sum_{r\in \Z_k} \Lambda^d_{j,r}(it).
\eeq

\

\noindent $\bullet$ {\em class 2 - class 2}

For the overlap of two class 2, B-branes we obtain
\bea
\label{BBcc}
\langle B;s_1,\alpha_1|e^{-\pi T H^c}|B;s_2,\alpha_2 \rangle&=&
\sum_{m\in Z_{2k}} \int_0^{\infty} ds' \bigg [
\rho_1(s';s_1|s_2)+\rho_2(s';s_1|s_2) \bigg ] \nonumber\\
& & (1+e^{\pi i (m+2\alpha_1+2\alpha_2)})
\Lambda^c_{\frac{1}{2}+is',\frac{m}{2}}(it)
~
\eea
where $\rho_1(s';s_1|s_2)$ and $\rho_2(s';s_1|s_2)$ are the previously defined
densities in eqs.\ (\ref{CD15}) and (\ref{CD16}) respectively.

\

\noindent $\bullet$ {\em class 3 - class 3}

We compute these overlaps in some detail.
Starting from the defining relations (\ref{BJ})-(\ref{BJ4}), we obtain
\bea
\label{B3B3one}
& &\langle B;j_1|e^{-\pi TH^c}|B;j_2\rangle=
\frac{k}{\sqrt{k(k-2)}}
\int_0^{\infty} ds
\frac{1}
{\sinh\big(\frac{2\pi s}{k-2}\big) \sinh(2\pi s)} \nonumber\\
& &\cosh\pi s(\frac{k-4j_1}{k-2}) \cosh\pi s(\frac{k-4j_2}{k-2})
\bigg(\Lambda^c_{\frac{1}{2}+is,0}(iT)+e^{2 \pi i (j_2-j_1)}
\Lambda^c_{\frac{1}{2}+is,\frac{k}{2}}(iT)\bigg) +\nonumber\\
& &
\frac{k}{2\sqrt{k(k-2)}}
\sum_{j=1}^{\frac{k}{2}-1}
\frac{e^{\frac{4 \pi i}{k-2}(j-\frac{1}{2})(j_2-j_1)}}
{\sin\pi\frac{2j-1}{k-2}}
\bigg(\Lambda^d_{j,-j}(iT) + e^{2\pi i (j_1-j_2)}
\Lambda^d_{j,\frac{k}{2}-j}(iT) \bigg).
\eea
In these relations $j$ is an integer by construction.

The discrete character terms can be recast into a simpler
form with the use of the character identity
$\Lambda^d_{j,r} =
\Lambda^d_{\frac{k}{2}-j,-r}$.
The resulting expression in the closed string channel is
\beq
\label{3b3bsf}
\frac{k}{\sqrt{k(k-2)}}
\sum_{j=1}^{\frac{k}{2}-1}
\frac{\cos\frac{2\pi}{k-2}(2j-1)(j_2-j_1)}
{\sin\pi\frac{2j-1}{k-2}}  \Lambda^d_{j,-j}(iT)
~.
\eeq
For simplicity let us consider in detail the consistency of the general self-overlap.
For $j_1=j_2$ the modular transformation of $\Lambda^d_{j,-j}(iT)$
in (\ref{3b3bsf}) yields several contributions. One of them involves the discrete characters
$\Lambda^d_{j',r'}(it)$, which appear with multiplicities
(using the character identity once again):
\beq
-\frac{2}{k-2} \sum_{j=1}^{\frac{k}{2}-1}
\frac{\sin\frac{\pi}{k-2}(2j-1)(2j'-1)}
{\sin\pi\frac{2j-1}{k-2}}
~.
\eeq
Note that there are values of $j'$ in this expression for which the multiplicities are negative.
Hence, we are led to conclude that the B-type, class 3 boundary states (\ref{BJ})-(\ref{BJ4})
are inconsistent.

The existence of consistent A-type class 3 branes in
$\NN=2$ Liouville theory \cite{Eguchi:2003ik} suggests a possible alternative
with boundary states labelled by a pair of discrete labels.
Thus, in analogy with the analysis of  \cite{Eguchi:2003ik}
we consider the class 3' branes:

{\bf Class 3'}
~ $|B;j_1,j_2\rangle$, with  $j_1,j_2=1,\frac{3}{2},...,\frac{k-1}{2}$ and
\bea
\label{Bjj}
&\Phi_{j_1,j_2}(s,0)= \frac{\sqrt{k}}{(k(k-2))^{1/4}}
\sqrt{\frac{|\cosh \pi s|^2}{\sinh\big (\frac{2\pi s}{k-2}\big) \sinh(2\pi s)}}
\frac{\cosh \pi \big ( s\frac{k-4j_1}{k-2}\big)+\cosh\pi \big( s\frac{k-4j_2}{k-2}\big)}
{\cosh \pi s} ~,\\
&\Phi_{j_1,j_2}(s,\frac{k}{2})=\frac{\sqrt{k}}{(k(k-2))^{1/4}}
\sqrt{\frac{|\cosh \pi s|^2}{\sinh\big (\frac{2\pi s}{k-2}\big) \sinh(2\pi s)}}
\frac{e^{2\pi i j_1}\cosh \pi \big ( s\frac{k-4j_1}{k-2}\big)+
e^{2\pi i j_2}\cosh\pi \big( s\frac{k-4j_2}{k-2}\big)}
{\cosh \pi s} ~,
\\
&D_{j_1,j_2}(j,-j)=\frac{i \sqrt{k}}{\sqrt{2}(k(k-2))^{1/4}}
\frac{1}{\sqrt{\sin \frac{\pi (2j-1)}{k-2}}}
\bigg ( e^{\frac{4\pi i}{k-2} (j_1-\frac{1}{2})(j-\frac{1}{2})}+
e^{\frac{4\pi i}{k-2} (j_2-\frac{1}{2})(j-\frac{1}{2})} \bigg)
~, \\
&D_{j_1,j_2}(j,\frac{k}{2}-j)=\frac{i \sqrt{k}}{\sqrt{2}(k(k-2))^{1/4}}
\frac{1}{\sqrt{\sin \frac{\pi (2j-1)}{k-2}}}
\bigg ( e^{-2\pi i j_1}e^{\frac{4\pi i}{k-2} (j_1-\frac{1}{2})(j-\frac{1}{2})}+
e^{-2\pi i j_2} e^{\frac{4\pi i}{k-2} (j_2-\frac{1}{2})(j-\frac{1}{2})} \bigg)
~. \nonumber\\
\eea
These are simple linear superpositions of the class 3 branes (\ref{BJ})-(\ref{BJ4}), i.e.\
\beq
|B;j_1,j_2\rangle =|B;j_1\rangle +|B;j_2 \rangle
~.
\eeq
The overlaps of these branes follow easily from the overlaps computed
above. In particular, for self-overlaps of the form
$\langle B;j_1,j_2|e^{-\pi TH^c}|B;j_1,j_2\rangle$
we simply have to sum over four overlaps of the type
$\langle B;j|e^{-\pi TH^c}|B;j'\rangle$ with
$(j,j')=(j_1,j_1)$, $(j_1,j_2)$, $(j_2,j_1)$ and $(j_2,j_2)$.
The contribution of discrete characters (in the closed string channel) now becomes
\beq
\frac{2k}{\sqrt{k(k-2)}}
\sum_{j=1}^{\frac{k}{2}-1}
\frac{\Big(1+\cos\frac{2\pi}{k-2}(2j-1)(j_1-j_2)\Big)}
{\sin\pi\frac{2j-1}{k-2}}  \Lambda^d_{j,-j}(iT)
~.
\eeq
After a modular transformation to the open string channel and the use of the
character identity $\Lambda^d_{j,r} =
\Lambda^d_{\frac{k}{2}-j,-r}$,
we obtain the following
multiplicity for the discrete character $\Lambda^d_{j',r'}(it)$:
\beq
\label{BBmulti}
N(j',r';j_1,j_2|j_1,j_2) = -\frac{4}{k-2} \sum_{j=1}^{\frac{k}{2}-1}
\frac
{\sin\frac{\pi}{k-2}(2j-1)(2j'-1)
 \sin^2\Big(\frac{\pi}{k-2}(2j-1)(j_1-j_2)-\frac{\pi}{2}\Big)}
{\sin\pi\frac{2j-1}{k-2}}
~.
\eeq

The last summation can be performed with the use of the
Verlinde formula for the $SU(2)_{k_0}$ affine algebra with $k_0=k-4\geq 0$.
Recall that the modular transformation matrix for the $SU(2)_{k_0}$ affine
characters  $\chi_{s}(\tau)$ is
\beq
\chi_r(-\frac{1}{\tau})=\sum_{2r=0}^{k_0} S_{r}^{s} \chi_{s}(\tau),
\;\;\;{\rm with}\;\;\;
S_{r}^{s} =\sqrt{\frac{2}{k_0+2}}\sin{\frac{\pi}{k_0+2}(2r+1)(2s+1)}
\eeq
and the Verlinde formula is the statement:
\beq
\label{versu2}
\sum_{2r=0}^{k_0} \frac{S_s^r S_t^r S_r^u}{S_0^r} = N_{s\;t}^u,
\eeq
with the fusion coefficients
\beq
N_{s\;t}^u =\left\{
\begin{array}{ll}
1 ~,
\;\;\;& s+t+u \in \Z \;{\rm and}\; |s-t|\leq u \leq{\rm min}\{s+t,k_0-s-t\}\\
0 ~, \;\;\;& {\rm otherwise}
\end{array}\right.
\eeq
In each of these relations\footnote{The present indices
$r$ and $s$ should not be confused
with the charge of the discrete characters or the Casimir of
the continuous ones in previous Sections.}
$r,s,t,u=0,\frac{1}{2},\ldots,\frac{k_0}{2}$.
In our case, however, the summation variable $r$ can take only integer
values. The appropriate modification of the Verlinde formula reads (c.f.  \cite{Maldacena:2001ky})
\beq
\label{versu2s}
\sum_{r=0}^{\frac{k_0}{2}} \frac{S_s^r S_t^r S_r^u}{S_0^r} =
\frac{1}{2}\Big(N_{s\;t}^u
+N_{s\;t}^{\frac{k_0}{2}-u}\Big)
~.
\eeq

Now, in (\ref{versu2s}) set $k_0=k-4$ as well as $2r=2j-2$, $2s=2j'-2$,
and $2t=2u=(j_1-j_2)+\frac{k_0}{2}$.
With this substitution eq.\ (\ref{BBmulti})
takes the simple form
\beq
\sum_{2j'=2}^{k-2}  \sum_{r' \in \Z_k} N(j',r';j_1,j_2|j_1,j_2)
\Lambda^d_{j',r'}(it) = -  \sum_{2s=0}^{k_0} \Big(N_{s
  \;u(j_1,j_2)}^{u(j_1,j_2)}+
N_{s \;u(j_1,j_2)}^{\frac{k_0}{2}-u(j_1,j_2)}\Big)
\Lambda^d_{s+1,-(s+1)}(it)
~,
\eeq
where $u(j_1,j_2)=\frac{j_1-j_2}{2}+\frac{k_0}{4}$.
Thus, we see that the only potentially consistent class 3 boundary states are
those labeled by  $(j_1,j_2)$ such that
$N_{s \;u(j_1,j_2)}^{u(j_1,j_2)}=N_{s \;u(j_1,j_2)}^{\frac{k_0}{2}-u(j_1,j_2)}=0$
for all values of $s$.
This is the case only when $u<0$ or $u>\frac{k_0}{2}$. Equivalently,
this requires $|j_1-j_2|>\frac{k-4}{2}$. Taking into account the range of
the parameters $j_1$ and $j_2$,
we conclude that there are two possibilities
\beq
j_1=\frac{1}{2}, ~ ~ j_2=\frac{k-1}{2},\;\; {\rm or}\;\;j_1=\frac{k-1}{2}, ~ ~ j_2=\frac{1}{2}~,
\eeq
which are in fact equivalent
since $|B;j_1,j_2\rangle \equiv |B;j_2,j_1\rangle$.
Note that for these values of $j_1$ and $j_2$
the class 3' boundary states (\ref{BJ}) simply reduce to the class 2 boundary states (\ref{Bsa})
with $s=0$ and $\alpha=\frac{k}{4}$ mod 1.
Hence, we are again led to conclude that the class 1 and class 2 are the
only B-type boundary states of subsection 3.3.2 that
satisfy the Cardy consistency conditions.
This should not be interpreted as the statement that
there are no B-type boundary states based on the discrete representations.
It may still be possible to construct appropriate linear combinations
of the above class 3 boundary states that satisfy the Cardy
consistency conditions. A similar case by case analysis was performed
in the $\NN=2$ Liouville case in \cite{Eguchi:2003ik}.

\section{Physical interpretation}\label{Semiclassical}

In this Section we discuss the physical properties of the
$SL(2,\R)/U(1)$ boundary states that satisfy the Cardy consistency condition and
their relation to the boundary states proposed in
\cite{Ribault:2003ss}.

As we have already stressed in various places
of the previous discussion, the modular bootstrap approach can determine uniquely only the
modulus of the one-point functions but not the overall phase. Although we will not do this
explicitly, for the couplings of the boundary states to continuous closed string modes
we can partially restrict this ambiguity by employing the "reflection property" of the one-point
functions as in \ci{Eguchi:2003ik, Ribault:2003ss}:
\beq
\label{reflection}
\Psi_{\xi}(-s,m,w)= R(s,m,w)\Psi_{\xi}(s,m,w)
~,
\eeq
where
\beq
\label{reflampl}
R(s,m,w)= (\nu_b)^{2is}{\Gamma(1+2is)\Gamma(2isb^2) \over
\Gamma(1-2is)\Gamma(-2isb^2)} {\Gamma(\ha+{m-kw \over 2} -is)
\Gamma(\ha+{m+kw \over 2} -is) \over \Gamma(\ha+{m-kw \over 2}
+is) \Gamma(\ha+{m+kw \over 2} +is)} \nonumber
\eeq
with
\beq
\label{nub}
b^2=\frac{1}{k-2}~, ~ ~ \nu_b={\Gamma(1-b^2) \over \Gamma(1+b^2)}~.
\eeq
For discrete representations the analytic continuation of
the reflection amplitude $R(s,m,w)$
exhibits a series of poles. These singularities
can be attributed to the specific normalization of the discrete primary fields.
An alternative normalization would lead to the same amplitudes, since the
divergent terms cancel in physical quantities \ci{Ribault:2003ss}.
Note that the approach we followed for the construction of boundary
states in this paper picks out a naturally divergent-free normalization for the one-point functions
of the discrete states.

For quick reference we begin with a short
summary of the results of the previous Sections.

\subsection{Summary of Cardy consistent boundary states}

The analysis of Sections 3 and 4 provided the following Cardy consistent
boundary states:

\

{\bf A-type,  class 2} ~ $|A; s,\frac{m}{2}\rangle, ~ s \geq 0, ~ m \in \Z_{2k}$
\beq
\label{C2A5}
\Psi_{s,m}(s',m')=
\frac{2}{(k (k-2))^{1/4}} e^{i \delta(m')}
e^{-\pi i {m m' \over k}}
\cos ({4 \pi s s' \over k-2})
\sqrt{\frac{|\cosh\pi (s'+i\frac{m'}{2})|^2}{\sinh\big(\frac{2\pi s'}{k-2}\big) \sinh(2\pi s')}}
~,
\eeq

{\bf A-type, class 2'} ~ $|A; s,\frac{m}{2}\rangle', ~ s \geq 0, ~ m \in \Z_{2k}$
\beq
\label{C2A5}
\Psi'_{s,m}(s',m')=
\frac{2}{(k(k-2))^{1/4}}
e^{-\pi i {m m' \over k}}
\frac{e^{4 \pi i s s' \over k-2}+ (-1)^{m'} e^{-4 \pi i s s' \over k-2}}{2}
\sqrt{\frac{|\cosh\pi (s'+i\frac{m'}{2})|^2}{\sinh\big(\frac{2\pi s'}{k-2}\big) \sinh(2\pi s')}}
~,
\eeq

{\bf B-type, class 1} ~ $|B\rangle$,
\bea
\label{Bphi5}
&&\Phi(s',0)=\sqrt{k}
\sqrt{S^c(s',0|r=0)}=\frac{\sqrt k}{(k(k-2))^{1/4}}
\sqrt{\frac{\sinh\big(\frac{2\pi s'}{k-2}\big)\sinh(2\pi s')}
{|\cosh(\pi s')|^2}}
~,\\
\label{Bphi6}
&&\Phi(s',k)=\sqrt{k}
\sqrt{S^c(s',\frac{k}{2}|r=0)}=
\frac{\sqrt k}{(k(k-2))^{1/4}}
\sqrt{\frac{\sinh\big(\frac{2\pi s'}{k-2}\big)\sinh(2\pi s')}
{|\cosh(\pi s')|^2}}
~,\\
\label{Bphi7}
&&D(j,-j)=\sqrt{k}
\sqrt{S^d(j,-j|r=0)}=\frac{\sqrt{2k}}{(k(k-2))^{1/4}}
\sqrt{\sin\frac{\pi (2j-1)}{k-2}}~, \\
\label{Bphi8}
&&D(j,\frac{k}{2}-j)=\sqrt{k}
\sqrt{S^d(j,\frac{k}{2}-j|r=0)}=\frac{\sqrt{2k}}{(k(k-2))^{1/4}}
\sqrt{\sin\frac{\pi (2j-1)}{k-2}}~. \eea

{\bf B-type, class 2} ~ $|B;s,\alpha \rangle,   ~ s \geq 0, ~ \alpha = 0,\frac{1}{2}$
\bea
\label{Bsa5}
&&\Phi_{s,\alpha}(s',0)=\frac{4\sqrt
k}{(k(k-2))^{1/4}} \cos \big (\frac{4\pi s s'}{k-2}\big)
\sqrt{\frac{|\cosh(\pi s')|^2}{\sinh\big( \frac{2\pi s'}{k-2}\big)\sinh(2\pi s')}}~,\\
&&\Phi_{s,\alpha}(s',k)=\frac{4\sqrt k}{(k(k-2))^{1/4}}
e^{2\pi i \alpha} \cos \big (\frac{4\pi s s'}{k-2}\big)
\sqrt{\frac{|\cosh(\pi s')|^2}{\sinh\big( \frac{2\pi s'}{k-2}\big)\sinh(2\pi s')}}~,\\
&&D_{s,\alpha}(j,-j)=0,\\
&&D_{s,\alpha}(j,\frac{k}{2}-j)=0~.
\eea

\subsection{D-branes on $SL(2,\R)/U(1)$ from $AdS_3$}

The geometry of D-branes in
the $SL(2,\R)$ WZW model, which
describes strings propagating in $AdS_3$,
has been studied semiclassically in
\cite{Bachas:2000fr}.
In that paper it was found that solutions of the DBI action
correspond to regular and twined conjugacy classes of $SL(2, \R)$.
More specifically, one finds
D1-branes with $AdS_2$ worldvolumes and two types of D2-branes with
either $H_2$ or $dS_2$ worldvolumes.
The first type of D2-branes has a D-instanton density
on its worldvolume and the second is
unphysical due to a supercritical worldvolume electric field.

D-branes in coset models $G/H$ have
worldvolumes localized on the projection of the
product of a conjugacy class of $G$ with a conjugacy class of $H$
onto the coset \cite{Fredenhagen:2001kw}.
In our case of interest, namely $SL(2,\R)/U(1)$,
we expect that the cigar D-branes will
be projections on a constant $\tau$ plane of the
$SL(2, \R)$ D-branes.
Here $\tau$ denotes the time coordinate of $AdS_3$
that is being gauged.
The semiclassical aspects of cigar D-branes have been
analyzed in \ci{Fotopoulos:2003vc, Ribault:2003ss}\footnote{See
also \ci{Nakayama:2004vy} for a recent analysis of the
supersymmetric coset in a similar spirit.}.
Here we review some of their features
for comparison with our D-branes.

\begin{figure}
\begin{center}
\includegraphics[width=12cm]{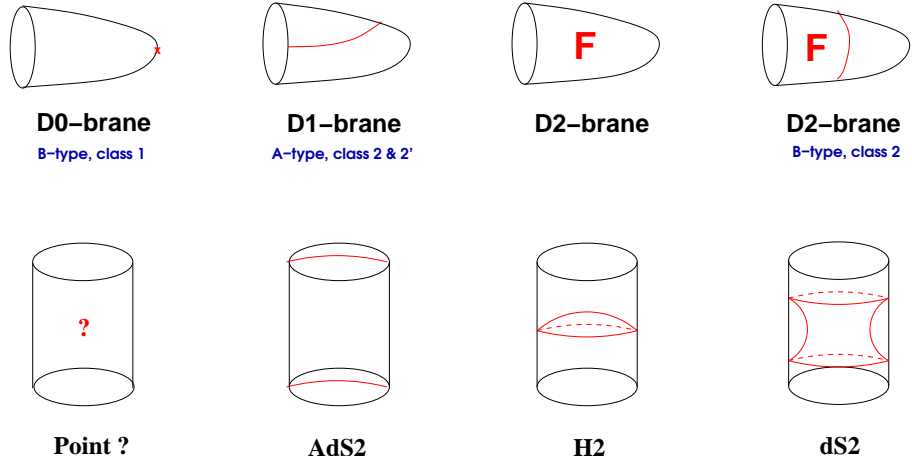}
\end{center}
\caption{D-branes on the cigar and corresponding $AdS_3$ branes}
\label{fig:branes}
\end{figure}

For D1-branes on the cigar the worldvolume can be written
as a function $\rho(\theta)$ and
we need to minimize the "energy" of the system $ E = \dot{\rho}
\frac{\partial {\cal L}}{\partial \dot{\rho}} - {\cal
L}$. Here the dot denotes a derivative with respect to
$\theta$ and ${\cal L}$ is the DBI Lagrangian.
For the D2-branes one has to solve the equations of
motion for the gauge field $F_{\rho \theta}$. The analysis is
straightforward and is summarized in Table 1. There
we give the embedding equations for the various branes
expected semiclassically as well as the $SL(2, \R)$
branes from which they descend.
$k$ is the level, $N$ is the quantized
D0-brane charge on the D2-branes, $\rho_{min}$ is the minimum distance from the tip,
and $\theta_0$, $\theta_0 + \pi$ are the points where the two branches
of the D1-branes reach the asymptotic circle.
Notice that by D$p$-brane here we mean a $p$-dimensional
object in the cigar geometry.
In \fig{fig:branes} we depict the geometry of these branes on
the cigar as well as the $SL(2, \R)$ branes from which
they descent.

\begin{eqnarray}
\label{T1}
\begin{array}{|c||c|c|c|}
\hline SL(2,\R) &{\rm Cigar} & {\rm Embedding\ equations} & {\rm
Moduli}

\\
\hline H_2 & {\rm D2}& 2\pi F_{\rho\theta}=k\frac{C\tanh
  \rho}{\sqrt{\cosh^2\rho-C^2}} & C=\sin \ang N, N \in \N
\\
\hline dS_2 & {\rm D2} & 2\pi F_{\rho\theta}=k\frac{C\tanh
  \rho}{\sqrt{\cosh^2\rho-C^2}} & C=\cosh \rho_{min}\geq 1,\ A_{\theta}
\\
\hline AdS_2 & {\rm D1}&  \sin (\theta -\theta_0)=\frac{\sinh \rho_{min}}{\sinh \rho} &
\rho_{min}\geq0,  \theta_0 \in [0,2\pi)
\\
\hline ? & {\rm D0} & \rho=0 & {\rm none}
\\
\hline
\end{array}
\nonumber
\end{eqnarray}
\begin{center}
{\bf Table 1}
\end{center}

We see that the cigar D1-branes, which
descend from $AdS_2$ branes in $AdS_3$, have two continuous classical moduli,
$\theta_0$ and $\rho_{min}$.
The DBI analysis for the D2-branes that cover
the whole cigar implies that they carry a D0-brane charge, which leads
to the quantization of the modulus $C$.
Since they have trivial topology, they do not admit a non-trivial Wilson line.
Finally, for the D2-branes that descend from $dS_2$
branes we expect a continuous modulus and the
possibility to have a Wilson line due to the nontrivial topology.
Notice that in this semiclassical analysis
we have not been able to identify the $SL(2,\R)$ brane
that yields the D0-branes on the cigar.
On the other hand, the analysis of \cite{Ribault:2003ss}
has a natural, although geometrically obscure, candidate: the spherical branes of
\ci{Ponsot:2001gt} with imaginary radius. These branes preserve the
rotational symmetry of the cigar.

The boundary state analysis of the previous sections yields a similar set of branes.
The class 1 branes have an open string spectrum that includes only the identity
module and they have Dirichlet boundary conditions along
$\rho$. The class 2 branes have Neumann boundary
conditions in the radial direction.
Now, we proceed to analyze in detail
each of these boundary states.

\subsection{A-type, class 2 and class 2'}

These boundary states describe D1-branes extending
along the radial direction of the cigar. They have Dirichlet boundary conditions in
the $\theta$ coordinate and the simple pole of their wavefunctions at $s'=0$
suggests Neumann boundary conditions on $\rho$. This is in line
with the interpretation of the $\NN=2$ A-type, class 2 boundary states
of \cite{Eguchi:2003ik} as analogues of the FZZT Liouville branes
that extend along the Liouville direction \cite{Fateev:2000ik, Teschner:2000md}.

As expected semiclassically, the type-A, class 2 and class 2' branes possess two branches.
In our exact conformal field theory analysis, this is manifest
from the computation of the open string spectrum (\ref{CD14}), (\ref{CD20}) with
 $s_1=s_2$ and $m_1=m_2$, i.e. for a single
D1-brane. This spectrum includes open string
states with either integer or half-integer winding $w$.
Indeed, these are precisely the spectra encoded in the
extended continuous characters $\Lambda^c_{\frac{1}{2}+is,0}$ and
$\Lambda^c_{\frac{1}{2}+is,\frac{k}{2}}$ in the annulus amplitudes
(\ref{CD14}), (\ref{CD20}).

The class 2' boundary states are by construction the ones appearing
in \cite{Ribault:2003ss}. The class 2 are closely related.
In order to make the comparison with the result of \cite{Ribault:2003ss} more
transparent,
we can rewrite the denominator in our class 2 (2') boundary states
in terms of $\Gamma$ functions. With the use of
(\ref{coshidentity}) and
the following identities
\bea \label{gamma}
& &\Gamma(1+x+iy)\Gamma(1-x+iy)\Gamma(1+x-iy)\Gamma(1-x-iy)= \frac{2
\pi^2(x^2+y^2)}{\cosh(2\pi y)-\cos(2 \pi x)}, \nonumber\\
& &\sinh(\pi x)= \frac{ \pi x}{\Gamma(1+ix)\Gamma(1-ix)},\nonumber
\eea
we obtain (up to an unimportant phase factor) the class 2 wavefunctions in the form
\beq
\label{ourd1}
\Psi_{s,m}(s',m')=
\frac{2}{(k (k-2)^3)^{1/4}} e^{i \delta(m')}
e^{-\pi i {m m' \over k}}
\frac{\Gamma(\frac{2is}{k-2})\Gamma(1+2is)}{\Gamma(\ha+{m \over
2}+is)\Gamma(\ha-{m \over 2}+is)}\cos ({4 \pi s s' \over k-2})
~.
\eeq
The corresponding wavefunctions of \cite{Ribault:2003ss},
which motivated our class 2' boundary states, read
\beq
\label{RSD1}
(\Psi^j_n)_{(r,\theta_0)}=
\frac{e^{i n \theta_0}}{(k (k-2)^3)^{1/4}}
\bigg( {\Gamma(1-b^2) \over \Gamma(1+b^2)}\bigg)^{is}
{\Gamma(1+2is)
\Gamma(\frac{2is}{k-2} ) \over \Gamma( \ha +is + {n \over 2})\Gamma( \ha
+is - {n \over 2})} \frac{e^{-2irs} + (-1)^n e^{+2irs}}{2}
~,
\eeq
where $b^2=\frac{1}{k-2}$ and $j=-\frac{1}{2}+is$.
Using the identifications
\bea
\label{thetam}
\theta_0&=&-\pi \frac{m}{k},\\
\label{rs}
r&=&2\pi \frac{s}{k-2},
\eea
and setting $n=m'+2kl, \; l \in \Z, \ m' \in \Z_{2k}$, we obtain full
agreement between our class 2' boundary states and (\ref{RSD1})
and partial agreement
between the class 2 expression (\ref{ourd1}) and (\ref{RSD1}).
Incidentally, notice that the
cylinder amplitude (4.17) in \cite{Ribault:2003ss}
can be recast in terms of the extended characters for integer levels $k$,
therefore pointing towards the extended
Ishibashi states we have been using throughout this paper.

The basic differences between our class 2 D1-branes
and those of \cite{Ribault:2003ss} are the following. First, recall that the
latter branes reach the asymptotic circle of the cigar at $\rho
\rightarrow \infty$ on two opposite points $\theta_0$ and
$\theta_0+\pi$ and the minimum distance from the tip is parametrized
by $r=\rho_{min} \geq 0$.
In \cite{Ribault:2003ss} both of these parameters are
continuous. Matching them to the parameters of our boundary
state (see eq.\ (\ref{thetam})) shows that our angle $\theta_0$ is discrete, because
$m \in \Z_{2k}$.  Hence,
our D1-branes can reach the circle at infinity only at $2k$
special points. This is a simple manifestation of the fact that our branes
preserve a larger chiral symmetry than the Virasoro, and it is precisely
what one expects from a construction based on extended characters.

The most important difference, however, concerns the
coupling of each brane to the closed string modes.
The wavefunctions agree for even momentum $n$, but they
disagree for odd momentum $n$.
This difference affects crucially the semiclassical
properties of the class 2 states. In order to make this point clear,
we can compare the spectral densities appearing on each class.

First of all, the densities
obtained in Section 4 for the class 2' -  class 2' overlaps
should be in perfect agreement with those appearing in  \cite{Ribault:2003ss},
if we choose the same regularization scheme. This should be true by construction.
Indeed, considering for simplicity the self-overlaps $s_1=s_2$ and $m_1=m_2$ and
subtracting the spectral densities of a
reference brane with $s=0, m=0$, yields the
relative spectral densities:
\beq
\label{2p5}
\rho_5(s';s_1|s_1)-\rho_5(s';0|0)=
\frac{1}{2}\Big(\rho_2(s';2s_1|0)-\rho_2(s';0|0)\Big)
\eeq
for states with integer winding, and
\beq
\label{2p6}
\rho_6(s';s_1|s_1)-\rho_6(s';0|0)=
\frac{1}{2}\Big(\rho_1(s';2s_1|0)-\rho_1(s';0|0)\Big)
\eeq
for states with half-integer winding. It is now a simple
exercise to check that under the identification (\ref{rs})
and the explicit expressions (\ref{rho1}) and (\ref{rho2})
the densities above match (up to a factor of 4
expected from the different normalization of the wavefunctions)
with the corresponding relative spectral densities of   \cite{Ribault:2003ss} (relations
(4.8), (4.9), (4.10), and (4.11) in that reference).
Of course, one needs to perform a summation over $l$ in the result
of  \cite{Ribault:2003ss} in order to rewrite it in terms
of extended characters.

As has been discussed in \cite{Ribault:2003ss}, the above
densities possess the expected
semiclassical behavior, i.e. $\rho_5(s';s_1|s_1)-\rho_5(s';0|0)$
is finite in the limit $k \rightarrow \infty$,
whereas $\rho_6(s';s_1|s_1)-\rho_6(s';0|0)$ is not well-defined,
because it receives contributions from purely stringy states that
decouple in the semiclassical limit.
The situation is reversed, however, when we consider
the class 2 boundary states. In particular, the
regularized density for zero winding (field theory) modes
$\rho_1(s';s_1|s_1)-\rho_1(s';0|0)$ is ill-defined
in the large $k$ limit, implying that the semiclassical
interpretation of these boundary states is problematic.
This pathology could be an indication
that the class 2 branes are unphysical.

In order to elucidate further the physical properties of the class 2 branes,
we can compute the semiclassical limit of the one-point
functions in the coordinate system introduced in
\cite{Ponsot:2001gt} for $H_3^+$. First,
we write the metric of $H_3^+$
as
\bea
\label{metricH3}
ds^2&=& d \psi^2 + \cosh ^2 \psi \;
(e^{2 \chi} d \nu^2 + d \chi^2) \\
&=& d\phi^2 + e^{2\phi} d\gamma d\bar{\gamma},
\eea
with $\psi$ an angular coordinate and
$\chi, \phi \in [0,\infty)$ radial coordinates.
In this coordinate system the $AdS_2$ branes are given by surfaces of
constant $\psi=r$.
They descend, as argued in \cite{Ponsot:2001gt,
Ribault:2003ss}, to D1-branes on the cigar. The one-point
function of the resulting branes, ignoring factors unimportant to the
present calculation, is conveniently written from (\ref{RSD1}) as
\beq
\label{RSD1b}
(\Psi^j_n)_{(r, \theta_0)} \sim
\Gamma(1+b^2(2j+1)) \; d^j_n \; \Big( \pi_n^0 \cosh r
(2j+1) - \pi^1_n \sinh r(2j+1) \Big),
\eeq
with the definitions:
\bea
\label{dpi} d^j_n&=& {\Gamma(2j+1)
\over \Gamma( 1+j + {n \over 2})\Gamma( 1+j - {n \over 2})},
\nonumber \\
\pi_n^{\epsilon}&=& \left\{
\begin{array}{lll}
1-\epsilon, & n&:{\rm even}, \\
\epsilon,   &n&:{\rm odd}.
\end{array}\right.
\eea

Now we can Fourier transform between the $j,n,w$ and
the $j,u, \bar{u}$ coordinates
using the following formulae \cite{Ponsot:2001gt}:
\bea
\label{Fourier}
F^j_{n,w}&=& \int_{\mathbb{C}} d^2u \;e^{-i\;n \;{\rm arg}(u)}
|u|^{2j+kw} f(u), \\
\label{invFourier}
f(u)&=& i k (2\pi)^{-2} \sum_{n \in \Z} \int_{\R}
dw \;e^{i \;n\; {\rm arg}(u)} |u|^{-2j-2-kw} F^j_{n,w}
~.
\eea
One can also show the identity
\beq
\label{intA}
d^j_n \; \pi^{\epsilon}_n \;\delta(w)= (2\pi)^{-2} \int_{\mathbb{C}}
d^2u \;e^{i\ n\;{\rm  arg}(u)} |u|^{-2j-2-kw} |u + \bar{u}|^{2j}
{\rm sign}^{\epsilon}(u+\bar{u})
~,
\eeq
which will be used in a moment.

The integral of the
Laplace operator eigenfunctions  $\Phi^j(u|\psi,\chi,\nu)$  \cite{Ponsot:2001gt}
over the semiclassical one-point
functions gives the geometrical shape of the corresponding D-brane.
For the $AdS_2$ branes and by using (\ref{intA}), this computation yields
\beq
\int dj \int d^2u \;
\Big(\Phi^j(u|\psi,\chi,\nu)\Big)^*
|u+\bar{u}|^{2j} \Big(\cosh r(2j+1)-
{\rm sign}(u+\bar{u}) \sinh r(2j+1)\Big) \sim \delta (\psi -r),
\eeq
therefore describing a
D1-brane as in Table 1.

One can determine the shape of the class 2
boundary states in the semiclassical limit in a similar way. First we need to
write down the corresponding wavefunctions in a form similar
to (\ref{RSD1b}):
\beq
\label{A2wf}
(\Psi^j_n)'_{(r, \theta_0)} \sim \Gamma(1+b^2(2j+1)) \;d^j_n \;
\Big( \pi_n^0 \cosh r (2j+1) -  i\pi^1_n \cosh r(2j+1) \Big).
\eeq
Integrating them as above with the use of the eqs.\ (\ref{Fourier}) and (\ref{intA})
gives
\bea
\label{A2b}
\int &dj& \int d^2u \;
\Big(\Phi^j(u|\psi,\chi,\nu)\Big)^* |u+\bar{u}|^{2j} \Big(\cosh r(2j+1)- i\;
{\rm sign}(u+\bar{u}) \cosh r(2j+1)\Big) \sim \nonumber\\
 \int_{0}^{\infty} &dP & \Big( \sin 2P(\psi+r) + \sin
2P(\psi-r)\Big) + \delta (\psi -r) +
\delta (\psi +r).
\eea
The integral in the second line comes from the term
involving ${\rm sign}(u+\bar{u})$ and it can be traced back to the
contribution of odd momenta to $\pi_n^1$. This integral is not well
defined and does not lead to a sensible geometrical interpretation of
the class 2 D1-branes.

A possible way out is to consider a linear
combination of the A-type class 2 wavefunction (\ref{C2A}) with its
complex conjugate, in particular $(\Psi^j_n)^{even}_{(r,
\theta_0)}=(\Psi^j_n)_{(r, \theta_0)}+(\Psi^j_n)_{(r,
\theta_0)}^*$. In this case, the only non-zero couplings to
closed strings are the ones with even momentum $n$. As a consequence,
we only get the $\delta$ function terms
in (\ref{A2b}). This boundary state describes two D1-branes with parameters
$(r, \theta_0)$ and $(-r, \theta_0)$.
This, however, is exactly the same boundary state that one would obtain from a linear
superposition of two D1-branes of \cite{Ribault:2003ss}.
Therefore, it seems that employing the Cardy ansatz
does not lead to D1-branes with nice semiclassical features and we are probably left
only with the class 2' branes of  \cite{Ribault:2003ss}.

Finally, we can check if the class 2 states satisfy
the $H_3$ factorization constraints derived in \cite{Ponsot:2001gt}.
It was shown there that the allowed  one-point functions for
the $AdS_2$ branes (and correspondingly for their
descendants on the cigar), have the following form
\beq
\label{Fact1}
(\Psi^j(u|{i \over 2}))_r= |u+\bar{u}|^{2j}
\nu_b^{j+\ha} \Gamma(1+b^2(2j+1)) E_{\sigma}(j|r)
\eeq
with
$\nu_b$ a normalization constant and $\sigma={\rm sign}(u+\bar{u})$.
The function $E_{\sigma}(j|r)$ is restricted by factorization
constrains to have the form
\beq
\label{Fact2}
E_{\sigma}(j|r)=
(e^0(j|r) + e^1(j|r) \sigma) e^{-r(2j+1)\sigma},
\eeq
with
$e^{\epsilon}(-1-j|r)=(-1)^{\epsilon} e^{\epsilon}(j|r)$.
The general A-type, class 2 boundary state
with Fourier transformed
wavefunction (\ref{A2wf})
is not of the above form and hence does not solve the $H_3$ factorization
constrains.
It would be an important
but challenging task to determine the consistency of these
boundary states by analyzing the factorization constraints
directly in the cigar conformal field theory.

\subsection{B-type, class 1}

We find a single B-type, class 1 brane. This boundary state
has Neumann boundary conditions on the angular direction of the
cigar (it is B-type) and is localized in the radial direction (it is class 1). A
circular D1-brane wrapped around the cigar will slip towards
the tip to lower its energy. Therefore, our B-type,
class 1 brane will eventually be localized at the tip of the cigar, where it should
rather be interpreted as a D0-brane. Perhaps it is useful to think
of this brane as the analog of the $SU(2)/U(1)$ B-type brane
with $j=0$ that sits at the center of the $SU(2)/U(1)$ disk
\cite{Maldacena:2001ky}. However, unlike those branes,
which are unstable and can move away from the center
of the disk, the D0-brane of the cigar is stable and cannot move
from the tip. Analogous branes
were formulated for the cigar CFT in \cite{Ribault:2003ss}. These
branes were parametrized by a single discrete parameter
$m=1,2,...$, but as explained in that paper, only the $m=1$ brane
appeared to be physical. A comparison of the properties of our
B-type, class 1 brane with those of the $m=1$ brane of
\cite{Ribault:2003ss} below indicates that these are the same
branes. Notice that we have restricted our analysis only to unitary
representations. As a result, our approach does not yield the $m>1$
boundary states of \cite{Ribault:2003ss}, which carry non-unitary open string spectra.

The B-type, class 1 boundary state couples to both continuous and
discrete closed string modes. The appropriate wavefunctions are
listed in eqs.\ (\ref{Bphi5})-(\ref{Bphi8}) and they are in
agreement
with the expressions appearing in
\cite{Ribault:2003ss}. For the continuous part the wavefunctions read:
\beq
\label{RS1B}
\Psi(j,w)_{m=1}^{D0}=(-)^w (k(k-2))^{1/4} \Big( {\Gamma(1-b^2) \over
\Gamma(1+b^2)}\Big)^{j+\frac{1}{2}}\frac{ \Gamma(-j+\frac{kw}{2})
\Gamma(-j-\frac{kw}{2})}{\Gamma( -b^2(2j+1))\Gamma(-2j)}
~.
\eeq
where $b^2={1\over k-2}$ and $j=-\ha +is$ in the notation of \cite{Ribault:2003ss}.
Indeed, up to a phase factor that does not affect the analysis
of the cylinder amplitudes, we find that these wavefunctions depend
on $w$ only through its parity (i.e.\ on whether $w$ is even or odd).
With a few straightforward algebraic manipulations one can show that
they can be brought into the form
\beq
\label{RS1B1}
\Phi(s,a)= (k(k-2))^{1/4} \Big( {\Gamma(1-b^2) \over
\Gamma(1+b^2)}\Big)^{is} \frac{ \Gamma(\ha+\frac{ak}{2}-is)
\Gamma(\ha-\frac{ak}{2}-is)}{\Gamma( -b^22is)\Gamma(1-2is)}
~,
\eeq
with $a=0,~ 1$.
Similarly, for the discrete representations one obtains the wavefunctions
\beq
\label{RS1B2}
D(j, {ak \over 2}-j)= \sqrt{2}\bigg(\frac{k}{k-2}\bigg)^{1/4}
\sqrt{\sin(-b^2\pi (2j+1))}
~.
\eeq
The above wavefunctions yield a cylinder amplitude $\langle B|e^{-\pi T H^c}|B\rangle$,
which agrees with the result following from eqs.\ (\ref{Bphi5})-(\ref{Bphi8}).
There is a small difference in the contribution of the discrete representations,
which we now flesh out.

The discrete part of the above cylinder amplitude involves a
summation over the quantum numbers $j$ and $w$. In \ci{Ribault:2003ss}
(see eq. (3.18)) this summation runs over the restricted
set $\JJ^d_{0,w} := \big[\frac{1}{2},\frac{k-1}{2} \big]
\cap \big( \frac{1}{2}|kw|-\N  \big)$. Our result involves a similar
summation over $j$ and $w$, but $\JJ^d_{0,w}$ should be
replaced by the unrestricted interval
$\big( \frac{1}{2},\frac{k-1}{2} \big)$. Notice, however,
that this difference affects only the $w=0$ contribution.
Indeed, we are dealing with even levels $k$, so can set
$k=2 \nu$, where $\nu$ is some positive integer. Then, the condition
that $j$ belongs in $\JJ^d_{0,w}$ requires the existence of a positive
integer $n$, such that
$j=-n+1/2|kw|= -n+\nu|w|$ and such that it satisfies the inequality
$1/2+ \nu(|w|-1)\leq n \leq \nu|w|-1/2$. $n$ will be necessarily positive for any $k > 2$.
It should be clear that for each $w \in \Z^*$ we can always find
an appropriate integer $n$ in the range defined above and for such $n$'s
we get the same values of $j$ for every possible non-vanishing value of $w$. The
value $w=0$ is the only exception. It is therefore excluded in the
computation of \ci{Ribault:2003ss}, but it is included in our computation.
Unfortunately, we cannot offer an explanation for this discrepancy.

Despite this minor difference, the cylinder amplitudes agree perfectly
in the open string channel. The result appearing in (\ref{BB}) can be recast
in terms of the standard $\lambda^d_{j,r}$ discrete characters,
using the identity (\ref{D11}). One finds:
\beq
\label{LIrrc}
\sum_{r \in \Z _k}\Lambda^I_{r}(\tau) = \sum_{l\in \Z}
(\lambda^d_{0,l}(\tau)-\lambda^d_{1,l-1}(\tau))=q^{-\frac{c_{cs}}{24}}\Big(
1+ 2q^{1+{1\over k}} + q^2+ O(q^{2+ {1\over k}}) \Big)
~.
\eeq
In this expression $\lambda^d_{0,n}$ denotes the extrapolation of the
discrete character (\ref{chardiscr}) to $j=0$ and not the
character (\ref{charid}) of the identity representation. This is
precisely the result found in eq.\ (3.6) of \ci{Ribault:2003ss}.

As we see, the open string spectrum of the B-type, class 1 brane is discrete
and there is no open string winding. This is a simple
manifestation of the fact that the B-type, class 1 brane is
rotationally invariant. For finite level $k$ the one-loop result in
(\ref{LIrrc}) suggests that the open string spectrum does not contain
any marginal modes. This is consistent with the semiclassical analysis, which
fixes the position of the D0-brane at the tip of the cigar. In the flat
space limit, however, two states with scaling dimension $h=1$ appear
signaling the free motion of a D0-brane in a flat 2-dimensional space.

\subsection{B-type, class 2}

These branes are new and have not been considered before
in the boundary conformal field theory context. They are
labeled by a continuous parameter $s \in [0,\infty)$
and they have Neumann boundary conditions both in
the angular and radial directions.
Hence, their shape is that of 2-dimensional D-branes.
Since they do not couple to any discrete modes, which are localized near
the tip of the cigar, it is natural to expect that in general these branes
do not reach the tip and hence they are candidates for the
D2-branes descending from  $dS_2$ branes of
$AdS_3$ in Table 1. Such branes develop
a circular opening at a distance $\rho_{min}$ away from
the tip and terminate there. The parameter $\rho_{min}$ is directly
related to the continuous parameter $s$. For example, we can easily verify
that both parameters scale to infinity when we send $k \rightarrow \infty$.
For finite $k$ notice that when we send $s \rightarrow 0$, the circular opening
of the D2-brane shrinks to zero radius and the brane covers the
whole cigar.

According to the semiclassical analysis of subsection 5.2 these D2-branes
can have a non-vanishing $\Z_2$ Wilson line because of the non-trivial cycle
on their worldvolume. The anticipation of a $\Z_2$-valued, instead of $\R$-valued,
Wilson line will be better clarified in a moment with the use of a T-duality argument.
The possibility of the Wilson line is realized explicitly by our B-type, class 2 branes.
The one-point functions in (\ref{Bsa}) depend on the extra $\Z_2$-valued
parameter $\alpha =0 , ~ \frac{1}{2}$. The Cardy computation in (\ref{BBcc})
shows clearly that the overlap of two branes with the same values
$\alpha_1=\alpha_2$ results in open string states with even index $m$.
On the other hand, the same computation for branes with different values of
$\alpha_1$ and $\alpha_2$ involves open string states with odd $m$. The explicit form of
the extended character $\Lambda^c_{\ha + is, \ {m \over 2}}$ in (\ref{D4}) reveals that
even $m$ indices correspond to integer momentum modes in the angular
direction, whereas odd $m$ indices correspond to half-integer momentum modes.
Therefore, this open string spectrum suggests that $\alpha$ parameterizes a
$\Z_2$ Wilson line. It is known (see e.g.\ \ci{Sen:1998tt}) that if we
have a circular D1 string, the presence of a $\Z_2$ Wilson line
results, under a $2 \pi$ translation around the $S^1$,
into a minus sign in the wavefunction of an open string ending on the D1-string.
In our case, that would imply half-integer angular momenta for open strings
stretching between an $\alpha=0$ and an $\alpha=\frac{1}{2}$ brane.
As a consequence, in the field theory limit one would expect, for example,
a tachyon field of the form \ci{Sen:1998tt}:
\beq
\label{tachyon}
T(\theta,\rho)= \sum_{n \in \Z} T_n(\rho) e^{i(n + \ha)\theta}
~.
\eeq
Indeed, this is what we observe in the open string spectra of our branes.
Eq.\ (\ref{D4}) with $s=0$ and $m=2n+1$ ($n \in \Z$) shows that
the open string spectrum includes $l=0$ modes with scaling dimensions
$h={(n+ \ha)^2 \over k}$. In the semiclassical limit of
large $k$ these are precisely the scaling dimensions of tachyon modes
with half-integer angular momentum on a circle of radius $\sqrt{k}$.

The $\Z_2$ Wilson line can also be derived by the following
T-duality argument.
As we briefly mentioned in Section 2, the trumpet geometry can be
obtained from the cigar by T-duality. This T-dual background can
also be viewed as a $\Z_k$ orbifold of the cigar. This is most apparent in the
asymptotic cylinder region, where the $Z_k$ orbifold reduces
the asymptotic radius $R_{cig}$ to $R_{cig}/k$. To
avoid the subtleties due to the singularity in the trumpet geometry we can
restrict our discussion only on the asymptotic cylinder region. On the trumpet we
can formulate D1-branes, which are the analogs of the A-type, class 2 (2')
branes on the cigar (see \cite{Fotopoulos:2003vc}).
Again, these branes will be parameterized by a continuous parameter $s$
and geometrically will correspond to D1-branes starting from the asymptotic
infinity and returning back to it. They will be perturbative and stay far
from the trumpet singularity for large parameters $s$.
The boundary states of these D1-branes can be deduced easily from
the analysis of the A-type, class 2 (2') branes on the cigar. They are
simply $\Z_k$ orbifold projections of those branes. In other words,
they can be obtained by summing over $k$ images of the
A-type, class 2 (2') boundary states\footnote{A similar argument was employed
in \cite{Maldacena:2001ky}
to uncover B-type branes in the parafermion model.}.
To be definite, let us consider only
the A-type, class 2' cigar branes, which have better semiclassical behavior.\footnote{It
turns out that for the purposes of the present argument we could also use
the class 2 branes and obtain the same result.}
There are two possible sums over $k$ images that we can write
and one can easily show that:
\bea
\label{Tdeven}
|A;s,1\rangle_{trumpet} \equiv
\frac{1}{\sqrt k} \sum_{l \in \Z_k} |A;s,l \rangle = \frac{1}{2}|B;s,\alpha =0\rangle ~,
\\
\label{Tdodd}
|A;s,2\rangle_{trumpet} \equiv
\frac{1}{\sqrt k} \sum_{l \in \Z_k} |A;s,l+\frac{1}{2} \rangle =
\frac{1}{2}|B;s,\alpha =\frac{1}{2}\rangle
~.
\eea
This result is precisely what we expect from T-duality. The A-type trumpet D1-branes
were expected to yield B-type D2-branes on the cigar and this is exactly what
we find.
The Wilson line is naturally $\Z_2$-valued, because there are only $2k$
D1-branes on the cigar and the $\Z_k$ orbifold leaves behind a remaining
$\Z_2$ degree of freedom. Hence, the fact that we do not have an $\R$-valued
Wilson line, as would be expected on geometric grounds, can be traced
back to the use of extended Ishibashi states that preserve more symmetry than just the
Virasoro.

As a final comment on the B-type class 2 branes notice that
the self overlaps in (\ref{BBcc}) yield an open string
spectrum that has a marginal deformation for $s=\sqrt{k-2}$ and $m=0$.
This indicates that our D2-branes are marginally stable to a change of the $s$ modulus.
Such a marginal mode is also expected from the semiclassical analysis. The DBI action
has a flat direction of solutions parameterized by $\rho_{min}$. Moreover,
one can compute the radial stress energy tensor component $T_{\rho  \rho}$:
\beq
\label{Trr} T_{\rho \rho}= {\partial {\cal L} \over \partial
F_{\rho \theta}} F_{\rho \theta} - {\cal L}= T_{D2} \Bigg( {\cosh \rho
(F_{\rho \theta})^2 \over \sqrt{\tanh ^2 \rho + (F_{\rho
\theta})^2 }} -\cosh \rho \sqrt{\tanh ^2 \rho + (F_{\rho
\theta})^2 } \Bigg)
\eeq
and show, using the equations of motion from Table 1,
that it is zero (and therefore continuous) at the minimum distance from the tip $\rho_{min}$.
This provides further confirmation for the existence of the above marginal deformation.

\subsection{B-type, class 3}

The analysis of Section \ref{CCC} revealed that the
class 3 (and most of the class 3') boundary states have
negative open string multiplicities and hence do not satisfy the
Cardy consistency conditions. Nevertheless, there are some striking similarities
between the class 3' boundary states of this paper and
the cigar covering D2-branes of \ci{Ribault:2003ss} and it is instructive to
compare the two results. For those branes we use the
form of the closed string couplings that appear in \ci{Israel:2004jt}\footnote{Again,
we specialize in the case of even $k$,
but similar manipulations should also go through for odd level $k$ with
minor modifications.}
\bea
\label{RSD2}
\langle \Phi^j_w({i \over 2}, -{i\over 2}) \rangle &=&
{b^2 \over 4} (\nu_b)^{j+ \ha} { \Gamma
(j+{kw \over 2})\Gamma (j-{kw \over 2})\over \Gamma( 2j-1) \Gamma
( 1-(2j-1)b^2)}
\nonumber\\
& &{ e^{i\sigma (2j-1)} \sin\pi(j+ {kw \over 2})+
e^{-i\sigma (2j-1)} \sin\pi(j- {kw \over 2}) \over \sin \pi (2j-1)
\sin \pi b^2 (2j-1) }=
\\
&=&2(-1)^{{kw \over 2}}{b^2 \over 4} (\nu_b)^{j+ \ha} {\Gamma
(j+{kw \over 2})\Gamma (j-{kw \over 2})\sin \pi j  \cos \sigma
(2j-1) \over \Gamma( 2j-1) \Gamma ( 1-(2j-1)b^2)\sin \pi (2j-1)
\sin \pi b^2 (2j-1)}
~. \nonumber
\eea
For $2j_1= (k-2) { \sigma \over \pi} +{k \over 2}$, and
$2j_2=-(k-2) {\sigma \over \pi}+{k\over 2}$, the class 3' boundary state wavefunctions
(\ref{BJ})-(\ref{BJ4}) become:
\bea
\label{B35}
\Phi_{j_1, j_2}(s,0)&=& 2\bigg({k \over k-2}\bigg)^{{1\over 4}} {\cosh 2\sigma s
\over \sqrt{ \sinh 2 \pi s b^2 \sinh 2 \pi s}} ~,\\
\Phi_{j_1, j_2}(s,k)&=& 2\bigg({k \over k-2}\bigg)^{{1\over 4}} e^{i \pi
{k\over 2}}{\cosh 2 \sigma s
\over \sqrt{ \sinh 2 \pi s b^2 \sinh 2 \pi s}} ~, \\
D_{j_1,j_2}(j,-j)&=& 2{i \over \sqrt{2}}\bigg({k \over k-2}\bigg)^{{1\over
4}}{1 \over \sqrt{\sin \pi b^2 (2j-1)}} e^{{i\pi \over
2}(2j-1)}\cos \sigma (2j-1) ~,\\
D_{j_1,j_2}(j,{k \over 2}-j)&=& 2{i \over \sqrt{2}}\bigg({k \over
k-2}\bigg)^{{1\over 4}}{1 \over \sqrt{\sin \pi b^2 (2j-1)}} e^{{i\pi
\over 2}(2j-1)}e^{i \pi {k\over 2}}\cos \sigma (2j-1)
~.
\eea
Notice that since $1 \leq 2j_i\leq k-1$ we can easily
obtain the quantization condition $\sigma=\pi {m \over k-2}$
with the bound $\sigma \in [0, {\pi \over 2})$.

It is not difficult to compare the above couplings
with those of (\ref{RSD2}). One finds that
the continuous couplings are the same except for several phase factors that do
not affect the cylinder amplitudes. The discrete couplings, on the other hand,
appear to be different. This should be expected. From
a similar comparison involving the B-type class 1 branes \cite{Ribault:2003ss}, we
know that for discrete representations properly normalized
one-point functions should be considered.
More precisely, in computing the
cylinder amplitudes using (\ref{RSD2}) one has to consider
discrete contributions of the form:
\beq
\label{RSD2disc}
\sum_{w, j} {\rm Res} \bigg({\Psi_{\sigma}(j,w) \Psi_{\sigma}(j,w)^*
\over \langle
\Phi^j_{0,w} \Phi^j_{0,w} \rangle} \bigg)\lambda^d_{j,\frac{kw}{2}} (iT)
~,
\eeq
where one uses the bulk two-point function of the operator
$\Phi^j_{0,w}$ to normalize the cylinder contributions for each
closed string mode.
It should be obvious from the last equation in (\ref{RSD2}) that the
dependence on $\sigma$ is the same as that in (\ref{B35}). Moreover, the $\Gamma$
functions in the numerator of (\ref{RSD2}) exhibit discrete poles
for $j \pm {kw \over 2}= -n, \ n\in \N$. For even levels $k$
this yields contributions from all the $integer$
values of $j \in \big[\frac{1}{2},\frac{k-1}{2} \big]$, precisely
as expected from the cylinder amplitude computation of section
\ref{CCC}.

Despite the apparent similarity of the above one-point functions,
the expression (\ref{BBmulti}) for the self-overlap differs
from the annulus amplitude computed in \ci{Ribault:2003ss,
Israel:2004jt}. In particular, these papers find that the contribution
of discrete open string states vanishes in the case of self-overlaps.
This result is in contrast to (\ref{BBmulti}) except for $\sigma=\pm \frac{\pi}{2}$,
in which case our class 3 B-type boundary states reduce to class 2 boundary states.
This vanishing, combined with the
negative open string multiplicities of the remaining cross-overlaps, led
the authors of \ci{Israel:2004jt} to conclude that only a single D2-brane should be
allowed.

We should point out, however, that the analysis of
\ci{Ribault:2003ss, Israel:2004jt} is, strictly speaking, valid only for
non-rational levels $k$. For rational levels there are additional subleties involving
contributions due to extra poles, which were not taken into account in the discussion of
\ci{Ribault:2003ss,Israel:2004jt}.
Hence, it is plausible that the results of those papers might not apply
directly to the case of integer levels $k$ that we study in this work.
If this is the case, then it would be
interesting to see if the extra pole contributions can explain the
discrepancy between our results and those appearing in
\ci{Ribault:2003ss,Israel:2004jt}.

\section{Summary and conclusions}

In this paper we formulated D-branes on the $SL(2,\R)_k/U(1)$ coset conformal field
theory at integer level $k$. The basic building blocks of our analysis were the
extended coset characters of \cite{Israel:2004xj}. With the use of the corresponding extended
Ishibashi states we proposed Cardy-like boundary states and verified the Cardy
consistency conditions. Although the discussion we presented here applies directly only to
integer levels $k$, the extension to rational levels should be straightforward.

Our analysis yielded the following results. We found two basic types of boundary states:
A- and B-type. We presented two A-type boundary states, which we called class 2 and
class 2'. Both of them are based on the continuous representations and geometrically
correspond to D1-branes starting and terminating at the asymptotic infinity.
The class 2 boundary states resulted from a direct application of
the Cardy ansatz and exhibit some puzzling semiclassical features. It is not completely
clear whether these branes are fully consistent.
The class 2' boundary states are a slight variant of the Cardy ansatz and they have
appeared previously in \cite{Ribault:2003ss}, where they were derived from
$AdS_2$ boundary states in the Euclidean $AdS_3$.
These branes have the expected semiclassical behavior.

We also presented two classes of B-type boundary states. These were called
class 1 and class 2. We found a single B-type class 1 boundary state,
which is based on the identity representation and corresponds to
a rotationally invariant D1-brane that wraps the circular direction of the cigar
and slips toward the tip in order to minimize its energy. In this way it
becomes effectively a pointlike object, i.e.\ a D0-brane.
The class 2 branes
are based only on continuous representations and they are D2-branes
(with a $\Z_2$-valued Wilson line) covering part of the cigar (see Fig. 1). An interesting
question concerns the formulation of D2-branes covering the whole cigar. These
are expected to be based on the discrete representations. We formulated
such branes in Section 3 as class 3 boundary states, but they were found in Section 4
to violate the Cardy consistency conditions and they had to be excluded.
Nevertheless, it would still be interesting to see if one can postulate
Cardy consistent class 3 boundary states as appropriate linear combinations of
the class 3 boundary states of Section 3, or with wavefunctions that do not follow
from the application of the Cardy ansatz. The simple linear superposition
that yielded the class 3' boundary states produced a set of
D2-branes that share many similarities with the cigar covering D2-branes
of refs.\ \ci{Ribault:2003ss, Israel:2004jt}. Only one of them was found, however,
to be Cardy consistent. A similar observation was made in \cite{Israel:2004jt}.

In conclusion we would like to mention a few open problems.
The analysis we presented was based exclusively on the study of
the Cardy consistency conditions. Hence, it should be supplemented by
a detailed study of the disk 1-point function factorization constraints.
This would provide additional evidence for the results presented in this paper and would
clarify important aspects of the open string theory on the D-branes we considered.
The analysis of the 4-point function conformal blocks is expected to be
the difficult part of this task. Another related task would
be the study of the boundary two- and three-point functions of the open
string theories presented here.

Moreover, it would be intriguing to study the implications of our results
for two-dimensional holography and the connection with the matrix model
of \cite{Kazakov:2000pm} in the spirit of
\cite{McGreevy:2003kb, Klebanov:2003km, McGreevy:2003ep}.
On a related context, one could also explore the possibility
of deriving the connection to the matrix model through an auxiliary
Riemann surface similar to the analysis that has been
performed for the minimal (super)string theory in
\cite{Seiberg:2003nm}.

Finally, it would be worthwhile to obtain boundary states
for the supersymmetric orbifold
\beq
\frac{SL(2,\R)_k/U(1) \times SU(2)_k/U(1)}{\Z_k}
\eeq
by extending the techniques used here.
This background appears in the exact CFT description of NS5-branes
(put symmetrically on a topologically trivial circle) and provides a
holographic dual of
Little String Theory in the double scaling limit
\cite{Giveon:1999px}. It is therefore of
obvious importance to understand the corresponding D-branes.
Furthermore, analyzing the dynamics of these branes along the lines of \cite{Elitzur:2000pq}
and in connection with Hanany-Witten type configurations
\cite{Hanany:1996ie, Ribault:2003sg} is another interesting open problem.

\section*{Acknowledgments}
We would like to thank D.~Israel, E.~Kiritsis, D.~Kutasov, J.~Troost, and
S.~Ribault for valuable discussions.
The work of A.~F. is supported in part by the  PPARC SPG grant 00613.
The work of V.~N. is supported in part by the DOE grant DE-FG02-90ER-40560.
The work of N.~P. is supported by the Deutsche Forschungsgemeinschaft under the
project number DFG Lu 419/7-2 and
in part by the European Community's Human Potential
Programme under contract HPRN--CT--2000--00131 Quantum Spacetime.

\newpage

\appendix

\section{The coset partition sum in terms of standard and extended characters}
\label{partition}

The spectrum of string theory on the axially-gauged (cigar) coset involves
only the continuous and discrete representations. For a semiclassical
discussion of this spectrum we refer the reader to \cite{Dijkgraaf:1991ba}.
A more recent discussion, from the point of view of the
torus partition sum, can be found in \cite{Hanany:2002ev}.
In this Appendix we review the basic features of that computation and
show explicitly how the torus partition sum can be recast in terms of the standard and
extended characters. The standard character form did not appear explicitly in \cite{Hanany:2002ev},
but it was already implicit in the considerations of that paper. The supersymmetric
version of the partition sum expression in terms
of extended characters appeared recently
in \cite{Eguchi:2004yi, Israel:2004ir}.

Our starting point is the torus partition
sum (3.9) of \cite{Hanany:2002ev}:
\bea
\label{39}
\ZZ (\tau,\bar{\tau}) &=& 2 (k(k-2))^{1/2} \int_0^1 ds_1 ds_2 \nonumber\\
& & \sum_{m,w \in \Z} \frac{(q\bar q)^{-2/24}}{|\sin(\pi(s_1\tau-s_2))|^2}
\frac{e^{-\frac{k\pi}{\tau_2}|(s_1+w)\tau-(s_2+m)|^2+\frac{2\pi}{\tau_2}(\Im
(s_1 \tau-s_2))^2}}{\displaystyle |\prod_{r=1}^{\infty}
(1-e^{2\pi i r\tau-2\pi i (s_1\tau-s_2)})(1-e^{2\pi i r\tau+2\pi i(s_1\tau-s_2)})|^2}
~. \nonumber\\
\eea
This result can be derived from
a direct path integral computation in the corresponding
gauged WZW theory.

With the use of the Poisson resummation formula in $m$
it is possible to recast (\ref{39}) into the form
\bea
\label{t1}
\ZZ(\tau,\bar{\tau}) &=& 8 \sqrt{(k-2)\tau_2} |\eta(\tau)|^2
\int_0^1 ds_1 ds_2 \frac{e^{2\pi \tau_2 s_1^2}}{|\theta_1(s_1\tau-s_2|\tau)|^2} \nonumber\\
& &\sum_{n,w \in\Z} e^{2\pi i n s_2} q^{\frac{1}{4k}(n-k(w+s_1))}
\bar{q}^{\frac{1}{4k}(n+k(w+s_1))^2}
~,
\eea
where
\beq
\label{theta1}
\theta_1(z|\tau)=-2e^{\pi i \tau/4} \sin \pi z \prod_{m=1}^{\infty}
(1-q^m) (1-e^{2\pi i z} q^m)(1-e^{-2\pi i z}q^m)~.
\eeq
We can expand this function in the denominator of the partition
sum, with the help of the formula ($y=e^{2\pi i (s_1\tau-s_2)}$)
\beq
\label{th1exp}
\frac{1}{\theta_1(s_1\tau-s_2|\tau)}=
\frac{i y^{1/2}}{\eta^3(\tau)}\sum_{r\in \Z} y^r S_r(\tau)
~,
\eeq
where
\beq
\label{Sr}
S_r(\tau)=\sum_{s=0}^{\infty} (-)^s q^{\frac{1}{2}s(s+2r+1)}
~.
\eeq
This expansion is valid only for $|q|<|z|<1$, which is indeed the case
for $0<s_1<1$.

Plugging the expansion (\ref{th1exp}) into the partition sum (\ref{t1})
and integrating out $s_2$ yields
\bea
\label{t2}
\ZZ(\tau,\bar{\tau})&=&8\sqrt{(k-2)\tau_2} \frac{1}{|\eta(\tau)|^4}
\int_0^1 ds_1 \sum_{r,\bar r, w \in \Z}
e^{2\pi \tau_2(\frac{k-2}{2}s_1^2+s_1(1+r+\bar r+kw))}
\nonumber\\
& & q^{\frac{1}{4k}(r-\bar r-kw)^2}\bar{q}^{\frac{1}{4k}(r-\bar r+kw)^2}
S_r(\tau)S_{\bar r}(\bar\tau)
~,
\eea
with the constraint  $n=r-\bar r$.
Following \cite{Hanany:2002ev} we can introduce an auxiliary variable
$s$ to linearize the $s_1$ integral. This can be achieved with the use of the identity
\beq
\label{saux}
\sqrt{(k-2)\tau_2}e^{-2\pi \tau_2(\frac{k-2}{2}s_1^2+s_1(1+r+\bar r+kw))}
=\int_{-\infty}^{\infty} ds ~ e^{-\frac{\pi s^2}{(k-2)\tau_2}-2\pi s_1(is+\tau_2(1+r+\bar r+kw))}
~.
\eeq
Then, we can easily perform the $s_1$ integral to obtain the partition sum in the form
\bea
\label{t3}
\ZZ(\tau,\bar{\tau})&=&\frac{8}{|\eta(\tau)|^4} \int_{-\infty}^{\infty}
ds ~ \sum_{r,\bar r,w\in \Z} q^{\frac{1}{4k}(r-\bar r-kw)^2}\bar{q}^{\frac{1}{4k}(r-\bar r+kw)^2}
S_r(\tau)S_{\bar r}(\bar\tau) \nonumber\\
& &\frac{e^{-\frac{\pi}{(k-2)\tau_2}s^2}}{2\pi (is+\tau_2(1+r+\bar r+kw))}
\bigg ( 1- e^{-2\pi (is+\tau_2(1+r+\bar r+kw))}\bigg)
~.
\eea

In the above expression, the contribution of the discrete representations comes
from the second term. The exponent of the exponential inside the parenthesis
can be completed to a square by shifting $s$ and by defining the new variable
$t=s+i(k-2)\tau_2$ we can write the second term as
\bea
\label{2ndterm}
&-&\frac{4}{i\pi |\eta(\tau)|^4} \int_{\CC}
dt ~ \sum_{r,\bar r, w\in \Z} q^{\frac{1}{4k}(r-\bar r-kw)^2}
\bar{q}^{\frac{1}{4k}(r-\bar r+kw)^2} S_r(\tau)S_{\bar r}(\bar{\tau})
\nonumber\\
& &e^{-\frac{\pi}{(k-2)\tau_2}t^2} e^{-2\pi \tau_2(\frac{k}{2}+r+\bar r+kw)}
\frac{1}{t-i\tau_2(k-1+r+\bar r+kw)}
~,
\eea
where $\CC$ denotes the contour $\CC=\{ t|t=s+i\tau_2 (k-2), s\in \R\}$.
We may close the contour at
$\pm \infty$ to form an infinite parallelogram between the real line and
$\CC$. Then,
the integrand has poles at the locations $t=i\tau_2(k-1+r+\bar r+kw)$
inside the parallelogram, i.e.\ it has poles for those integers $r$, $\bar r$
and $w$ for which $0<k-1+r+\bar r+kw<k-2$. The contribution of these poles
gives the contribution $\ZZ_d$ of the discrete representations
to the torus partition sum.
After setting $2j\equiv -kw-r-\bar r$ it is straightforward to
show that
\beq
\label{tordiscrete}
\ZZ_d(\tau,\bar{\tau})=8 \sum_{\frac{1}{2}<j<\frac{k-1}{2}}
\sum_{w,r,\bar r\in \Z} \delta(2j+r+\bar r+kw) \lambda^d_{j,r}(\tau)
\lambda^d_{j,\bar r}(\bar{\tau})
~.
\eeq
Since we assume that $k$ is an (even) integer we see that $2j \in \Z$.
Notice that in principle we can replace $\lambda^d_{j,r}$ by $\lambda^d_{\frac{k}{2}-j,-r}$
since they share the same conformal weights. However, their respective $J_0^3$ charges
are different and choosing the second one would not lead to a sensible
interpretation of these charges as left- and right-moving momenta
of a free compact boson, which is required by the BRST constraint
of the corresponding gauged WZW model. In general one can check
that the discrete part cannot be written in terms of characters
with $j \neq \bar j$. This will be important when we discuss
the allowed coset Ishibashi states.

The contribution of the continuous representations can be obtained in
a similar way. The integral over the first term in (\ref{t3}) plus the
integral along the real line that remains from (\ref{2ndterm})
give\footnote{The integral in  (\ref{2ndterm}) can be written
as  $\displaystyle
\int_{\CC} dt = \int_{-\infty}^{\infty} ds - 2 \pi i {\rm \;Res}$
with $s\in \R$.}
\bea
\ZZ_c(\tau,\bar{\tau})&=&\frac{16}{|\eta(\tau)|^4}   \int_{-\infty}^{\infty} \frac{ds}{2\pi}
\sum_{r, \bar r \in \Z} \sum_{n,w \in \Z} \delta_{r-\bar r, n}\;
q^{\frac{s^2}{k-2}+\frac{1}{4k}(n-kw)^2}
\bar{q}^{\frac{s^2}{k-2}+\frac{1}{4k}(n+kw)^2} S_r(\tau)S_{\bar r}(\bar{\tau})
\nonumber \\
& &\Big(\frac{1}{2 i s + r + \bar r + k w +1}-
\frac{(q{\bar q})^{\frac{1}{2}(\frac{k}{2}+r+ \bar r + k w)}}
{2 i s + r +\bar r +k (w+1) -1}
\Big) .
\eea
Compared to  (\ref{t3}) and (\ref{2ndterm}) we have rescaled
the integration variable as $s \rightarrow 2 \tau_2 s$.
This expression can also be written as
\bea
\label{lefteqn}
& &\ZZ_c(\tau,\bar{\tau})=\frac{16}{|\eta(\tau)|^4}   \int_{-\infty}^{\infty} \frac{ds}{2\pi}
\sum_{r, \bar r \in \Z} \sum_{n,w \in \Z} \delta_{r-\bar r, n} \;
S_r(\tau)S_{\bar r}(\bar{\tau}) \nonumber\\
& &\Bigg(\frac{q^{\frac{s^2}{k-2}+\frac{1}{4k}(n-kw)^2}
\bar{q}^{\frac{s^2}{k-2}+\frac{1}{4k}(n+kw)^2}}
{2 i s + r + \bar r + k w +1}-
\frac{q^{\frac{s^2}{k-2}+\frac{1}{4k}(n-k(w+1))^2+r}+
\bar{q}^{\frac{s^2}{k-2}+\frac{1}{4k}(n+k(w+1))^2+\bar r}}
{2 i s + r +\bar r +k (w+1) -1}
\Bigg) . \nonumber\\
\eea
Shifting $w \rightarrow w-1$ in the second term and using the relation
$q^r S_r = S_{-r}$ yields
\bea
\ZZ_c(\tau,\bar{\tau})&=&\frac{16}{|\eta(\tau)|^4}   \int_{-\infty}^{\infty} \frac{ds}{2\pi}
\sum_{r, \bar r \in \Z} \sum_{n,w \in \Z} \delta_{r-\bar r, n}\;
 q^{\frac{s^2}{k-2}+\frac{1}{4k}(n-kw)^2}
\bar{q}^{\frac{s^2}{k-2}+\frac{1}{4k}(n+kw)^2}
\nonumber \\
& &\Big(\frac{S_r S_{\bar r}}{2 i s + r + \bar r + k w +1}-
\frac{S_{-r} S_{-\bar r}}
{2 i s + r +\bar r +k w -1}
\Big) .
\eea
This result can be simplified further with the use of
the identity $S_{-r}=1-S_{r-1}$ and by shifting
$r\rightarrow r+1, \bar r \rightarrow \bar r + 1$ in the second term.
This allows us to bring the above expression into the form
\bea
\ZZ_c(\tau,\bar{\tau})
&=&\frac{16}{|\eta(\tau)|^4}   \int_{-\infty}^{\infty} \frac{ds}{2\pi}
\sum_{r, \bar r \in \Z} \sum_{n,w \in \Z} \delta_{r-\bar r, n} \nonumber\\
& &q^{\frac{s^2}{k-2}+\frac{1}{4k}(n-kw)^2}
\bar{q}^{\frac{s^2}{k-2}+\frac{1}{4k}(n+kw)^2}
\frac{S_r + S_{\bar r}-1}{2 i s + r + \bar r + k w +1} .
\eea
In terms of the continuous characters appearing in
(\ref{charconti}) the previous expression can be written
as
\beq
\label{cont}
\ZZ_c(\tau,\bar{\tau})=\frac{16}{2\pi}  \int_{-\infty}^{\infty} ds
\sum_{n,w \in \Z} \QQ(s;n,w|\tau,\bar{\tau})
\lambda^c_{\frac{1}{2}+is,\frac{n-kw}{2}}(\tau)
\lambda^c_{\frac{1}{2}+is,-\frac{n+kw}{2}}(\bar \tau)
~,
\eeq
where $\QQ(s;n,w|\tau,\bar{\tau})$ is a function given by the formula
\bea
\label{genden}
\QQ(s;n,w|\tau,\bar{\tau}) &=& \sum_{r, \bar r \in \Z}  \delta_{r-\bar r, n}\;
\frac{S_r + S_{\bar r}-1}{2 i s + r + \bar r + k w +1} \nonumber\\
 & = &
\sum_{r\in \Z} \frac{S_r-\frac{1}{2}}{2 i s + 2r + k w +1 -n}
+ \sum_{\bar r \in \Z} \frac{S_{\bar r}-\frac{1}{2}}{2 i s + 2{\bar r}
+ k w +1 +n}
\eea

Let us focus on the first term of this function. We have
\bea
\label{Sr}
& &\sum_{r\in \Z} \frac{S_r-\frac{1}{2}}{2 i s + 2r + k w +1 -n} =
\sum_{r=0}^{\infty} \frac{S_r-\frac{1}{2}}{2 i s + 2r + k w +1 -n}
+\sum_{r=-\infty}^{-1} \frac{S_r-\frac{1}{2}}{2 i s + 2r + k w +1 -n}
\nonumber\\
&=&
\sum_{r=0}^{\infty} \Bigg(\frac{S_r-\frac{1}{2}}{2 i s + 2r + k w +1 -n}
+ \frac{S_{-r-1}-\frac{1}{2}}{2 i s - 2r + k w -1 -n}\Bigg) \nonumber\\
&=&
\sum_{r=0}^{\infty} (S_r-\frac{1}{2})
\Bigg(\frac{1}{2r + 2 i s+ k w +1 -n}
+ \frac{1}{2r - 2 i s -k w +1 +n}\Bigg) ,\nonumber\\
\eea
where in the last line we made use of the identity $S_{-r-1}=1-S_r$.
For $r>0$ the leading term of $S_r$ is $\OO(1)$ and the above
sum is divergent. It can be regularized with the introduction of a damping
exponential $e^{-\epsilon r}$
\beq
\label{regulator}
\sum_{r=0}^{\infty} \frac{e^{-\epsilon r}}{r+A} = \log\epsilon -
\frac{d}{dA} \log\Gamma(A)+\OO(\epsilon).
\eeq
Hence, for the summations appearing in (\ref{Sr}) this regularization gives
\beq
\sum_{r\in \Z} e^{-\epsilon r}
\frac{S_r-\frac{1}{2}}{2 i s + 2r + k w +1 -n} \cong
\frac{1}{4} \Bigg(2\log\epsilon+\frac{1}{i}\frac{d}{ds}\log
\frac{\Gamma(-is+\frac{1}{2}+\frac{n-kw}{2})}
{\Gamma(is+\frac{1}{2}-\frac{n-kw}{2})}\Bigg).
\eeq
Similarly, for the second term of (\ref{genden}) we obtain
\beq
\sum_{\bar r\in \Z} e^{-\epsilon \bar r}
\frac{S_{\bar r}-\frac{1}{2}}{2 i s + 2{\bar r} + k w +1 +n} \cong
\frac{1}{4} \Bigg(2\log\epsilon+\frac{1}{i}\frac{d}{ds}\log
\frac{\Gamma(-is+\frac{1}{2}-\frac{n+kw}{2})}
{\Gamma(is+\frac{1}{2}+\frac{n+kw}{2})}\Bigg).
\eeq
Therefore, the leading contribution to the function $\QQ(s;n,w|\tau,\bar{\tau})$
is $\tau, \bar{\tau}$ independent and
reads
\bea
\QQ(s;n,w|\tau,\bar{\tau}) &\cong& \log\epsilon+\frac{1}{4i}\frac{d}{ds}\log
\frac{\Gamma(-is+\frac{1}{2}+\frac{n-kw}{2})
\Gamma(-is+\frac{1}{2}-\frac{n+kw}{2})}
{\Gamma(is+\frac{1}{2}-\frac{n-kw}{2})
\Gamma(is+\frac{1}{2}+\frac{n+kw}{2})}\nonumber\\
&\cong&
\log\epsilon+\frac{1}{4i}\frac{d}{ds}\log
\frac{\Gamma(-is+\frac{1}{2}+m)
\Gamma(-is+\frac{1}{2}+\bar m )}
{\Gamma(is+\frac{1}{2}-m)
\Gamma(is+\frac{1}{2}-\bar m)}
\eea
where
the $J_0^3$ and ${\bar J}_0^3$ charges
$m, \bar m$ are defined as
\beq
\label{mmnw}
m=\frac{n-kw}{2},\;\;\;
\bar m =-\frac{n+kw}{2}.
\eeq

We conclude that since $\lambda^c_{\frac{1}{2}+is,m}= \lambda^c_{\frac{1}{2}+is,-m}$
and  $\lambda^c_{\frac{1}{2}+is,m}=\lambda^c_{\frac{1}{2}-is,m}$,
the leading contribution to (\ref{cont}) can be written
in terms of the density of states $\rho(s;m,\bar m)$:
\beq
\ZZ_c(\tau,\bar{\tau})=8 \int_{0}^{\infty} ds
\sum_{n,w \in \Z} 2\rho(s;m,\bar m)
\lambda^c_{\frac{1}{2}+is,m}(\tau)
\lambda^c_{\frac{1}{2}+is,\bar m}(\bar \tau)
~,
\eeq
where
\beq
\rho(s;m,\bar m)=2\Bigg(
\frac{1}{2\pi}\log\epsilon+\frac{1}{2\pi i}\frac{d}{4ds}\log
\frac{\Gamma(-is+\frac{1}{2}-m)
\Gamma(-is+\frac{1}{2}+\bar m )}
{\Gamma(+is+\frac{1}{2}-m)
\Gamma(+is+\frac{1}{2}+\bar m)}\Bigg)
~.
\eeq
This result is in perfect agreement with \cite{Hanany:2002ev}.

To summarize, we have shown that the torus partition sum
of the coset $SL(2,\R)/U(1)$ can be written in terms of the
standard characters $\lambda^c_{\frac{1}{2}+is,m}$ and
$\lambda^d_{j,r}$ in the form
\bea
\label{torus}
\ZZ(\tau,\bar{\tau})&=&16\sum_{n, w \in \Z} \int_0^{\infty}
ds ~ \rho(s;m,\bar m) \lambda^c_{\frac{1}{2}+is,m}(\tau)
\lambda^c_{\frac{1}{2}+is,\bar m}(\bar{\tau})
\nonumber\\
&+&8\sum_{\frac{1}{2}<j<\frac{k-1}{2}}
\sum_{w,r,\bar r \in \Z} \delta(2j+r+\bar r+kw)
\lambda^d_{j,r}(\tau) \lambda^d_{j,\bar r}(\bar{\tau})
~.
\eea
Notice that by keeping only the $\tau,\bar \tau$-independent part of
the function (\ref{genden}), we discard a set of states from
(\ref{cont}) whose interpretation is unclear \cite{Israel:2004ir},
as they do not seem to fit in any of the known $SL(2,\R)/U(1)$ representations.

For integer level $k$ it is possible to rewrite (part of) the above partition sum
in terms of the extended coset characters. This is actually quite straightforward
for the discrete part. Starting from
(\ref{tordiscrete}) we can write
\bea
\label{discreteextended}
\ZZ_d(\tau,\bar{\tau})&=& 8 \sum_{j=\frac{1}{2}}^{\frac{k-1}{2}}
\sum_{w,n,\bar n\in \Z} \sum_{r,\bar r\in \Z_k}
\delta(2j+r+\bar r+k(n+\bar n+w)) \lambda^d_{j,r+kn}(\tau)
\lambda^d_{j,\bar r+k\bar n}(\bar{\tau}) \nonumber\\
&=&8 \sum_{j=\frac{1}{2}}^{\frac{k-1}{2}}
\sum_{w,n,\bar n\in \Z} \sum_{r,\bar r\in \Z_k}
\delta(2j+r+\bar r+kw) \lambda^d_{j,r+kn}(\tau)
\lambda^d_{j,\bar r+k\bar n}(\bar{\tau}) \nonumber\\
&=&8 \sum_{j=\frac{1}{2}}^{\frac{k-1}{2}}
\sum_{w\in \Z} \sum_{r,\bar r\in \Z_k}
\delta(2j+r+\bar r+kw) \Lambda^d_{j,r}(\tau)
\Lambda^d_{j,\bar r}(\bar{\tau})
~.
\eea

The continuous contribution, on the other hand, involves a series of subtle steps.
Initially, in eqs.\ (\ref{cont}) and (\ref{genden}) we can set
$n-kw=2l_1 k + g$, and $n+kw= 2 l_2 k + g$ with $l_1,
l_2 \in \Z$ and $g \in Z_{2k}$.
Then, by shifting $r\rightarrow r+l_1 k$ the first term in (\ref{genden})
can be written as
\bea
\label{r1}
\sum_{r\in \Z} \frac{S_r-\frac{1}{2}}{2 i s + 2r + k w +1 -n} & = &
\sum_{r\in \Z}
 \frac{S_{r+l_1 k} -\frac{1}{2}}{2 i s + 2r +1 -g}.
\eea
In a similar way, the second term gives
\bea
\label{r2}
\sum_{\bar r \in \Z} \frac{S_{\bar r}-\frac{1}{2}}{2 i s + 2 \bar r + k w +1 +n} & = &
\sum_{\bar r \in \Z}
 \frac{S_{{\bar r}+l_2 k} -\frac{1}{2}}{2 i s + 2 \bar r +1 +g}.
\eea

Recall that in the treatment of the standard characters we kept
only the constant, $\tau, \bar{\tau}$ independent, terms in
$S_r-\frac{1}{2}$. A similar approach will not work for the
extended characters because of the following reason. From the expansion of the
functions $S_r$ in (\ref{Sr}) it is clear that the constant term
of $S_r$ is dependent on the sign of $r$. More specifically, for $r>0$
the constant term arises only from the $s=0$ term in the expansion
and it gives $S_r-\frac{1}{2} \sim \frac{1}{2}$. For $r<0$
the constant term arises from two terms with opposite signs, $s=0$ and $s=-2r-1$. So,
it gives $S_r-\frac{1}{2} \sim 1-1-\frac{1}{2}=-\frac{1}{2}$.
This makes the r.h.s.\ of eqs.\ (\ref{r1}) and (\ref{r2}) $l_1$-
and $l_2$-dependent and spoils the decomposition of the partition
sum in terms of the continuous extended characters $\Lambda^c_{\frac{1}{2}+is,m}$.
This complication does not arise for the part of the contributions
(\ref{r1}) and (\ref{r2}) that comes only from the $s=0$ term in the expansion of
the functions $S_{r+kl_1}$ and $S_{r+kl_2}$. This part is $\tau, \bar{\tau}$-
and $l_1, l_2$-independent. As before, it gives divergent terms
and the sums over $r$ and $\bar r$ should be regularized. With the regulator
appearing in eq.\ (\ref{regulator}) we can obtain
\beq
\sum_{r\in \Z} e^{-\epsilon r}
\frac{S_{r+l_1 k}-\frac{1}{2}}{2 i s + 2r +1 -g } \bigg|_{s=0} \cong
\frac{1}{4} \Bigg(2\log\epsilon+\frac{1}{i}\frac{d}{ds}\log
\frac{\Gamma(-is+\frac{1}{2}+\frac{g}{2})}
{\Gamma(is+\frac{1}{2}-\frac{g}{2})}\Bigg)
\eeq
and
\beq
\sum_{\bar r\in \Z} e^{-\epsilon \bar r}
\frac{S_{\bar r+ l_2 k}-\frac{1}{2}}{2 i s + 2{\bar r} +1 +g} \bigg|_{s=0} \cong
\frac{1}{4} \Bigg(2\log\epsilon+\frac{1}{i}\frac{d}{ds}\log
\frac{\Gamma(-is+\frac{1}{2}-\frac{g}{2})}
{\Gamma(is+\frac{1}{2}+\frac{g}{2})}\Bigg).
\eeq
We have inserted the notation $\big |_{s=0}$ to keep track of the
fact that we keep only the $s=0$ constant
term in the expansions of the $S_{r+k l_1}$ and $S_{r+kl_2}$ functions.
For these contributions the corresponding terms in the torus partition sum (\ref{cont})
give
\beq
\label{conts0}
\ZZ_c^{s=0}(\tau,\bar{\tau})=8\int_{0}^{\infty} ds
 \sum_{g \in \Z_{2k}} 2\rho(s;g)
\Lambda^c_{\frac{1}{2}+is,\frac{g}{2}}(\tau)
 \Lambda^c_{\frac{1}{2}+is,-\frac{g}{2}}(\bar \tau)
\eeq
with spectral density
\beq
\rho(s;g)=2\Bigg(
\frac{1}{2\pi}\log\epsilon+\frac{1}{2\pi i}\frac{d}{4ds}\log
\frac{\Gamma(-is+\frac{1}{2}-\frac{g}{2})
\Gamma(-is+\frac{1}{2}-\frac{g}{2} )}
{\Gamma(+is+\frac{1}{2}-\frac{g}{2})
\Gamma(+is+\frac{1}{2}-\frac{g}{2})}\Bigg) .
\eeq
The remaining contributions in the torus partition sum cannot be written in terms
of the extended characters. Similar partition function
decompositions in terms of extended coset characters for
the supersymmetric $SL(2,\R)/U(1)$
were obtained in \cite{Israel:2004ir}.

\section{Derivation of the $\Lambda^I_r(\tau)$
$\SS$-modular transformation}\label{modid}

The authors of \cite{Israel:2004xj} discussed the coset extended characters
for the continuous and discrete representations.
The identity representation does not appear in the closed string Hilbert space, but it
does appear in the open string sector. In this Appendix we discuss the
corresponding extended characters, and determine how they transform under
the $\SS$-transformation $\tau \rightarrow -\frac{1}{\tau}$. Our computation
starts with the known modular transformation properties of the corresponding
$\NN=2$ extended characters $\chi_I(r;\tau)$ and utilizes the
character decompositions (\ref{idddd}), (\ref{disccosetdecom}) and
(\ref{conticosetdecom}) that appear in Section 2.

The $\NN=2$ identity characters transform in the following
manner \cite{Eguchi:2003ik}:
\bea
\label{eq:twosix}
\chi_I(r;-\frac{1}{\tau},\frac{z}{\tau})&=& e^{\pi i \hat c z^2/\tau}
\bigg [ \sum_{j' \in \Zi_{2(k-2)}} \int_0^{\infty} dp' \SS^{p',j'}_r
\chi_c(p',j';\tau,z)
\nonumber\\
&+& \sum_{r'\in \Zi_{k-2}} \sum_{s'=2}^{k-2} \SS^{r',s'}_r
\chi_d(r',s';\tau,z) \bigg ] ~.
\eea
In this expression, we have set
\bea
\label{eq:oneseventhree}
\SS^{p',j'}_r&=& \frac{1}{\sqrt{2(k-2)}}e^{-2\pi i r j'/(k-2)} \frac{\sinh(\pi \QQ p')
\sinh (2\pi p'/\QQ)}{\big | \cosh \pi \big ( \frac{p'}{\QQ}+i \frac{j'}{2} \big ) \big |^2}
~ , \\
\label{eq:onesevenfour}
\SS^{r',s'}_r &=& \frac{2}{k-2} \sin \frac{\pi (s'-1)}{k-2} e^{-2\pi i
\frac{r(s'+2r')}{k-2}}
~,
\eea
with $\QQ = \sqrt{\frac{2}{k-2}}$. The substitution
of the character expansions of Section 2
in (\ref{eq:twosix}) gives (for simplicity we set $z=0$)
\bea
\label{eq:threefour}
\frac{1}{\sqrt{k(k-2)}}& &\sum_{n\in \Zi} \sum_{m' \in \Zi_{k(k-2)}}
e^{-2\pi i m' \frac{n(k-2)+2r}{k(k-2)}} \Theta_{m',\frac{k(k-2)}{2}}(\tau)
\lambda^I_{-n+r}(-\frac{1}{\tau})
\nonumber\\
&=&\sum_{n\in\Zi}\sum_{j'\in \Zi_{2(k-2)}} \int_0^{\infty} dp' \SS^{p',j'}_r
\Theta_{n(k-2)+j',\frac{k(k-2)}{2}}(\tau) \lambda^c_{1/2+is(p'),n-j'/2}(\tau)
\nonumber\\
&+& \sum_{n\in\Zi}\sum_{r'\in \Zi_{k-2}}\sum_{s'=2}^{k-2} \SS^{r',s'}_r
\Theta_{n(k-2)+s'+2r',\frac{k(k-2)}{2}}(\tau) \lambda^d_{s'/2,-n+r'}(\tau)
~.
\eea
Next we find the linearly independent terms. On the l.h.s.\ of (\ref{eq:threefour})
this can be achieved easily by setting $n=-k \tilde n+p$, with $\tilde n \in \Zi$
and $p\in \Zi_k$. On the r.h.s.\ we have two terms. For the first term, involving
the continuous characters, again set $n=k\tilde n+p$, with $\tilde n \in \Zi$
and $p\in \Zi_k$ and split $j'$ as $j'=(k-2)q'+t$, where $q'\in \Zi_2$ and
$t\in \Zi_{k-2}$. Then shift $p+q' \rightarrow p$. For the moment
we leave the second term as it is.
In this way, eq.\ (\ref{eq:threefour}) takes the form
\bea
\label{eq:tfour}
& &\frac{1}{\sqrt{k(k-2)}} \sum_{\tilde n\in \Zi}\sum_{p\in \Zi_k} \sum_{m'\in \Zi_{k(k-2)}}
e^{-2\pi i m'\frac{(k-2)p+2r}{k(k-2)}} \Theta_{m',\frac{k(k-2)}{2}}(\tau)
\lambda^I_{k\tilde n+r-p}\big (-\frac{1}{\tau}\big)
\nonumber\\
& &=\sum_{\tilde n \in \Zi} \bigg[ \sum_{\tilde p\in \Zi_k}
 \sum_{q'\in \Zi_2}\sum_{t\in \Zi_{k-2}}\int_0^{\infty} dp' \SS^{p',(k-2)q'+t}_r
\Theta_{\tilde p(k-2)+t,\frac{k(k-2)}{2}}(\tau)\lambda^c_{\frac{1}{2}+is(p'),k\tilde n+
\tilde p-\frac{k}{2}q'-\frac{t}{2}}(\tau)
\nonumber\\
& &+ \sum_{p'\in \Zi_k}
\sum_{r'\in \Zi_{k-2}}\sum_{s'=2}^{k-2}
\SS^{r',s'}_r
\Theta_{p'(k-2)+s'+2r',\frac{k(k-2)}{2}}(\tau)
\lambda^d_{\frac{s'}{2},k\tilde n-p'+
+r'}(\tau)\bigg ]
~.
\nonumber\\
\eea
We notice that the sums over $\tilde n$ convert
the unextended characters to extended ones and they give
\bea
\label{eq:tfive}
& &\frac{1}{\sqrt{k(k-2)}}
\sum_{p\in \Zi_k} \sum_{m'\in \Zi_{k(k-2)}}
e^{-2\pi i m'\frac{(k-2)p+2r}{k(k-2)}} \Theta_{m',\frac{k(k-2)}{2}}(\tau)
\Lambda^I_{r-p}\big (-\frac{1}{\tau}\big)
\nonumber\\
& &= \sum_{\tilde p\in \Zi_k}
 \sum_{q'\in \Zi_2}\sum_{t\in \Zi_{k-2}}\int_0^{\infty} dp' \SS^{p',(k-2)q'+t}_r
\Theta_{\tilde p(k-2)+t,\frac{k(k-2)}{2}}(\tau)
\Lambda^c_{\frac{1}{2}+is(p'),\tilde p-\frac{k}{2}q'-\frac{t}{2}}(\tau)
\nonumber\\
& &+ \sum_{p'\in \Zi_k}
\sum_{r'\in \Zi_{k-2}}\sum_{s'=2}^{k-2}
\SS^{r',s'}_r
\Theta_{p'(k-2)+s'+2r',\frac{k(k-2)}{2}}(\tau)
\Lambda^d_{\frac{s'}{2},-p'+r'}(\tau)
~.
\nonumber\\
\eea

Now we can fix $m'$ and equate those terms that multiply the same
$\Theta$-functions. From the second line in (\ref{eq:tfive})
this entails setting $m'=\tilde p(k-2)+t$.
In the third line, we keep all the terms
with $p'(k-2)+s'+2r'=m'=\tilde p (k-2)+t \;\;\mod
\;\;k(k-2)$.
Hence, we get
\bea
\label{eq:tsix}
& &\frac{1}{\sqrt{k(k-2)}}
\sum_{p\in \Zi_k}
e^{-2\pi i m'\frac{p}{k}}
\Lambda^I_{r-p}\big (-\frac{1}{\tau}\big)
\nonumber\\
& &= e^{2\pi i m'\frac{2r}{k(k-2)}} \Bigg(
 \sum_{q'\in \Zi_2}
\int_0^{\infty} dp' \SS^{p',(k-2)q'+t}_r
\Lambda^c_{\frac{1}{2}+is(p'),\tilde p-\frac{k}{2}q'-\frac{t}{2}}(\tau)
\nonumber\\
& &+ \sum_{p'\in \Zi_k}
\sum_{r'\in \Zi_{k-2}}\sum_{s'=2}^{k-2}\sum_{n \in \Zi}
\SS^{r',s'}_r
\delta_{p'(k-2)+s'+2r',m'+n k (k-2)}
\Lambda^d_{\frac{s'}{2},-p'+r'}(\tau) \Bigg)
~.
\nonumber\\
\eea
In this expression, we use the standard
Kronecker delta and the extra summation over an integer $n$ takes
into account the $k(k-2)$-periodicity of the $\Theta$ functions.
We have also moved to the right the part of the exponential that depends on $r$.

Now, set $m'=\tilde p k +t-2\tilde p = \tilde p k+\sigma$
and sum over $\sigma \in \Zi_k$. In the left side this keeps $k$ copies of
the term with $p=0$.
Plugging in the explicit form of the modular $\SS$-matrices for the $\NN=2$
characters  (\ref{eq:oneseventhree}), (\ref{eq:onesevenfour}):
\beq
\label{eq:ssigma}
\SS^{p',(k-2)q'+t}_r=\frac{1}{\sqrt{2(k-2)}} e^{-2\pi i \frac{r t}{k-2}}
\frac{\sinh(\pi \QQ p') \sinh(2\pi p'/\QQ)}{\big | \cosh\pi( \frac{p'}{\QQ}
+i\frac{(k-2)q'+t}{2}) \big |^2} ~,
\eeq
\beq
\label{eq:smsigma}
\SS^{r',s'}_r=\frac{2}{k-2} \sin \frac{\pi (s'-1)}{k-2} e^{-2\pi i r \frac{s'+2r'}{k-2}}
~ .
\eeq
we see that in the part corresponding to the continuous characters at the
right,
the ${\tilde p}$ dependence drops out. So far we have obtained the equation
\bea
\label{eq:tseven}
& &\frac{k}{\sqrt{k(k-2)}}
\Lambda^I_{r}\big (-\frac{1}{\tau}\big) = \nonumber\\
& &
 \sum_{q'\in \Zi_2} \sum_{\sigma \in \Zi_k}
\int_0^{\infty} dp'
\frac{1}{\sqrt{2(k-2)}} e^{-2 \pi i \frac{r\sigma}{k}}
\frac{\sinh(\pi \QQ p') \sinh(2\pi p'/\QQ)}{\big | \cosh\pi( \frac{p'}{\QQ}
+i\frac{kq'+\sigma}{2}) \big|^2}
\Lambda^c_{\frac{1}{2}+is(p'),-\frac{\sigma+k q'}{2}}(\tau) +
\nonumber\\
& & \sum_{\sigma\in \Zi_k}
\sum_{p'\in \Zi_k}
\sum_{r'\in \Zi_{k-2}}\sum_{s'=2}^{k-2}\sum_{n \in \Zi}
 e^{2\pi i m'\frac{2r}{k(k-2)}} \SS^{r',s'}_r
\delta_{p'(k-2)+s'+2r',\tilde p k +\sigma +n k (k-2)}
\Lambda^d_{\frac{s'}{2},-p'+r'}(\tau)
~.
\nonumber\\
\eea
With the definition $l=\sigma+k q' \in \Zi_{2k}$ we can recast it into the form
\bea
\label{eq:teight}
& &\Lambda^I_{r}\big (-\frac{1}{\tau}\big) =
\frac{1}{\sqrt{2k}} \sum_{l \in Z_{2k}}
\int_0^{\infty} dp'
e^{-2 \pi i \frac{rl }{k}}
\frac{\sinh(\pi \QQ p') \sinh(2\pi p'/\QQ)}{\big | \cosh\pi( \frac{p'}{\QQ}
+i\frac{l}{2}) \big|^2}
\Lambda^c_{\frac{1}{2}+is(p'),-\frac{l}{2}}(\tau) +
\nonumber\\
& & \frac{2}{\sqrt{k(k-2)}}
\sum_{\sigma\in \Zi_k}
\sum_{p'\in \Zi_k}
\sum_{r'\in \Zi_{k-2}}\sum_{s'=2}^{k-2}\sum_{n \in \Zi}
\sin \frac{\pi (s'-1)}{k-2}  e^{2\pi i (\tilde p k + \sigma) \frac{2r}{k(k-2)}}
e^{-2\pi i r \frac{s'+2r'}{k-2}} \nonumber\\
& &
\delta_{p'(k-2)+s'+2r',\tilde p k +\sigma +n k (k-2)}
\Lambda^d_{\frac{s'}{2},-p'+r'}(\tau)
~.
\nonumber\\
\eea

We can further manipulate the second term at the right as follows.
For simplicity, we may  fix $\tilde p =0$. From the Kronecker delta we deduce
that the charge $-p'+r'$ in the character can be replaced by
$(\sigma-p'k-s')/2$ (remember that the extended characters have periodicity $k$).
It is crucial that $k$ is even and that the term $n k (k-2)$ is
an even multiple of $k$. The exponents can be treated in a similar manner.
Neglecting, for the moment, some multiplicative constants, the second term
in (\ref{eq:teight}) can be written as
\beq
\sum_{\sigma\in \Zi_k}
\sum_{p'\in \Zi_k}
\sum_{r'\in \Zi_{k-2}}\sum_{s'=2}^{k-2}\sum_{n \in \Zi}
\sin \frac{\pi (s'-1)}{k-2} e^{-2 \pi i \frac{r \sigma}{k}}
\delta_{p'(k-2)+s'+2r'+nk(k-2),\sigma}
\Lambda^d_{\frac{s'}{2},(\sigma-p'k-s')/2}(\tau)
~.
\eeq
Since $k$ is even, we set $k=2l$ and we may write
$p'(k-2)+2r'+nk(k-2)=2(n 2l(l-1)+p'(l-1)+r')$. Furthermore, since
$p' \in \Z_k$ we can write $p'=2n'+q$ with $n' \in \Z_l, q \in \Z_2$
and eventually we get $p'(k-2)+2r'+nk(k-2)=2m+2q(l-1)$ with $m\in\Z$.
The summations over $n, n'$ and $r'$ can be replaced by a single summation
over the integer $m$ and then we get
\beq
\sum_{\sigma\in \Zi_k} \sum_{s'=2}^{k-2} \sum_{q\in \Zi_2}
\sum_{m\in\Zi}
\sin \frac{\pi (s'-1)}{k-2} e^{-2 \pi i \frac{r \sigma}{k}}
\delta_{2m+2q(l-1)+s',\sigma}
\Lambda^d_{\frac{s'}{2},(\sigma-qk-s')/2}(\tau)
~.
\eeq
Because of the Kronecker delta, which implies $\sigma=2m+qk-2q+s'$, this can
also be written as
\beq
\sum_{\sigma\in \Zi_k} \sum_{s'=2}^{k-2} \sum_{q\in \Zi_2}
\sum_{m\in\Zi}
\sin \frac{\pi (s'-1)}{k-2} e^{-2 \pi i \frac{r (2m-2q+s')}{k}}
\delta_{2m+2q(l-1)+s',\sigma}
\Lambda^d_{\frac{s'}{2},m-q}(\tau)
~.
\eeq
By shifting $m \rightarrow m+q$
and then decomposing it as $m=n k + t$, with $n\in\Z,t\in\Z_k$ we find
\beq
\sum_{\sigma\in \Zi_k} \sum_{s'=2}^{k-2} \sum_{q\in \Zi_2}
\sum_{t\in \Zi_k} \sum_{n\in\Zi}
\sin \frac{\pi (s'-1)}{k-2}  e^{-2 \pi i \frac{r (2t+s')}{k}}
\delta_{2nk+2t+qk+s',\sigma}
\Lambda^d_{\frac{s'}{2},t}(\tau)
~.
\eeq
Finally, notice that $2nk+qk-\sigma$ recombines into a new integer
$m \in \Z$ and the summations over $q, \sigma$ and $n$
can be substituted with a single summation over $m$:
\beq
 \sum_{s'=2}^{k-2} \sum_{t\in \Zi_k} \sum_{m\in\Zi}
\sin \frac{\pi (s'-1)}{k-2}  e^{-2 \pi i \frac{r (2t+s')}{k}}
\delta_{2t+s'+m,0}
\Lambda^d_{\frac{s'}{2},t}(\tau)
~.
\eeq
For each pair $(t,s')$ there is always a unique integer
$m$ such that $2t+s'+m=0$. Hence, we can just drop the delta function and
the summation over $m$ and the final result reads
\beq
 \sum_{s'=2}^{k-2} \sum_{t\in \Zi_k}
\sin \frac{\pi (s'-1)}{k-2}  e^{-2 \pi i \frac{r (2t+s')}{k}}
\Lambda^d_{\frac{s'}{2},t}(\tau)
\eeq

Recombining everything in (\ref{eq:teight}) gives
the modular identity
\bea
\label{eq:teliki}
& &\Lambda^I_{r}\big (-\frac{1}{\tau}\big) =
\frac{1}{\sqrt{2k}} \sum_{l \in Z_{2k}}
\int_0^{\infty} dp'
e^{-2 \pi i \frac{rl }{k}}
\frac{\sinh(\pi \QQ p') \sinh(2\pi p'/\QQ)}{\big | \cosh\pi( \frac{p'}{\QQ}
+i\frac{l}{2}) \big|^2}
\Lambda^c_{\frac{1}{2}+is(p'),-\frac{l}{2}}(\tau) +
\nonumber\\
& & \frac{2}{\sqrt{k(k-2)}}
 \sum_{s'=2}^{k-2} \sum_{t\in \Zi_k}
\sin \frac{\pi (s'-1)}{k-2}  e^{-2 \pi i \frac{r (2t+s')}{k}}
\Lambda^d_{\frac{s'}{2},t}(\tau)
~.
\nonumber\\
\eea

\section{Explicit expressions for the extended characters $\Lambda^I_r$}
\label{lambdag}

In this Appendix we derive an explicit expression for the extended coset character
$\Lambda^I_r$. This is needed for the identification of various open string
spectra. Our derivation starts from the defining equation (\ref{idcosetdecom}):
\beq
\label{D1}
\chi_{I}(r;\tau)=\eta^{-1}(\tau) \sum_{n\in \Z} q^{\frac{k-2}{2k}(\frac{2r}{k-2}+n)^2}
\Lambda^I_{-n+r}(\tau)
~.
\eeq
To proceed we could either write down $\chi_I$ explicitly
and then expand the denominators using the geometric series expansion, or
we could use an identity that relates $\chi_I$ with the continuous and
discrete characters and then deduce the expansion of the identity character
$\Lambda^I_r$ from the known expansions of the coset characters
$\Lambda^c_{\frac{1}{2}+is,\frac{m}{2}}$ and $\Lambda^d_{j,r}$.
We follow the latter approach\footnote{A quick
derivation of the identity extended characters $\Lambda^I_r$ also follows directly
from the expression of the standard identity character, see eq.\ (\ref{charid}) in Section 2.
The alternative derivation of this Appendix gives a further consistency
check of our formulae.} .

The $\NN=2$ character identity we would like to use can be found in
eq.\ (2.24) of \cite{Eguchi:2003ik}. It has the form:
\beq
\label{D2}
\chi_I(r;\tau)=\chi_c(\pm\frac{i}{2},2r;\tau)-\chi_d(r,k-2;\tau)-
\chi_d(r-1,2;\tau)
~.
\eeq
The $\pm$ values in this expression
make no difference in the final result.
Plugging on the r.h.s.\ the expansions (\ref{disccosetdecom}) and (\ref{conticosetdecom})
we find the coset character identity
\beq
\label{D3}
\Lambda^I_r(\tau)=\Lambda^c_{\frac{1}{2}\mp \frac{1}{2},-r}(\tau)-
\Lambda^d_{\frac{k-2}{2},r+1}(\tau)-\Lambda^d_{1,r-1}(\tau)
~.
\eeq
The next step is to use the explicit expansions of the continuous and
discrete characters that appear in \cite{Israel:2004xj}. They have the
following form:
\beq
\label{D4}
\Lambda^c_{\frac{1}{2}+is,\frac{m}{2}}=\eta^{-2}(\tau) q^{-\frac{1}{4(k-2)}}
\sum_{l\in \Z} q^{k(l+\frac{m}{2k})^2}
~,
\eeq
\beq
\label{D5}
\Lambda^d_{j,r}(\tau)=\eta^{-2}(\tau) q^{-\frac{(j-\frac{1}{2})^2}{k-2}}
\sum_{l\in \Z} q^{\frac{(j+r+kl)^2}{k}} S_{r+kl}(\tau)
~,
\eeq
where
\beq
\label{D6}
S_r(\tau)=\sum_{s=0}^{\infty} (-1)^s q^{\frac{1}{2}s(s+2r+1)}
~.
\eeq
Plugging them into (\ref{D3}) we find:
\beq
\label{D7}
\Lambda^I_r(\tau)= \eta^{-2}(\tau) q^{-\frac{1}{4(k-2)}}
\sum_{l \in \Z} q^{\frac{1}{k}(r+kl)^2} \bigg ( 1- q^{r+kl+1} S_{r+kl+1}(\tau)
- S_{r+kl-1}(\tau) \bigg )
~. \nonumber\\
\eeq
By using the identities
\beq
\label{D8}
q^{r}S_r(\tau)=S_{-r}(\tau)
~,
\eeq
and
\beq
\label{D9}
S_{-r}(\tau)=1-S_{r-1}(\tau)
~
\eeq
we can write
\beq
\label{D10}
\Lambda^I_r(\tau)= \eta^{-2}(\tau) q^{-\frac{1}{4(k-2)}}
\sum_{l \in \Z} q^{\frac{1}{k}(r+kl)^2} \bigg (S_{-r-kl}(\tau)-S_{-r-kl-1}(\tau)
\bigg)
~.
\eeq
Finally, after summing over $r$ we obtain
\bea
\label{D11}
\sum_{r\in \Z_k} \Lambda^I_r(\tau)&=& \eta^{-2}(\tau) q^{-\frac{1}{4(k-2)}}
\sum_{n\in \Z} q^{\frac{1}{k}n^2} \bigg (S_n(\tau)-S_{n-1}(\tau)\bigg)
~.
\eea

\end{document}